%% file: main.tex
\PassOptionsToPackage{numbers,compress}{natbib}
\PassOptionsToPackage{table}{xcolor}
\documentclass{article}
\usepackage{latexml}
\iflatexml
  \input{html-layout}
\else
  \usepackage[singlecolumn]{simple}
\fi
\input{manuscript-layout}
\input{preamble}
\usepackage{hyperref}
\iflatexml\else\input{navigation}\fi
\title{ExtremePhysBench: Do AI Weather Models Capture the Physical Structure of Extreme Weather Events?}
\newcommand{\PaperAuthorList}{%
  \PaperAuthorName{Yujie Wang}\textsuperscript{1},
  \PaperAuthorName{Yang Liu}\textsuperscript{2},
  \PaperAuthorName{Guowen Li}\textsuperscript{3},
  \PaperAuthorName{Haoyuan Liang}\textsuperscript{3},
  \PaperAuthorName{Shilei Cao}\textsuperscript{1},
  \PaperAuthorName{Zhaocheng Yang}\textsuperscript{1},
  \PaperAuthorName{Junqi Hu}\textsuperscript{1},
  \PaperAuthorName{Juepeng Zheng}\textsuperscript{1,4,\S},
  \PaperAuthorName{Hong Cheng}\textsuperscript{2},
  \PaperAuthorName{Yutong Lu}\textsuperscript{1,4},
  \PaperAuthorName{Haohuan Fu}\textsuperscript{3,4}%
}
\newcommand{\PaperAffiliationList}{%
  \textsuperscript{1}Sun Yat-sen University\\
  \textsuperscript{2}The Chinese University of Hong Kong\\
  \textsuperscript{3}Tsinghua University\\
  \textsuperscript{4}National Supercomputing Center in Shenzhen%
}
\newcommand{\PaperAuthorNotes}{%
  \textsuperscript{\S}Corresponding author: \href{mailto:zhengjp8@mail.sysu.edu.cn}{\textcolor{black}{\texttt{zhengjp8@mail.sysu.edu.cn}}}%
}
\author{%
  \PaperAuthorList\\[6pt]
  {\small\PaperAffiliationList\endgraf}
  \smallskip
  {\footnotesize\PaperAuthorNotes\endgraf}}
\date{}
\begin{document}
\maketitle
\input{sections/00_abstract}
\PaperHighlight{%
  Similar field errors can mask substantial differences in physical structure.
  In TC MEARI, two forecasts differ by at most 10\% across each of 15 field RMSEs, yet their Event Physics Scores differ by 9.51 points.
  Physical diagnostics reveal which processes account for such discrepancies, turning an overall accuracy comparison into a more targeted assessment of model strengths and weaknesses.
  This supports evaluating how well a forecast reproduces atmospheric variables relationships, alongside how closely its individual fields match the reference.%
}

\input{sections/01_introduction}

\input{sections/02_related_work}

\input{sections/03_dataset}

\input{sections/04_benchmark}
\input{sections/05_experimental_setup}
\input{sections/06_results}
\input{sections/07_discussion}

\input{sections/08_statements}
\bibliographystyle{ieee_fullname}
\bibliography{ExtremePhysBench}
\PaperAppendix
\PaperAppendixTOC
\input{appendices/A_related_frameworks}
\input{appendices/B_dataset_details}

\input{appendices/C_inventory_tables}
\input{appendices/D_physics_metrics}
\input{appendices/E_forecast_systems}
\input{appendices/F_additional_results}
\end{document}

%% file: html-layout.tex
\usepackage{amsmath,amssymb,graphicx,xcolor,booktabs,enumitem,etoolbox}
\usepackage[super,sort&compress,numbers]{natbib}

%% file: manuscript-layout.tex
\usepackage{etoolbox,placeins}
\makeatletter
\newcommand{\PaperAuthorName}[1]{\mbox{\@PaperNameSpaces#1 \@nil}}
\def\@PaperNameSpaces#1 #2\@nil{%
  #1\def\@PaperNameRest{#2}%
  \ifx\@PaperNameRest\@empty\else
    \nobreakspace\@PaperNameSpaces#2\@nil
  \fi}
\makeatother
\newcommand{\PaperHighlight}[1]{\par\noindent\textbf{Key finding.} #1\par\medskip}
\iflatexml
  
  \newcommand{\PaperAppendix}{\appendix}
  \newcommand{\PaperAppendixTOC}{\section*{Appendix Contents}\input{html-appendix-contents}}
\else
  \AtBeginDocument{%
    \pretocmd{\section}{\FloatBarrier}{}{}%
  }
  \titleformat{\paragraph}[runin]{\normalsize\bfseries}{}{0pt}{}
  \titleformat{\section}{\large\bfseries}{\thesection}{0.6em}{}
  \titleformat{\subsection}{\normalsize\bfseries}{\thesubsection}{0.6em}{}
  
  \makeatletter
  \newcommand{\PaperAppendix}{\PaperAppendixContentstrue\appendix}
  \makeatother
\fi

%% file: html-appendix-contents.tex
\begin{description}
\item[\ref{app:related-work-comparison}] \hyperref[app:related-work-comparison]{Comparison with Related Evaluation Frameworks}
\item[\ref{app:dataset-details}] \hyperref[app:dataset-details]{Extreme Weather Events Dataset Details}
\begin{description}
\item[\ref{htmltoc:b-dataset-details-1}] \hyperref[htmltoc:b-dataset-details-1]{Source Data and Common Conventions}
\item[\ref{htmltoc:b-dataset-details-2}] \hyperref[htmltoc:b-dataset-details-2]{Gridded Event Construction}
\item[\ref{htmltoc:b-dataset-details-3}] \hyperref[htmltoc:b-dataset-details-3]{Landfalling Tropical Cyclones}
\item[\ref{htmltoc:b-dataset-details-4}] \hyperref[htmltoc:b-dataset-details-4]{Catalog and Task Records}
\item[\ref{htmltoc:b-dataset-details-5}] \hyperref[htmltoc:b-dataset-details-5]{Catalog Summary}
\end{description}
\item[\ref{app:benchmark-inventory}] \hyperref[app:benchmark-inventory]{Benchmark Inventory, Scales, and Aggregation}
\begin{description}
\item[\ref{app:diagnostic-inventory}] \hyperref[app:diagnostic-inventory]{Diagnostic and Task Inventory}
\item[\ref{app:score-scales}] \hyperref[app:score-scales]{Fixed Physics Score Scales}
\item[\ref{app:score-aggregation}] \hyperref[app:score-aggregation]{Explicit Hierarchical Estimator}
\end{description}
\item[\ref{app:physics-metrics}] \hyperref[app:physics-metrics]{Diagnostic Operators and Error Geometry}
\begin{description}
\item[\ref{htmltoc:d-physics-metrics-1}] \hyperref[htmltoc:d-physics-metrics-1]{Notation and Spherical Operators}
\item[\ref{app:derivations}] \hyperref[app:derivations]{From Governing Relations to Observable Physical Structure}
\item[\ref{htmltoc:d-physics-metrics-2}] \hyperref[htmltoc:d-physics-metrics-2]{Reference-Only Support}
\item[\ref{htmltoc:d-physics-metrics-3}] \hyperref[htmltoc:d-physics-metrics-3]{Universal Diagnostics U1--U10}
\item[\ref{htmltoc:d-physics-metrics-4}] \hyperref[htmltoc:d-physics-metrics-4]{Thermal-Stress Diagnostics E1--E5}
\item[\ref{htmltoc:d-physics-metrics-5}] \hyperref[htmltoc:d-physics-metrics-5]{Precipitation Diagnostics E1--E5}
\item[\ref{htmltoc:d-physics-metrics-6}] \hyperref[htmltoc:d-physics-metrics-6]{Tropical-Cyclone Diagnostics E1--E5}
\item[\ref{app:diagnostic-error-geometries}] \hyperref[app:diagnostic-error-geometries]{Diagnostic Error Geometries}
\end{description}
\item[\ref{app:forecast-systems}] \hyperref[app:forecast-systems]{Forecast Systems and Inference Details}
\begin{description}
\item[\ref{htmltoc:e-forecast-systems-1}] \hyperref[htmltoc:e-forecast-systems-1]{Earth2Studio Execution Layer}
\item[\ref{htmltoc:e-forecast-systems-2}] \hyperref[htmltoc:e-forecast-systems-2]{Initial Conditions and Forecast Outputs}
\end{description}
\item[\ref{app:additional-results}] \hyperref[app:additional-results]{Additional Experimental Results}
\begin{description}
\item[\ref{app:rmse-results}] \hyperref[app:rmse-results]{Conventional RMSE Comparison}
\item[\ref{app:aggregate-model-comparisons}] \hyperref[app:aggregate-model-comparisons]{Aggregate Model Comparisons}
\item[\ref{app:event-comparisons}] \hyperref[app:event-comparisons]{Event Distributions and Paired Comparisons}
\item[\ref{app:diagnostic-results}] \hyperref[app:diagnostic-results]{Physical Structure Profiles by Diagnostic}
\item[\ref{app:lead-results}] \hyperref[app:lead-results]{Detailed Views by Forecast Lead}
\item[\ref{app:gfs-reference}] \hyperref[app:gfs-reference]{Sensitivity to the GFS Verifying Analysis}
\item[\ref{app:score-sensitivity}] \hyperref[app:score-sensitivity]{Sensitivity to Score Mapping and Aggregation}
\item[\ref{app:case-studies}] \hyperref[app:case-studies]{Case Studies: Field Errors and Physical Structure}
\end{description}
\end{description}

%% file: preamble.tex
\input{math_commands.tex}

\usepackage{xurl}

\usepackage{mathtools}
\usepackage{longtable,array,tabularx}
\usepackage{siunitx}
\usepackage{threeparttable}
\usepackage{makecell,multirow}
\usepackage{float}
\usepackage{tikz}
\usetikzlibrary{arrows.meta,positioning,calc}

\allowdisplaybreaks

\newcolumntype{Y}{>{\raggedright\arraybackslash}X}
\newlength{\DiagnosticIndent}
\newcommand{\DiagnosticEntry}[2]{%
  \begingroup
  \settowidth{\DiagnosticIndent}{#1:\ }
  \hangindent=\DiagnosticIndent
  \hangafter=1
  \noindent#1:\ #2\par
  \endgroup}

\newcommand{\TableNoteFont}{\scriptsize}
\newcommand{\TableNum}[1]{%
  \num[round-mode=places,round-precision=2]{#1}}
\newcommand{\TableScore}[2]{{%
  \ifnum#2=1\relax\bfseries\fi\TableNum{#1}}}
\newcommand{\TableSignedNum}[1]{%
  \ifdim#1pt>0pt +\fi\TableNum{#1}}
\newcommand{\TablePairedInline}[4]{%
  \ifnum#1=1\relax
    {\scriptsize\TableSignedNum{#2}\,[\TableNum{#3},\TableNum{#4}]}%
  \else --\fi}
\newcommand{\TableDeltaInline}[2]{%
  \ifnum#1=1\relax
    {\TableSignedNum{#2}}%
  \else --\fi}
\newcommand{\FC}{\mathrm{F}}
\newcommand{\ERA}{\mathrm{E}}

%% file: math_commands.tex
\usepackage{amsmath,amsfonts,bm}

\def\Figref#1{Figure~\ref{#1}}

\def\Secref#1{Section~\ref{#1}}

\def\eqref#1{equation~\ref{#1}}
\def\Eqref#1{Equation~\ref{#1}}

\newcommand{\Tabref}[1]{Table~\ref{#1}}
\newcommand{\Appref}[1]{Appendix~\ref{#1}}
\newcommand{\Figrefs}[1]{Figures~\ref{#1}}
\newcommand{\Tabrefs}[1]{Tables~\ref{#1}}

\def\1{\bm{1}}

\DeclareMathAlphabet{\mathsfit}{\encodingdefault}{\sfdefault}{m}{sl}
\SetMathAlphabet{\mathsfit}{bold}{\encodingdefault}{\sfdefault}{bx}{n}



%% file: sections/00_abstract.tex
\begin{abstract}
Extreme weather depends on relationships among wind, temperature, and moisture across locations and heights. Conventional field errors and hazard metrics alone do not reveal which of these physical structures a forecast captures or misses. We introduce ExtremePhysBench, a benchmark for evaluating physical structure agreement during extreme weather events. Its dataset fixes 820 heat, cold, precipitation, and landfalling tropical cyclone events during 2022--2025, defining 6,317 forecast tasks independently of model outputs. The benchmark applies the same 25 physical diagnostics to forecast and ERA5 reanalysis fields under shared applicability rules. Physical diagnostic errors are normalized using fixed scales and converted to Physics Scores, with aggregation by physical family supporting both overall and process-level comparisons. Evaluating six AI weather models and the operational GFS reveals strengths that aggregate rankings conceal. FengWu leads all physical families for heat and the thermal processes for cold, while Aurora leads cold event hydrostatic structure. For precipitation, GraphCast leads water vapor content, whereas FengWu leads moisture transport and convergence. For tropical cyclones, Aurora leads inner-core circulation, while GraphCast leads warm-core structure. Case study further show that better conventional metric accuracy does not necessarily imply closer agreement in physical structure. ExtremePhysBench reveals physical structure differences beyond conventional metrics.
\end{abstract}

%% file: sections/01_introduction.tex
\section{Introduction}
\label{sec:introduction}

\begin{figure}[!htbp]
  \centering
  \includegraphics[width=\linewidth]{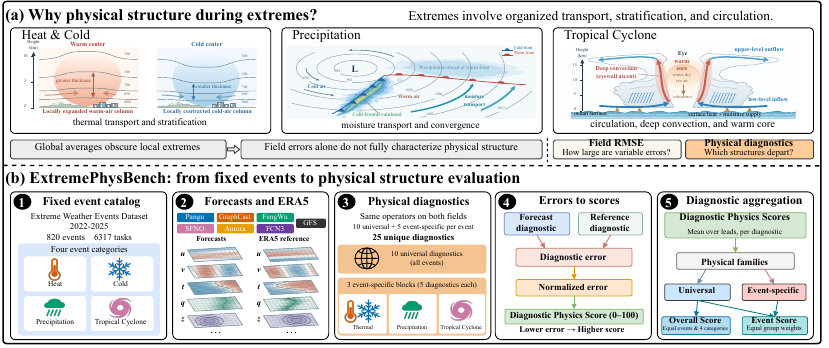}
  \caption{Motivation and workflow of ExtremePhysBench. \textbf{(a)} Thermal stratification, moisture transport, and cyclone circulation and warmth motivate evaluation beyond field errors. \textbf{(b)} A forecast-independent catalog fixes tasks. Shared diagnostics compare forecasts with reference; normalized errors become Physics Scores, aggregated through physical families.}
  \label{fig:framework}
\end{figure}

Extreme weather events, such as heavy heat, cold, precipitation, and tropical cyclone (TC), pose substantial risks to societies and critical infrastructure, making their accurate and timely prediction a central goal of weather forecasting. Yet forecasting them involves more than reproducing the magnitude of individual atmospheric variables. Extreme weather depends on how wind, temperature, and moisture interact across locations and heights. Examples include winds carrying moisture into regions of heavy precipitation, and winds circulating around a TC with a warm center and outward flow at upper levels \citep{ipcc2021extremes,holton2013dynamic}.  Therefore, a reliable forecast should capture these relationships between atmospheric variables, making physical structure agreement an essential part of forecast evaluation during extremes.

Existing evaluations measure forecast field accuracy using conventional metrics such as root mean squared error (RMSE) and anomaly correlation coefficient (ACC), alongside hazard characteristics across extreme weather events \citep{ran2025hrextreme,mcgovern2026extremeweatherbench,gomez2026tcbench,pasche2025extremes}. However, these outcomes alone do not reveal which underlying physical structures a model captures or misses. For example, precipitation error alone does not distinguish departures in moisture supply from those in local water content, while TC track and intensity errors do not separately characterize circulation and core warmth. A shared evaluation of these structures across extreme weather categories is therefore needed to identify model strengths and weaknesses by physical process.

Evaluating physical structure across extreme weather categories presents \textbf{three challenges}.
\textbf{(i) Event comparability.} Heat, cold, precipitation, and TCs differ in duration, extent, and motion. Evaluation regions and target times must support consistent comparisons without letting individual forecasts determine which samples are scored.
\textbf{(ii) Diagnostic validity.} Physical structure comparisons must respect each diagnostic's atmospheric applicability and separate storm structure from position error. The reference atmosphere can itself depart from idealized balance, so a smaller balance residual does not necessarily indicate a more faithful forecast.
\textbf{(iii) Score aggregation.} Diagnostic errors have different physical units and scales, while physical families and events contain different numbers of diagnostics and forecast leads. Aggregation must account for these differences while retaining information about individual physical processes.

We introduce \textbf{ExtremePhysBench} (\Figref{fig:framework}) to address these challenges. Its \textbf{Extreme Weather Events Dataset} fixes 820 heat, cold, precipitation, and landfalling TC events during 2022--2025, yielding 6,317 tasks independently of forecast outputs. We compute the same physical diagnostics separately from forecast fields and the fifth-generation reanalysis (ERA5) from the European Centre for Medium-Range Weather Forecasts (ECMWF) under shared applicability rules, comparing TC structures relative to their respective storm centers. Differences between physical diagnostic outputs measure departures from the atmospheric physical structures represented in ERA5, including its departures from idealized balance. We normalize diagnostic errors using fixed scales and map them to Physics Scores. Related diagnostics are grouped into \textbf{physical families}, such as moisture transport or geostrophic flow, for aggregation that accounts for unequal numbers of diagnostics and forecast leads while preserving process-level reports.

Our evaluation reveals how model strengths vary across atmospheric physical processes in all four event categories. For heat, FengWu leads all seven physical families. For cold, FengWu leads advection, stability, and frontogenesis, whereas Aurora leads hydrostatic structure. For precipitation, GraphCast leads water vapor content, while FengWu leads moisture transport and convergence. For TC, Aurora leads inner-core circulation, whereas GraphCast leads warm-core structure. These physical diagnostic views distinguish broad advantages from strengths in specific physical processes, revealing differences in physical structure agreement that a single aggregate score can conceal.

Our contributions are threefold. \textbf{First}, we build a dataset covering four extreme weather categories, with fixed forecast tasks for consistent model comparison. \textbf{Second}, we provide a reusable evaluation protocol comprising 25 physical diagnostics, shared applicability rules, and normalized Physics Scores aggregated by physical family. \textbf{Third}, we evaluate seven forecast systems, identifying distinct strengths across physical processes and showing through selected cases that better conventional metric accuracy does not necessarily imply closer physical structure agreement.

\input{tables/extreme_benchmarks}

\input{tables/event_catalog}

%% file: tables/extreme_benchmarks.tex
\begin{table}[!htbp]
  \centering
  \begin{threeparttable}
    \caption{Comparison of extreme weather benchmarks by coverage and evaluation target. ExtremePhysBench uses physical diagnostics to assess relationships among atmospheric variables, complementing field accuracy, hazard verification, and meteorological reasoning.}
    \label{tab:extreme-benchmarks}
    \scriptsize
    \setlength{\tabcolsep}{4pt}
    \begin{tabularx}{\linewidth}{@{}p{0.14\linewidth}
      >{\centering\arraybackslash}p{0.14\linewidth}
      >{\centering\arraybackslash}p{0.14\linewidth}
      >{\centering\arraybackslash}p{0.14\linewidth}
      >{\raggedright\arraybackslash}X@{}}
      \toprule
      \textbf{Benchmark} & \makecell{\textbf{Multiple}\\\textbf{hazards}} & \makecell{\textbf{Forecast field}\\\textbf{evaluation}} & \makecell{\textbf{Diagnostic}\\\textbf{structure} \textbf{evaluation}} & \textbf{Main question} \\
      \midrule
      HR-Extreme & $\checkmark$ & $\checkmark$ & $\times$ & Are forecast variables accurate within extreme events? \\
      ExtremeWeatherBench & $\checkmark$ & $\checkmark$ & $\times$ & Are high-impact weather events predicted accurately? \\
      TCBench & $\times$ & $\checkmark$ & $\times$ & Are storm tracks and intensities predicted accurately? \\
      WeatherQA & $\checkmark$ & $\times$ & $\times$ & Can models predict severe-weather threats from weather maps? \\
      \rowcolor{gray!12}
      \textbf{ExtremePhysBench} & $\checkmark$ & $\checkmark$ & $\checkmark$ & \textbf{Do forecasts reproduce physical relationships among variables?} \\
      \bottomrule
    \end{tabularx}
  \end{threeparttable}
\end{table}

%% file: tables/event_catalog.tex
\begin{table}[!htbp]
  \centering
  \begin{threeparttable}
    \caption{Event catalog and forecast task inventory. The dataset contains 820 events across four categories during 2022--2025, yielding 6,317 forecast tasks selected independently of the evaluated forecasts.}
    \label{tab:event-catalog}
    \scriptsize
    \setlength{\tabcolsep}{2pt}
    \begin{tabularx}{\linewidth}{@{}>{\raggedright\arraybackslash}p{0.12\linewidth}
      >{\raggedright\arraybackslash}X
      S[table-format=3.0]
      S[table-format=4.0]
      >{\centering\arraybackslash}p{0.1\linewidth}@{}}
      \toprule
      \textbf{Event category} & \textbf{Catalog rule} & {\textbf{Events}} &
      {\textbf{Tasks}} & \textbf{Leads (h)} \\
      \midrule

      Heat & ERA5-HEAT UTCI; daily max $>p_{95}$; tracked; $\geq3$ d and $\geq1600$ cells & 314 & 2198 & $24{:}24{:}168$ \\
      Cold & ERA5-HEAT UTCI; daily min $<p_{05}$; tracked; $\geq3$ d and $\geq1600$ cells & 106 & 742 & $24{:}24{:}168$ \\
      Precipitation & GPM IMERG V07B; daily total $\geq p_{95}$ and 10 mm d$^{-1}$; tracked; $\geq3$ d and $\geq300$ cells & 275 & 1925 & $24{:}24{:}168$ \\
      TC & IBTrACS v04r01; offshore wind $\geq34$ kt; target 24 h before selected offshore fix & 125 & 1452 & $6{:}6{:}120$ \\
      \midrule
      \bfseries Total & \multicolumn{1}{l}{--} & {\bfseries 820} & {\bfseries 6317} & -- \\
      \bottomrule
    \end{tabularx}
  \end{threeparttable}
\end{table}

%% file: sections/02_related_work.tex
\section{Related Work}
\label{sec:related-work}

WeatherBench, WeatherBench2, and WeatherBench-X provide foundational benchmarks and evaluation infrastructure for AI weather forecasting models, with an emphasis on conventional forecast skill metrics \citep{rasp2020weatherbench,rasp2024weatherbench2,weatherbenchx2025}. Complementing these broad assessments, work on extreme weather evaluation spans two complementary directions: verifying forecast fields and hazard characteristics, and assessing meteorological reasoning from weather information. These efforts establish whether models predict or interpret extreme weather events successfully, motivating a further question: \textbf{Do forecast fields capture the thermodynamic and dynamical structures that link wind, temperature, and moisture across locations and heights during extremes?} 

\textbf{Benchmarks for extreme weather forecasts.}
HR-Extreme evaluates forecast field errors within masks for 17 extreme weather types \citep{ran2025hrextreme}. ExtremeWeatherBench standardizes high-impact cases, reference data, and hazard-specific metrics, including derived quantities such as integrated vapor transport for atmospheric rivers \citep{mcgovern2026extremeweatherbench}. TCBench evaluates TC track and intensity \citep{gomez2026tcbench}. 
These benchmarks establish systematic comparisons of field accuracy and hazard characteristics.

\textbf{Reasoning about extreme weather.}
WeatherQA tests whether multimodal models can identify affected areas and severe convection potential from weather maps and contextual information \citep{ma2024weatherqa}. It evaluates meteorological reasoning, whereas ExtremePhysBench derives and verifies physical diagnostics from forecast fields.

\Tabref{tab:extreme-benchmarks} summarizes benchmarks about extreme weather. Our ExtremePhysBench complements these efforts with a systematic set of physical diagnostics covering balance, transport, stability, and vortex structure. A common protocol fixes the forecast tasks, reference-based validity rules, diagnostic scales, and aggregation by physical family. This enables comparisons of specific physical processes alongside aggregate rankings. \Appref{app:related-work-comparison} reviews broader evaluation infrastructure.

%% file: sections/03_dataset.tex
\section{Extreme Weather Events Dataset}
\label{sec:dataset}

We construct the 2022--2025 catalog from thermal stress and precipitation products and TC best track records, independently of the evaluated forecasts. \Tabref{tab:event-catalog} summarizes the events and tasks; \Appref{app:dataset-details} specifies source versions, tracking parameters, and selection counts.

\input{tables/physical_structures}

\textbf{Heat and cold events.}
ERA5-HEAT provides daily maximum and minimum Universal Thermal Climate Index (UTCI) fields \citep{dinapoli2021era5heat,copernicus2020era5heat}. Heat masks exceed the local calendar day 95th percentile of daily maximum UTCI, and cold masks fall below the 5th percentile of daily minimum UTCI. Thresholds use 1991--2020 and a nominal centered window of 15 days. Tracking these masks yields 314 extreme heat stress and 106 extreme cold stress events. The masks identify thermal stress events; the benchmark then evaluates physical structure of the atmosphere inside them.

\textbf{Precipitation events.}
Daily precipitation comes from Version 07B of the Global Precipitation Measurement (GPM) mission's Integrated Multi-satellitE Retrievals for GPM (IMERG) product \citep{huffman2020imerg,huffman2023imergdaily}. We remap the product from $0.1^\circ$ to $0.25^\circ$ and require both the local calendar day 95th percentile over 2001--2020 and 10 mm day$^{-1}$. The resulting catalog contains 275 events. These precipitation observations define where evaluation takes place; ERA5 supplies the atmospheric reference for evaluating water vapor, transport, and convergence.

\textbf{From daily masks to events.}
For heat, cold, and precipitation, masks cover land between $60^\circ$S and $60^\circ$N on a common $0.25^\circ$ grid. TempestExtremes links connected regions across days and calendar boundaries \citep{ullrich2021tempestextremes}. Retained regions persist for at least three days and meet the declared size thresholds. Each event has one fixed target at 00 (UTC) on its second retained day and seven forecasts with leads from 24 to 168 h. This common anchor avoids selecting a peak using forecast performance. Daily masks define the region of an extreme weather event, while the scored fields represent the prescribed instantaneous state; the target need not coincide with the daily extremum.

\textbf{Landfalling tropical cyclone events.}
We select 125 storms from the International Best Track Archive for Climate Stewardship (IBTrACS), version 04r01 \citep{knapp2010ibtracs,gahtan2024ibtracs}. Each has an eligible transition from an offshore fix with reported wind of at least 34 kt to a landed fix 6 h later. One transition per storm defines a target 24 h before the offshore fix. Leads span 6--120 h, but initializations before the first synoptic fix with wind of at least 25 kt are excluded. This gives 1,452 tasks, with one to 20 per storm. All 125 targets remain in the catalog.

%% file: tables/physical_structures.tex
\begin{table}[!htbp]
  \centering
  \begin{threeparttable}
    \caption{Physical structure diagnostics for extreme weather events. Ten universal diagnostics assess atmospheric relationships shared across event categories, while five event-specific diagnostics per category add the processes characteristic of each extreme. Physical families group diagnostics by the structures or processes they describe.}
    \label{tab:physical-structures}
    \scriptsize
    \setlength{\tabcolsep}{4pt}
    \begin{tabularx}{\linewidth}{@{}>{\raggedright\arraybackslash}p{0.16\linewidth}p{0.055\linewidth}>{\raggedright\arraybackslash}p{0.375\linewidth}Y@{}}
      \toprule
      \textbf{\itshape Physical family} & \textbf{Code} & \textbf{Diagnostic} & \textbf{Evaluated physical structure} \\
      \midrule
\multicolumn{4}{@{}l}{\textbf{Universal}}\\[2pt]
\textit{Hydrostatic structure} & U1 & Normalized signed hydrostatic residual components & Thickness--temperature consistency by sublayer \\
\specialrule{0.25pt}{0.5pt}{0.5pt}
\multirow[c]{2}{=}{\itshape Geostrophic flow} & U2 & Geostrophic wind components, 500 & Flow implied by geopotential gradients \\
 & U3 & Geostrophic shear components, 850 and 500 & Vertical shear of inferred geostrophic flow \\
\specialrule{0.25pt}{0.5pt}{0.5pt}
\multirow[c]{5}{=}{\itshape Flow kinematics} & U4 & Relative vorticity, 850 & Lower tropospheric rotation \\
 & U5 & Horizontal divergence, 850 & Lower tropospheric convergence/divergence \\
 & U6 & Relative vorticity, 500 & Middle tropospheric rotation \\
 & U7 & Horizontal divergence, 500 & Middle tropospheric convergence/divergence \\
 & U8 & Deformation components, 850 & Horizontal stretching and shearing \\
\specialrule{0.25pt}{0.5pt}{0.5pt}
\multirow[c]{2}{=}{\itshape Thermodynamics} & U9 & Dry $N^2$, 850--500 & Layer dry static stability \\
 & U10 & Moist static energy (MSE), 850 & Thermal, potential, and latent energy state \\
\midrule
\multicolumn{4}{@{}l}{\textbf{Heat and cold}}\\[2pt]
\multirow[c]{3}{=}{\itshape Transport} & E1 & Temperature advection, 850 & Horizontal temperature transport \\
 & E2 & Moisture advection, 850 & Horizontal humidity transport \\
 & E3 & Horizontal MSE advection, 850--500 & Partial column MSE advection \\
\specialrule{0.25pt}{0.5pt}{0.5pt}
\textit{Stability} & E4 & Dry $N^2$, 850--700 & Lower tropospheric dry stability \\
\specialrule{0.25pt}{0.5pt}{0.5pt}
\textit{Frontogenesis} & E5 & Kinematic temperature frontogenesis, 850 & Flow-induced temperature gradient change \\
\midrule
\multicolumn{4}{@{}l}{\textbf{Precipitation}}\\[2pt]
\textit{Water amount} & E1 & Water vapor, 850--300 & Partial column water vapor content \\
\specialrule{0.25pt}{0.5pt}{0.5pt}
\multirow[c]{3}{=}{\itshape Moisture transport and convergence} & E2 & Vapor transport components, 850--300 & Integrated vapor transport vector \\
 & E3 & Vapor flux convergence, 850--300 & Integrated vapor flux convergence \\
 & E4 & Moisture flux convergence, 850 & Lower tropospheric moisture convergence \\
\specialrule{0.25pt}{0.5pt}{0.5pt}
\textit{Thermodynamics} & E5 & MSE difference, 850 and 500 & Signed vertical MSE contrast \\
\midrule
\multicolumn{4}{@{}l}{\textbf{Tropical cyclones}}\\[2pt]
\textit{Inner core circulation} & E1 & Tangential wind profile, 850 & Circulation within 150 km of the center \\
\specialrule{0.25pt}{0.5pt}{0.5pt}
\multirow[c]{2}{=}{\itshape Potential vorticity and rotation} & E2 & Total Ertel potential vorticity, 700 & Rotation, stratification, and tilting \\
 & E3 & Radial vorticity profile, 500 & Radial distribution of rotation \\
\specialrule{0.25pt}{0.5pt}{0.5pt}
\textit{Upper outflow} & E4 & Radial divergence profile, 200 & Outflow within 500 km of the center \\
\specialrule{0.25pt}{0.5pt}{0.5pt}
\textit{Warm core structure} & E5 & Thickness anomaly, 300--700 & Core thickness relative to the environment \\
      \bottomrule
    \end{tabularx}
  \end{threeparttable}
\end{table}

%% file: sections/04_benchmark.tex
\section{ExtremePhysBench}
\label{sec:benchmark}

Using the events and forecast tasks defined in the Extreme Weather Events Dataset, we evaluate how well forecasts reproduce physical structures within each event. We compute the same physical diagnostics from the forecast and ERA5 fields, then measure their differences as diagnostic errors. We normalize these errors and map them to Physics Scores between 0 and 100. Finally, we aggregate the scores to summarize performance by physical family and event category, with an Overall score based on diagnostics shared across categories.

\subsection{Evaluation within Extreme Weather Events}
\label{sec:evaluation-principle}

The benchmark evaluates the physical structure of the atmosphere within a prescribed extreme weather event. Each task fixes an event, an initialization time, a forecast lead, a valid time, and an evaluation region. For heat, cold, and precipitation, the region is the catalog mask. For TCs, the catalog position seeds separate searches for the forecast and ERA5 centers; physical structures are compared relative to their own centers. This lets us compare storm physical structure separately from errors in storm location.

\textbf{Physical diagnostics are quantities computed from atmospheric variables to describe physical processes and structures.} For example, moisture convergence combines wind, humidity, and their spatial derivatives to describe where the flow concentrates moisture. These diagnostics reveal relationships between variables that individual variable errors do not directly assess. To evaluate that physical structure, we apply the same operator $\mathcal D_i$ to the forecast state $X_k^{\FC}$ and ERA5 state $X_k^{\ERA}$ for task $k$, then compare its outputs:
\begin{equation}
  \Delta D_{i,k}=\mathcal D_i(X_k^{\FC})-\mathcal D_i(X_k^{\ERA}).
  \label{eq:diagnostic-difference}
\end{equation}
This comparison retains relationships between variables before reducing them to an error. Its reference is the physical structure represented in ERA5. A balance residual measures how far the variables depart from a balance equation. We compare this departure with ERA5, where it may also be nonzero.

\textbf{A common region for comparison.}
Each model must be evaluated on the same reference samples. The grid points used for evaluation depend on the catalog region, ERA5 surface pressure, diagnostic availability, and the neighboring points needed to compute spatial derivatives. A forecast cannot remove difficult samples by producing missing values; such output is recorded as invalid. Geostrophic diagnostics exclude the equatorial band, and pressure levels below the reference surface are omitted. \Appref{app:physics-metrics} specifies these rules and the geometry used for TC profiles.

\subsection{Physical Diagnostics}
\label{sec:diagnostics}

\Tabref{tab:physical-structures} lists the physical diagnostics and evaluated processes and \Appref{app:diagnostic-inventory} gives the detailed inventory. These physical diagnostics follow standard atmospheric science theories. The universal metrics cover structures shared across event categories, while the event-specific metrics capture processes characteristic of each event category. There are 25 definitions: ten universal diagnostics and five event-specific diagnostics for each of thermal stress, precipitation, and TCs. Physical families group diagnostics by physical process for reporting and aggregation, preventing processes with more diagnostics from receiving greater weight.

\textbf{Physical structures shared across extreme weather events.}
The universal physical families describe several relationships. Hydrostatic structure (U1) relates the thickness of atmospheric layers to their temperature. Geostrophic flow and shear (U2--U3) describe winds inferred from a balance between pressure differences and Earth's rotation, and how these winds change with height. Flow kinematics (U4--U8) describe how winds rotate, spread, and stretch. Thermodynamic diagnostics (U9--U10) describe dry static stability, or resistance to vertical motion without condensation, and moist static energy (MSE), which combines contributions from temperature, height, and moisture.

\textbf{Heat and cold.}
Transport diagnostics (E1--E3) measure advection: how horizontal winds move heat, moisture, and MSE. Dry static stability (E4) describes resistance to vertical motion without condensation. Frontogenesis (E5) measures how horizontal flow strengthens or weakens temperature differences between neighboring locations. These distinguish processes that temperature error alone does not resolve.

\textbf{Precipitation.}
Water content (E1), vapor transport (E2), and moisture convergence (E3--E4) distinguish available water from its supply to a region. The MSE difference between two pressure levels (E5) describes how this energy changes with height. These diagnostics evaluate the event environment, not rainfall accumulation.

\textbf{Tropical cyclones.}
Tangential wind (E1) describes winds around the storm center. Potential vorticity (E2) combines rotation and atmospheric stratification (how air is layered vertically), while the radial vorticity profile (E3) describes how rotation varies with distance from the center. Upper-level divergence (E4) describes air spreading outward aloft. Thickness anomaly (E5) measures how much thicker the core's atmospheric layer is than its surroundings, indicating relative core warmth. Comparing structures relative to each storm's own center separates structure errors from location errors.

\subsection{Diagnostic Errors and Physics Scores}
\label{sec:scoring}

We summarize \Eqref{eq:diagnostic-difference} as diagnostic RMSE $E_{i,k}$. We use three steps to make these errors comparable. The \textbf{physical scale} $C_i$ expresses each error relative to a characteristic magnitude of the diagnostic. The \textbf{design tolerance} $\tau_i$ controls how strongly that relative error is penalized. The \textbf{bounded mapping} converts the normalized error into a score between 0 and 100, where higher is better, and limits the influence of very large errors when averaging scores. Following the exponential mapping used by Hodson et al.~\citep{hodson2021mse}, we define
\begin{equation}
  q_{i,k}=100\exp\!\left[-(E_{i,k}/S_i)^2\right],
  \qquad S_i=C_i\tau_i.
  \label{eq:physics-score}
\end{equation}
Scales and tolerances are fixed across models, events, and leads within each regime. \Appref{app:score-scales} provides the scales, tolerances, and mapping details; \Appref{app:diagnostic-error-geometries} explains how diagnostic errors are computed.

\subsection{Physical Family and Aggregate Reports}
\label{sec:aggregation}

Diagnostic and physical family reports identify processes hidden by aggregate scores. We map errors to scores before averaging leads for each diagnostic, diagnostics within each physical family, and physical families within each universal or event group. The Event Physics Score weights the two groups equally; Category Physics Scores weight events equally. This hierarchy prevents physical families with more diagnostics or events with more forecast leads from receiving greater weight solely because they contain more terms. The weighting scheme and score mapping are explicit benchmark design choices, consistent with guidance on composite indicators \citep{oecd2008composite}.

The Overall Physics Score has a narrower scope: it averages only the universal group (U1--U10), giving heat, cold, precipitation, and TC equal weight. The event-specific diagnostics describe different physical objects across event categories and are retained in their own category and physical family reports. Arithmetic averaging permits stronger agreement in one physical structure to offset weaker agreement in another, which is why the physical family report is part of the main result. Complete aggregation equations and applicability rules appear in \Appref{app:score-aggregation}.

%% file: sections/05_experimental_setup.tex
\begin{figure}[!htbp]
  \centering
  \includegraphics[width=\linewidth]{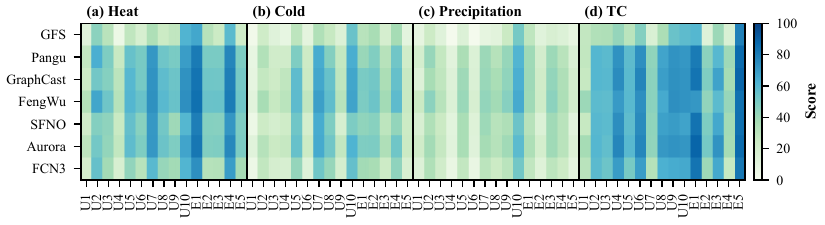}
  \caption{Physics Scores by event category. Higher scores indicate closer agreement with ERA5 under fixed diagnostic scales. Profiles reveal shared weaknesses and model-specific strengths across physical structures that aggregate scores can conceal. U1--U10 and E1--E5 denote universal and event-specific diagnostics, respectively (\Tabref{tab:physical-structures}).}
  \label{fig:main-physics-fingerprint}
\end{figure}

\input{tables/main_scoreboard}

\section{Experimental Setup}
\label{sec:experiments}

\subsection{Forecast Systems and Shared Fields}
\label{sec:models}

We evaluate six AI weather models: Pangu-Weather (\textbf{Pangu}) \citep{bi2023pangu}, GraphCast Operational (\textbf{GraphCast}) \citep{lam2023graphcast,deepmind2023graphcastrepo}, \textbf{FengWu} \citep{chen2023fengwu}, Spherical Fourier Neural Operator (\textbf{SFNO}) \citep{bonev2023sfno}, \textbf{Aurora} \citep{bodnar2025aurora}, and FourCastNet~3 (\textbf{FCN3}) \citep{bonev2025fcn3}. The operational Global Forecast System (\textbf{GFS}) provides the numerical weather prediction (NWP) baseline \citep{ncepGFS}. We do not include ECMWF's Integrated Forecasting System because a temporally continuous $0.25^\circ$ forecast archive covering the full 2022--2025 evaluation period was not available to us. Earth2Studio runs the released AI model checkpoints and retrieves archived GFS forecasts \citep{earth2studio2024}. \Appref{app:forecast-systems} identifies the model artifacts and inference configurations.

The AI models use initial states sampled every 6 h from the hourly Analysis-Ready Cloud-Optimized ERA5 (ARCO-ERA5) dataset \citep{hersbach2020era5,carver2023arcoera5}; GFS retains its operational initial state. Forecast fields are linearly interpolated to the valid time ERA5 grid at $0.25^\circ$ before evaluating diagnostics. All inputs to scored diagnostics belong to the 69-variable intersection of outputs available from the seven systems. The same operators, reference-defined evaluation points, scales, and aggregation apply to every system. Differences in initialization sources between the AI models and GFS remain part of this comparison.

\subsection{Tasks and Evaluation Protocol}
\label{sec:task-scope}
\label{sec:comparison-protocol}

Evaluation begins in 2022, after the latest disclosed data exposure of the selected AI models, including GraphCast's HRES fine tuning through 2021. All systems use the tasks and lead ranges in \Tabref{tab:event-catalog}, comparing forecasts initialized at different times for each fixed event target.

Every system has a terminal evaluation record for every requested task. Terminal coverage records whether evaluation finished with a declared status; it does not guarantee a finite score for every diagnostic. Formal scoreboard eligibility requires at least 95\% terminal coverage in each category. Physical family reports retain their corresponding event counts rather than treating an inapplicable diagnostic as a score of zero.
For model differences in the main scoreboard, paired event bootstrap intervals are reported at the nominal 95\% level without adjustment for multiple comparisons. 

%% file: tables/main_scoreboard.tex
\begin{table}[!htbp]
  \centering
  \begin{threeparttable}
    \caption{Aggregate Physics Scores by forecast system. Overall summarizes diagnostics shared across event categories, while Category scores also include event-specific diagnostics. Higher scores indicate closer physical structure agreement with ERA5.}
    \label{tab:scoreboard}
    \scriptsize
    \setlength{\tabcolsep}{2.4pt}
    \begin{tabular*}{\linewidth}{@{\extracolsep{\fill}}l
      *{5}{S[table-format=2.2]}
      r r@{}}
      \toprule
      \textbf{System} & {\textbf{Overall}} & {\textbf{Heat}} & {\textbf{Cold}} &
      {\textbf{Precipitation}} & {\textbf{TC}} &
      \textbf{Score $-$ GFS} &
      \textbf{$\Delta$ vs. FengWu [95\% CI]} \\
      \midrule

      FengWu & \TableScore{46.03}{1} & \TableScore{59.64}{1} & \TableScore{39.85}{1} & \TableScore{34.80}{1} & \TableScore{57.80}{0} & \TableDeltaInline{1}{19.21} & \TablePairedInline{0}{}{}{} \\
      Aurora & \TableScore{42.59}{0} & \TableScore{54.37}{0} & \TableScore{36.82}{0} & \TableScore{29.81}{0} & \TableScore{62.13}{1} & \TableDeltaInline{1}{15.77} & \TablePairedInline{1}{-3.44}{-3.90}{-2.93} \\
      GraphCast & \TableScore{42.42}{0} & \TableScore{56.22}{0} & \TableScore{37.57}{0} & \TableScore{31.98}{0} & \TableScore{60.91}{0} & \TableDeltaInline{1}{15.59} & \TablePairedInline{1}{-3.61}{-4.20}{-2.95} \\
      Pangu & \TableScore{42.05}{0} & \TableScore{55.55}{0} & \TableScore{34.85}{0} & \TableScore{31.29}{0} & \TableScore{58.36}{0} & \TableDeltaInline{1}{15.22} & \TablePairedInline{1}{-3.98}{-4.52}{-3.28} \\
      SFNO & \TableScore{37.96}{0} & \TableScore{50.73}{0} & \TableScore{31.03}{0} & \TableScore{26.99}{0} & \TableScore{58.60}{0} & \TableDeltaInline{1}{11.14} & \TablePairedInline{1}{-8.07}{-8.66}{-7.47} \\
      FCN3 & \TableScore{35.51}{0} & \TableScore{45.11}{0} & \TableScore{28.72}{0} & \TableScore{23.50}{0} & \TableScore{54.36}{0} & \TableDeltaInline{1}{8.69} & \TablePairedInline{1}{-10.52}{-11.16}{-9.79} \\
      GFS & \TableScore{26.82}{0} & \TableScore{35.22}{0} & \TableScore{22.86}{0} & \TableScore{18.18}{0} & \TableScore{41.66}{0} & \TableDeltaInline{0}{} & \TablePairedInline{0}{}{}{} \\
      \bottomrule
    \end{tabular*}
  \end{threeparttable}
\end{table}

%% file: sections/06_results.tex
\section{Results}
\label{sec:results}

Model strengths differ between water availability and moisture supply in precipitation, and between circulation and core warmth in TCs (\Tabref{tab:family-scores}). We examine these differences from aggregate scores to physical families and individual diagnostics.

\subsection{Aggregate Scores and Verification Sensitivity}
\label{sec:main-results}

\Tabref{tab:scoreboard} compares Overall scores from universal diagnostics and Category scores that include event-specific diagnostics. Overall is therefore not the mean of the four Category scores. \Appref{app:aggregate-model-comparisons} provides complementary aggregate views.

FengWu has the highest Overall score (46.03), but strengths differ by event category. The comparison between the AI models and GFS uses ERA5 initial states for the AI models and operational initial states for GFS, with all systems verified against ERA5; it is therefore specific to these settings. Event distributions overlap substantially (\Appref{app:event-comparisons}), so a higher category mean does not imply an advantage on every event.

\textbf{GFS verification sensitivity.}
We also rescore GFS against its own archived 0-h analyses, keeping the tasks, operators, scales, and aggregation rules fixed. Its Overall score rises from 26.82 to 30.33, with the largest category increase in TC ($+8.55$). The alternative GFS score remains numerically below the ERA5-verified AI model score range of 35.51--46.03. This check shows the size of the reference effect in our experiment, but does not remove initialization differences or define a new ranking across different references; each analysis also determines its own valid comparison samples (\Appref{app:gfs-reference}).

\subsection{Physical Families Reveal Different Strengths}
\label{sec:family-results}

\Tabref{tab:family-scores} uses events with complete scores for all models, giving eligible events equal weight within each physical family. Applicability varies, notably for geostrophy and TC warm core structure, so averaging the displayed physical family means need not reproduce Category scores. Comparisons use fixed diagnostic scales and are descriptive; no physical family confidence intervals were computed. \Figref{fig:main-physics-fingerprint} resolves physical family differences into individual diagnostics; \Appref{app:diagnostic-results} provides detailed diagnostic views.

\input{tables/family_scores}

\textbf{Heat: a broad advantage across processes.}
The heat results show a consistent advantage across transport, stability, and frontogenesis: FengWu leads all seven physical families. Its category advantage therefore spans how winds move heat and moisture, how temperature changes with height, and how horizontal flow changes temperature contrasts. Agreement across these physical families shows that its heat ranking is not driven by a single physical family.

\textbf{Cold: thermal and balance diagnostics give different views.}
Thermal-process and hydrostatic diagnostics favor different systems: FengWu leads the three physical families for thermal processes, whereas Aurora leads hydrostatic structure (12.26 versus 11.95). Frontogenesis scores are close for FengWu and GraphCast (26.55 versus 26.43). All systems score lower on the physical families for thermal processes in cold than in heat under shared scales; this sample comparison alone does not establish a difference in predictability.

\begin{figure}[!htbp]
  \centering
  \includegraphics[width=\linewidth]{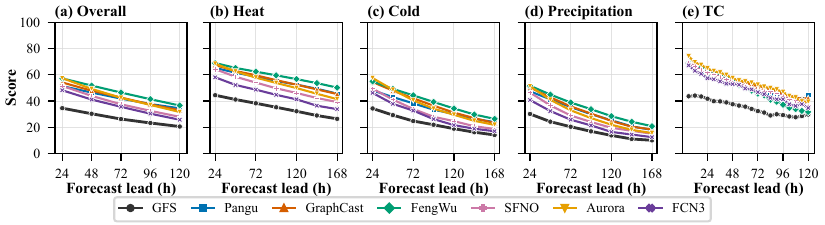}
  \caption{Physics Scores by forecast lead. Agreement with ERA5 generally decreases at longer leads, with differences across models and event categories. A smaller decline does not imply better performance because initial score levels differ and the mapping is nonlinear.}
  \label{fig:main-lead-scores}
\end{figure}

\textbf{Precipitation: water amount and supply differ.}
Water availability and moisture supply favor different systems: GraphCast leads water content (40.23), whereas FengWu leads moisture transport and convergence (34.74) and vertical energy contrast (20.81). A category score thus conceals differences between available vapor, its supply, and thermodynamic structure. The event-specific physical family for thermodynamics measures vertical energy contrast; the universal physical family measures stability and energy state.

\textbf{Tropical cyclones: circulation, outflow, and warmth separate.}
Circulation and core warmth favor different systems: Aurora leads circulation (87.23) and potential vorticity and rotation (61.12), whereas GraphCast leads warm-core structure (86.34). Outflow scores are close for FengWu (44.16), Pangu (44.10), and Aurora (44.03). Each physical family uses events with valid scores for all models. These comparisons measure structures relative to each storm's own center and complement track and intensity verification.

\subsection{Supplementary Analyses}
\label{sec:rmse-relationship}

\textbf{Relationship to field accuracy.}
We compare Physics Scores with RMSE for $u,v,t,q,z$ at 850, 700, and 500 hPa using the same eligible events and evaluation regions. Comparing the formal Physics Score ranks with ranks from the mean RMSE ranks over these fields gives an Overall correlation of $\rho=0.893$, showing that the two evaluations are related. The lower correlation for TC ($\rho=0.703$) shows greater rank disagreement for storms. \Appref{app:rmse-results} gives the RMSE composite definition and detailed rank comparisons.

Within models, events with lower field errors also tend to have higher Physics Scores. All 28 Spearman associations between RMSE composite skill and Event Physics Score are positive, and their nominal bootstrap 95\% intervals exclude zero (\Appref{app:rmse-results}). Physical diagnostics add a different level of information: they separate agreement in transport, stability, rotation, and other physical structures that a single RMSE composite does not identify. The physical-family reports make these differences visible within each extreme weather category.

\textbf{Case studies.}
Eight selected cases illustrate how field errors can conceal structural differences (\Appref{app:case-studies}). In TC MEARI, Pangu and GraphCast differ by at most 10\% in each of 15 field RMSEs, yet their Event Physics Scores differ by 9.51 points. In heat event 198, Aurora has larger field errors than FCN3 (RMSE ratio 1.260) but a higher Physics Score (58.97 versus 46.11). Family scores and diagnostic maps or profiles reveal contrasts in frontogenesis, moisture convergence, and cyclone circulation and outflow. These examples demonstrate that lower field RMSE need not imply better physical structure agreement.

\textbf{Score mapping and aggregation sensitivity.}
We test tolerances at $0.5$, $1$, and $2$ times their default values. Each setting uses either equal weights for physical families or equal weights for diagnostics within each group. All six sensitivity experiments preserve the rankings in all four event categories. Only Aurora and GraphCast exchange second and third place Overall ($\rho\geq0.964$). \Appref{app:score-sensitivity} provides the full results.

\textbf{Variation with forecast lead.}
All systems score lower at 120 h than at 24 h (\Figref{fig:main-lead-scores}). A smaller score decline does not necessarily indicate better physical structure agreement, because systems start from different score levels and the score mapping is nonlinear.

%% file: tables/family_scores.tex
\begin{table}[!htbp]
  \centering
  \begin{threeparttable}
    \caption{Physics Scores by event category and physical family. Water content and moisture supply in precipitation, and circulation and core warmth in TCs, show different patterns across models. These physical-family reports distinguish agreement with ERA5 in different physical structures beyond a single aggregate score. Higher scores indicate closer physical structure agreement with ERA5.}
    \label{tab:family-scores}
    \scriptsize
    \setlength{\tabcolsep}{0pt}
    \begin{tabular}{@{}>{\raggedright\arraybackslash}p{0.18\linewidth}*{4}{>{\centering\arraybackslash}p{0.1025\linewidth}}*{12}{>{\centering\arraybackslash}p{0.034166\linewidth}}@{}}
      \toprule
      \textbf{Model} & \multicolumn{4}{c}{\textbf{Universal}} & \multicolumn{12}{c}{\textbf{Event-specific}} \\
      \cmidrule(lr){2-5}\cmidrule(lr){6-17}
      \textbf{Heat} & Hyd. & Geos. & Kinem. & Thermo. & \multicolumn{4}{>{\centering\arraybackslash}p{0.136664\linewidth}}{Adv.} & \multicolumn{4}{>{\centering\arraybackslash}p{0.136664\linewidth}}{Stab.} & \multicolumn{4}{>{\centering\arraybackslash}p{0.136664\linewidth}}{Frontog.} \\
      \addlinespace[2pt]
      \hspace*{1em}GFS & \ifnum 1=1\relax\TableScore{12.75}{0}\else --\fi & \ifnum 1=1\relax\TableScore{38.39}{0}\else --\fi & \ifnum 1=1\relax\TableScore{22.72}{0}\else --\fi & \ifnum 1=1\relax\TableScore{45.20}{0}\else --\fi & \multicolumn{4}{>{\centering\arraybackslash}p{0.136664\linewidth}}{\ifnum 1=1\relax\TableScore{41.07}{0}\else --\fi} & \multicolumn{4}{>{\centering\arraybackslash}p{0.136664\linewidth}}{\ifnum 1=1\relax\TableScore{58.62}{0}\else --\fi} & \multicolumn{4}{>{\centering\arraybackslash}p{0.136664\linewidth}}{\ifnum 1=1\relax\TableScore{22.74}{0}\else --\fi} \\
      \hspace*{1em}Pangu & \ifnum 1=1\relax\TableScore{28.10}{0}\else --\fi & \ifnum 1=1\relax\TableScore{56.30}{0}\else --\fi & \ifnum 1=1\relax\TableScore{51.41}{0}\else --\fi & \ifnum 1=1\relax\TableScore{59.82}{0}\else --\fi & \multicolumn{4}{>{\centering\arraybackslash}p{0.136664\linewidth}}{\ifnum 1=1\relax\TableScore{60.40}{0}\else --\fi} & \multicolumn{4}{>{\centering\arraybackslash}p{0.136664\linewidth}}{\ifnum 1=1\relax\TableScore{76.81}{0}\else --\fi} & \multicolumn{4}{>{\centering\arraybackslash}p{0.136664\linewidth}}{\ifnum 1=1\relax\TableScore{49.10}{0}\else --\fi} \\
      \hspace*{1em}GraphCast & \ifnum 1=1\relax\TableScore{24.09}{0}\else --\fi & \ifnum 1=1\relax\TableScore{49.56}{0}\else --\fi & \ifnum 1=1\relax\TableScore{54.40}{0}\else --\fi & \ifnum 1=1\relax\TableScore{63.32}{0}\else --\fi & \multicolumn{4}{>{\centering\arraybackslash}p{0.136664\linewidth}}{\ifnum 1=1\relax\TableScore{62.75}{0}\else --\fi} & \multicolumn{4}{>{\centering\arraybackslash}p{0.136664\linewidth}}{\ifnum 1=1\relax\TableScore{78.93}{0}\else --\fi} & \multicolumn{4}{>{\centering\arraybackslash}p{0.136664\linewidth}}{\ifnum 1=1\relax\TableScore{51.51}{0}\else --\fi} \\
      \hspace*{1em}FengWu & \ifnum 1=1\relax\TableScore{33.18}{1}\else --\fi & \ifnum 1=1\relax\TableScore{60.01}{1}\else --\fi & \ifnum 1=1\relax\TableScore{56.22}{1}\else --\fi & \ifnum 1=1\relax\TableScore{64.28}{1}\else --\fi & \multicolumn{4}{>{\centering\arraybackslash}p{0.136664\linewidth}}{\ifnum 1=1\relax\TableScore{64.55}{1}\else --\fi} & \multicolumn{4}{>{\centering\arraybackslash}p{0.136664\linewidth}}{\ifnum 1=1\relax\TableScore{80.20}{1}\else --\fi} & \multicolumn{4}{>{\centering\arraybackslash}p{0.136664\linewidth}}{\ifnum 1=1\relax\TableScore{52.52}{1}\else --\fi} \\
      \hspace*{1em}SFNO & \ifnum 1=1\relax\TableScore{21.88}{0}\else --\fi & \ifnum 1=1\relax\TableScore{45.87}{0}\else --\fi & \ifnum 1=1\relax\TableScore{50.18}{0}\else --\fi & \ifnum 1=1\relax\TableScore{50.20}{0}\else --\fi & \multicolumn{4}{>{\centering\arraybackslash}p{0.136664\linewidth}}{\ifnum 1=1\relax\TableScore{57.81}{0}\else --\fi} & \multicolumn{4}{>{\centering\arraybackslash}p{0.136664\linewidth}}{\ifnum 1=1\relax\TableScore{71.53}{0}\else --\fi} & \multicolumn{4}{>{\centering\arraybackslash}p{0.136664\linewidth}}{\ifnum 1=1\relax\TableScore{48.49}{0}\else --\fi} \\
      \hspace*{1em}Aurora & \ifnum 1=1\relax\TableScore{30.13}{0}\else --\fi & \ifnum 1=1\relax\TableScore{47.42}{0}\else --\fi & \ifnum 1=1\relax\TableScore{52.06}{0}\else --\fi & \ifnum 1=1\relax\TableScore{57.91}{0}\else --\fi & \multicolumn{4}{>{\centering\arraybackslash}p{0.136664\linewidth}}{\ifnum 1=1\relax\TableScore{59.60}{0}\else --\fi} & \multicolumn{4}{>{\centering\arraybackslash}p{0.136664\linewidth}}{\ifnum 1=1\relax\TableScore{75.95}{0}\else --\fi} & \multicolumn{4}{>{\centering\arraybackslash}p{0.136664\linewidth}}{\ifnum 1=1\relax\TableScore{49.42}{0}\else --\fi} \\
      \hspace*{1em}FCN3 & \ifnum 1=1\relax\TableScore{20.09}{0}\else --\fi & \ifnum 1=1\relax\TableScore{47.27}{0}\else --\fi & \ifnum 1=1\relax\TableScore{38.81}{0}\else --\fi & \ifnum 1=1\relax\TableScore{49.94}{0}\else --\fi & \multicolumn{4}{>{\centering\arraybackslash}p{0.136664\linewidth}}{\ifnum 1=1\relax\TableScore{46.54}{0}\else --\fi} & \multicolumn{4}{>{\centering\arraybackslash}p{0.136664\linewidth}}{\ifnum 1=1\relax\TableScore{69.22}{0}\else --\fi} & \multicolumn{4}{>{\centering\arraybackslash}p{0.136664\linewidth}}{\ifnum 1=1\relax\TableScore{37.63}{0}\else --\fi} \\
      \midrule
      \textbf{Cold} & Hyd. & Geos. & Kinem. & Thermo. & \multicolumn{4}{>{\centering\arraybackslash}p{0.136664\linewidth}}{Adv.} & \multicolumn{4}{>{\centering\arraybackslash}p{0.136664\linewidth}}{Stab.} & \multicolumn{4}{>{\centering\arraybackslash}p{0.136664\linewidth}}{Frontog.} \\
      \addlinespace[2pt]
      \hspace*{1em}GFS & \ifnum 1=1\relax\TableScore{3.51}{0}\else --\fi & \ifnum 1=1\relax\TableScore{18.01}{0}\else --\fi & \ifnum 1=1\relax\TableScore{22.10}{0}\else --\fi & \ifnum 1=1\relax\TableScore{34.73}{0}\else --\fi & \multicolumn{4}{>{\centering\arraybackslash}p{0.136664\linewidth}}{\ifnum 1=1\relax\TableScore{29.29}{0}\else --\fi} & \multicolumn{4}{>{\centering\arraybackslash}p{0.136664\linewidth}}{\ifnum 1=1\relax\TableScore{38.13}{0}\else --\fi} & \multicolumn{4}{>{\centering\arraybackslash}p{0.136664\linewidth}}{\ifnum 1=1\relax\TableScore{10.79}{0}\else --\fi} \\
      \hspace*{1em}Pangu & \ifnum 1=1\relax\TableScore{8.33}{0}\else --\fi & \ifnum 1=1\relax\TableScore{28.03}{0}\else --\fi & \ifnum 1=1\relax\TableScore{41.53}{0}\else --\fi & \ifnum 1=1\relax\TableScore{45.89}{0}\else --\fi & \multicolumn{4}{>{\centering\arraybackslash}p{0.136664\linewidth}}{\ifnum 1=1\relax\TableScore{40.81}{0}\else --\fi} & \multicolumn{4}{>{\centering\arraybackslash}p{0.136664\linewidth}}{\ifnum 1=1\relax\TableScore{52.55}{0}\else --\fi} & \multicolumn{4}{>{\centering\arraybackslash}p{0.136664\linewidth}}{\ifnum 1=1\relax\TableScore{22.58}{0}\else --\fi} \\
      \hspace*{1em}GraphCast & \ifnum 1=1\relax\TableScore{9.16}{0}\else --\fi & \ifnum 1=1\relax\TableScore{25.49}{0}\else --\fi & \ifnum 1=1\relax\TableScore{46.62}{0}\else --\fi & \ifnum 1=1\relax\TableScore{48.01}{0}\else --\fi & \multicolumn{4}{>{\centering\arraybackslash}p{0.136664\linewidth}}{\ifnum 1=1\relax\TableScore{46.04}{0}\else --\fi} & \multicolumn{4}{>{\centering\arraybackslash}p{0.136664\linewidth}}{\ifnum 1=1\relax\TableScore{55.57}{0}\else --\fi} & \multicolumn{4}{>{\centering\arraybackslash}p{0.136664\linewidth}}{\ifnum 1=1\relax\TableScore{26.43}{0}\else --\fi} \\
      \hspace*{1em}FengWu & \ifnum 1=1\relax\TableScore{11.95}{0}\else --\fi & \ifnum 1=1\relax\TableScore{32.07}{1}\else --\fi & \ifnum 1=1\relax\TableScore{48.73}{1}\else --\fi & \ifnum 1=1\relax\TableScore{49.56}{1}\else --\fi & \multicolumn{4}{>{\centering\arraybackslash}p{0.136664\linewidth}}{\ifnum 1=1\relax\TableScore{47.35}{1}\else --\fi} & \multicolumn{4}{>{\centering\arraybackslash}p{0.136664\linewidth}}{\ifnum 1=1\relax\TableScore{58.14}{1}\else --\fi} & \multicolumn{4}{>{\centering\arraybackslash}p{0.136664\linewidth}}{\ifnum 1=1\relax\TableScore{26.55}{1}\else --\fi} \\
      \hspace*{1em}SFNO & \ifnum 1=1\relax\TableScore{5.79}{0}\else --\fi & \ifnum 1=1\relax\TableScore{21.85}{0}\else --\fi & \ifnum 1=1\relax\TableScore{41.88}{0}\else --\fi & \ifnum 1=1\relax\TableScore{34.11}{0}\else --\fi & \multicolumn{4}{>{\centering\arraybackslash}p{0.136664\linewidth}}{\ifnum 1=1\relax\TableScore{40.14}{0}\else --\fi} & \multicolumn{4}{>{\centering\arraybackslash}p{0.136664\linewidth}}{\ifnum 1=1\relax\TableScore{43.15}{0}\else --\fi} & \multicolumn{4}{>{\centering\arraybackslash}p{0.136664\linewidth}}{\ifnum 1=1\relax\TableScore{24.87}{0}\else --\fi} \\
      \hspace*{1em}Aurora & \ifnum 1=1\relax\TableScore{12.26}{1}\else --\fi & \ifnum 1=1\relax\TableScore{28.95}{0}\else --\fi & \ifnum 1=1\relax\TableScore{44.88}{0}\else --\fi & \ifnum 1=1\relax\TableScore{44.89}{0}\else --\fi & \multicolumn{4}{>{\centering\arraybackslash}p{0.136664\linewidth}}{\ifnum 1=1\relax\TableScore{44.25}{0}\else --\fi} & \multicolumn{4}{>{\centering\arraybackslash}p{0.136664\linewidth}}{\ifnum 1=1\relax\TableScore{52.94}{0}\else --\fi} & \multicolumn{4}{>{\centering\arraybackslash}p{0.136664\linewidth}}{\ifnum 1=1\relax\TableScore{25.16}{0}\else --\fi} \\
      \hspace*{1em}FCN3 & \ifnum 1=1\relax\TableScore{5.72}{0}\else --\fi & \ifnum 1=1\relax\TableScore{23.97}{0}\else --\fi & \ifnum 1=1\relax\TableScore{33.81}{0}\else --\fi & \ifnum 1=1\relax\TableScore{38.03}{0}\else --\fi & \multicolumn{4}{>{\centering\arraybackslash}p{0.136664\linewidth}}{\ifnum 1=1\relax\TableScore{33.41}{0}\else --\fi} & \multicolumn{4}{>{\centering\arraybackslash}p{0.136664\linewidth}}{\ifnum 1=1\relax\TableScore{44.46}{0}\else --\fi} & \multicolumn{4}{>{\centering\arraybackslash}p{0.136664\linewidth}}{\ifnum 1=1\relax\TableScore{18.01}{0}\else --\fi} \\
      \midrule
      \textbf{Precipitation} & Hyd. & Geos. & Kinem. & Thermo. & \multicolumn{4}{>{\centering\arraybackslash}p{0.136664\linewidth}}{Vapor} & \multicolumn{4}{>{\centering\arraybackslash}p{0.136664\linewidth}}{Moist.} & \multicolumn{4}{>{\centering\arraybackslash}p{0.136664\linewidth}}{Thermo.} \\
      \addlinespace[2pt]
      \hspace*{1em}GFS & \ifnum 1=1\relax\TableScore{9.17}{0}\else --\fi & \ifnum 1=1\relax\TableScore{18.85}{0}\else --\fi & \ifnum 1=1\relax\TableScore{7.75}{0}\else --\fi & \ifnum 1=1\relax\TableScore{38.27}{0}\else --\fi & \multicolumn{4}{>{\centering\arraybackslash}p{0.136664\linewidth}}{\ifnum 1=1\relax\TableScore{29.34}{0}\else --\fi} & \multicolumn{4}{>{\centering\arraybackslash}p{0.136664\linewidth}}{\ifnum 1=1\relax\TableScore{14.75}{0}\else --\fi} & \multicolumn{4}{>{\centering\arraybackslash}p{0.136664\linewidth}}{\ifnum 1=1\relax\TableScore{9.13}{0}\else --\fi} \\
      \hspace*{1em}Pangu & \ifnum 1=1\relax\TableScore{19.79}{0}\else --\fi & \ifnum 1=1\relax\TableScore{34.62}{0}\else --\fi & \ifnum 1=1\relax\TableScore{27.44}{0}\else --\fi & \ifnum 1=1\relax\TableScore{51.82}{0}\else --\fi & \multicolumn{4}{>{\centering\arraybackslash}p{0.136664\linewidth}}{\ifnum 1=1\relax\TableScore{39.57}{0}\else --\fi} & \multicolumn{4}{>{\centering\arraybackslash}p{0.136664\linewidth}}{\ifnum 1=1\relax\TableScore{30.77}{0}\else --\fi} & \multicolumn{4}{>{\centering\arraybackslash}p{0.136664\linewidth}}{\ifnum 1=1\relax\TableScore{16.59}{0}\else --\fi} \\
      \hspace*{1em}GraphCast & \ifnum 1=1\relax\TableScore{14.09}{0}\else --\fi & \ifnum 1=1\relax\TableScore{33.32}{0}\else --\fi & \ifnum 1=1\relax\TableScore{29.71}{0}\else --\fi & \ifnum 1=1\relax\TableScore{53.60}{0}\else --\fi & \multicolumn{4}{>{\centering\arraybackslash}p{0.136664\linewidth}}{\ifnum 1=1\relax\TableScore{40.23}{1}\else --\fi} & \multicolumn{4}{>{\centering\arraybackslash}p{0.136664\linewidth}}{\ifnum 1=1\relax\TableScore{33.49}{0}\else --\fi} & \multicolumn{4}{>{\centering\arraybackslash}p{0.136664\linewidth}}{\ifnum 1=1\relax\TableScore{19.67}{0}\else --\fi} \\
      \hspace*{1em}FengWu & \ifnum 1=1\relax\TableScore{25.22}{1}\else --\fi & \ifnum 1=1\relax\TableScore{38.51}{1}\else --\fi & \ifnum 1=1\relax\TableScore{31.31}{1}\else --\fi & \ifnum 1=1\relax\TableScore{55.96}{1}\else --\fi & \multicolumn{4}{>{\centering\arraybackslash}p{0.136664\linewidth}}{\ifnum 1=1\relax\TableScore{39.47}{0}\else --\fi} & \multicolumn{4}{>{\centering\arraybackslash}p{0.136664\linewidth}}{\ifnum 1=1\relax\TableScore{34.74}{1}\else --\fi} & \multicolumn{4}{>{\centering\arraybackslash}p{0.136664\linewidth}}{\ifnum 1=1\relax\TableScore{20.81}{1}\else --\fi} \\
      \hspace*{1em}SFNO & \ifnum 1=1\relax\TableScore{17.31}{0}\else --\fi & \ifnum 1=1\relax\TableScore{29.88}{0}\else --\fi & \ifnum 1=1\relax\TableScore{26.72}{0}\else --\fi & \ifnum 1=1\relax\TableScore{44.50}{0}\else --\fi & \multicolumn{4}{>{\centering\arraybackslash}p{0.136664\linewidth}}{\ifnum 1=1\relax\TableScore{32.04}{0}\else --\fi} & \multicolumn{4}{>{\centering\arraybackslash}p{0.136664\linewidth}}{\ifnum 1=1\relax\TableScore{28.69}{0}\else --\fi} & \multicolumn{4}{>{\centering\arraybackslash}p{0.136664\linewidth}}{\ifnum 1=1\relax\TableScore{11.91}{0}\else --\fi} \\
      \hspace*{1em}Aurora & \ifnum 1=1\relax\TableScore{22.39}{0}\else --\fi & \ifnum 1=1\relax\TableScore{32.88}{0}\else --\fi & \ifnum 1=1\relax\TableScore{26.48}{0}\else --\fi & \ifnum 1=1\relax\TableScore{48.77}{0}\else --\fi & \multicolumn{4}{>{\centering\arraybackslash}p{0.136664\linewidth}}{\ifnum 1=1\relax\TableScore{35.27}{0}\else --\fi} & \multicolumn{4}{>{\centering\arraybackslash}p{0.136664\linewidth}}{\ifnum 1=1\relax\TableScore{30.29}{0}\else --\fi} & \multicolumn{4}{>{\centering\arraybackslash}p{0.136664\linewidth}}{\ifnum 1=1\relax\TableScore{14.87}{0}\else --\fi} \\
      \hspace*{1em}FCN3 & \ifnum 1=1\relax\TableScore{14.68}{0}\else --\fi & \ifnum 1=1\relax\TableScore{28.35}{0}\else --\fi & \ifnum 1=1\relax\TableScore{19.78}{0}\else --\fi & \ifnum 1=1\relax\TableScore{41.67}{0}\else --\fi & \multicolumn{4}{>{\centering\arraybackslash}p{0.136664\linewidth}}{\ifnum 1=1\relax\TableScore{28.64}{0}\else --\fi} & \multicolumn{4}{>{\centering\arraybackslash}p{0.136664\linewidth}}{\ifnum 1=1\relax\TableScore{22.86}{0}\else --\fi} & \multicolumn{4}{>{\centering\arraybackslash}p{0.136664\linewidth}}{\ifnum 1=1\relax\TableScore{10.66}{0}\else --\fi} \\
      \midrule
      \textbf{Tropical Cyclone} & Hyd. & Geos. & Kinem. & Thermo. & \multicolumn{3}{>{\centering\arraybackslash}p{0.1025\linewidth}}{Circ.} & \multicolumn{3}{>{\centering\arraybackslash}p{0.1025\linewidth}}{PV} & \multicolumn{3}{>{\centering\arraybackslash}p{0.1025\linewidth}}{Outflow} & \multicolumn{3}{>{\centering\arraybackslash}p{0.1025\linewidth}}{Warm} \\
      \addlinespace[2pt]
      \hspace*{1em}GFS & \ifnum 1=1\relax\TableScore{26.03}{0}\else --\fi & \ifnum 1=1\relax\TableScore{39.26}{0}\else --\fi & \ifnum 1=1\relax\TableScore{35.48}{0}\else --\fi & \ifnum 1=1\relax\TableScore{56.65}{0}\else --\fi & \multicolumn{3}{>{\centering\arraybackslash}p{0.1025\linewidth}}{\ifnum 1=1\relax\TableScore{62.84}{0}\else --\fi} & \multicolumn{3}{>{\centering\arraybackslash}p{0.1025\linewidth}}{\ifnum 1=1\relax\TableScore{30.05}{0}\else --\fi} & \multicolumn{3}{>{\centering\arraybackslash}p{0.1025\linewidth}}{\ifnum 1=1\relax\TableScore{21.18}{0}\else --\fi} & \multicolumn{3}{>{\centering\arraybackslash}p{0.1025\linewidth}}{\ifnum 1=1\relax\TableScore{76.08}{0}\else --\fi} \\
      \hspace*{1em}Pangu & \ifnum 1=1\relax\TableScore{27.01}{0}\else --\fi & \ifnum 1=1\relax\TableScore{58.86}{0}\else --\fi & \ifnum 1=1\relax\TableScore{62.79}{0}\else --\fi & \ifnum 1=1\relax\TableScore{69.97}{0}\else --\fi & \multicolumn{3}{>{\centering\arraybackslash}p{0.1025\linewidth}}{\ifnum 1=1\relax\TableScore{79.98}{0}\else --\fi} & \multicolumn{3}{>{\centering\arraybackslash}p{0.1025\linewidth}}{\ifnum 1=1\relax\TableScore{51.77}{0}\else --\fi} & \multicolumn{3}{>{\centering\arraybackslash}p{0.1025\linewidth}}{\ifnum 1=1\relax\TableScore{44.10}{0}\else --\fi} & \multicolumn{3}{>{\centering\arraybackslash}p{0.1025\linewidth}}{\ifnum 1=1\relax\TableScore{83.47}{0}\else --\fi} \\
      \hspace*{1em}GraphCast & \ifnum 1=1\relax\TableScore{27.15}{0}\else --\fi & \ifnum 1=1\relax\TableScore{61.83}{1}\else --\fi & \ifnum 1=1\relax\TableScore{65.17}{1}\else --\fi & \ifnum 1=1\relax\TableScore{71.60}{1}\else --\fi & \multicolumn{3}{>{\centering\arraybackslash}p{0.1025\linewidth}}{\ifnum 1=1\relax\TableScore{83.40}{0}\else --\fi} & \multicolumn{3}{>{\centering\arraybackslash}p{0.1025\linewidth}}{\ifnum 1=1\relax\TableScore{59.44}{0}\else --\fi} & \multicolumn{3}{>{\centering\arraybackslash}p{0.1025\linewidth}}{\ifnum 1=1\relax\TableScore{42.84}{0}\else --\fi} & \multicolumn{3}{>{\centering\arraybackslash}p{0.1025\linewidth}}{\ifnum 1=1\relax\TableScore{86.34}{1}\else --\fi} \\
      \hspace*{1em}FengWu & \ifnum 1=1\relax\TableScore{38.19}{1}\else --\fi & \ifnum 1=1\relax\TableScore{57.87}{0}\else --\fi & \ifnum 1=1\relax\TableScore{62.06}{0}\else --\fi & \ifnum 1=1\relax\TableScore{69.95}{0}\else --\fi & \multicolumn{3}{>{\centering\arraybackslash}p{0.1025\linewidth}}{\ifnum 1=1\relax\TableScore{69.72}{0}\else --\fi} & \multicolumn{3}{>{\centering\arraybackslash}p{0.1025\linewidth}}{\ifnum 1=1\relax\TableScore{49.60}{0}\else --\fi} & \multicolumn{3}{>{\centering\arraybackslash}p{0.1025\linewidth}}{\ifnum 1=1\relax\TableScore{44.16}{1}\else --\fi} & \multicolumn{3}{>{\centering\arraybackslash}p{0.1025\linewidth}}{\ifnum 1=1\relax\TableScore{79.35}{0}\else --\fi} \\
      \hspace*{1em}SFNO & \ifnum 1=1\relax\TableScore{25.90}{0}\else --\fi & \ifnum 1=1\relax\TableScore{61.21}{0}\else --\fi & \ifnum 1=1\relax\TableScore{63.45}{0}\else --\fi & \ifnum 1=1\relax\TableScore{65.65}{0}\else --\fi & \multicolumn{3}{>{\centering\arraybackslash}p{0.1025\linewidth}}{\ifnum 1=1\relax\TableScore{81.81}{0}\else --\fi} & \multicolumn{3}{>{\centering\arraybackslash}p{0.1025\linewidth}}{\ifnum 1=1\relax\TableScore{55.77}{0}\else --\fi} & \multicolumn{3}{>{\centering\arraybackslash}p{0.1025\linewidth}}{\ifnum 1=1\relax\TableScore{43.46}{0}\else --\fi} & \multicolumn{3}{>{\centering\arraybackslash}p{0.1025\linewidth}}{\ifnum 1=1\relax\TableScore{80.95}{0}\else --\fi} \\
      \hspace*{1em}Aurora & \ifnum 1=1\relax\TableScore{36.79}{0}\else --\fi & \ifnum 1=1\relax\TableScore{59.91}{0}\else --\fi & \ifnum 1=1\relax\TableScore{64.79}{0}\else --\fi & \ifnum 1=1\relax\TableScore{69.28}{0}\else --\fi & \multicolumn{3}{>{\centering\arraybackslash}p{0.1025\linewidth}}{\ifnum 1=1\relax\TableScore{87.23}{1}\else --\fi} & \multicolumn{3}{>{\centering\arraybackslash}p{0.1025\linewidth}}{\ifnum 1=1\relax\TableScore{61.12}{1}\else --\fi} & \multicolumn{3}{>{\centering\arraybackslash}p{0.1025\linewidth}}{\ifnum 1=1\relax\TableScore{44.03}{0}\else --\fi} & \multicolumn{3}{>{\centering\arraybackslash}p{0.1025\linewidth}}{\ifnum 1=1\relax\TableScore{83.14}{0}\else --\fi} \\
      \hspace*{1em}FCN3 & \ifnum 1=1\relax\TableScore{29.86}{0}\else --\fi & \ifnum 1=1\relax\TableScore{57.62}{0}\else --\fi & \ifnum 1=1\relax\TableScore{56.16}{0}\else --\fi & \ifnum 1=1\relax\TableScore{61.94}{0}\else --\fi & \multicolumn{3}{>{\centering\arraybackslash}p{0.1025\linewidth}}{\ifnum 1=1\relax\TableScore{81.74}{0}\else --\fi} & \multicolumn{3}{>{\centering\arraybackslash}p{0.1025\linewidth}}{\ifnum 1=1\relax\TableScore{51.75}{0}\else --\fi} & \multicolumn{3}{>{\centering\arraybackslash}p{0.1025\linewidth}}{\ifnum 1=1\relax\TableScore{29.19}{0}\else --\fi} & \multicolumn{3}{>{\centering\arraybackslash}p{0.1025\linewidth}}{\ifnum 1=1\relax\TableScore{77.58}{0}\else --\fi} \\
      \bottomrule
    \end{tabular}
    \begin{tablenotes}[flushleft]
      \TableNoteFont
      \item Higher is better; bold marks each physical family's maximum; \textbf{Hyd.}: hydrostatic balance; \textbf{Geos.}: geostrophic wind and shear; \textbf{Kinem.}: flow kinematics; \textbf{Thermo.}: thermodynamics; \textbf{Adv.}: heat and moisture advection; \textbf{Stab.}: static stability; \textbf{Frontog.}: frontogenesis; \textbf{Vapor}: water vapor content; \textbf{Moist.}: moisture transport and convergence; \textbf{Circ.}: inner-core circulation; \textbf{PV}: potential vorticity and rotation; \textbf{Outflow}: upper-level outflow; \textbf{Warm}: warm-core structure. '\textbf{Thermo.}' denotes stability and energy state in the universal group, but the energy difference between atmospheric layers in the precipitation-specific group.
    \end{tablenotes}
  \end{threeparttable}
\end{table}

%% file: sections/07_discussion.tex
\section{Discussion and Conclusion}
\label{sec:discussion}
\label{sec:conclusion}

ExtremePhysBench provides a reusable protocol for evaluating physical structure within extreme weather events. The comparison between the AI models and GFS is conditional on the available initialization and verification settings. The GFS self-analysis check makes the reference sensitivity explicit; it does not establish a general advantage of AI models over NWP systems. Positive associations with RMSE, alongside differences in model ranking, support reporting physical structure agreement together with conventional field accuracy.

\textbf{Physical families connect scores to atmospheric processes.}
Different models lead water content and moisture transport for precipitation, and circulation and warm-core structure for TCs. Tracing aggregate scores through physical families to individual diagnostics identifies the processes associated with these differences. Universal diagnostics support comparisons across event categories; event-specific diagnostics retain each hazard's characteristic processes. Scores summarize physical structure agreement with ERA5, while diagnostic errors in their original units preserve its physical magnitude. Together, these reports complement field accuracy and hazard verification.

\textbf{Evaluation target and reference.}
We evaluate agreement with ERA5 in how temperature, humidity, and wind relate across locations and heights. This requires complete three-dimensional fields, which separate observations do not directly provide. ERA5 supplies these fields \citep{hersbach2020era5} and is a practical compromise for comparing the available model outputs on a common $0.25^\circ$ grid. At this scale, small TC inner-core features and peak winds cannot be fully resolved \citep{ecmwf2024era5uncertainty}. The scores therefore measure agreement with ERA5's resolved structures, rather than accuracy in real TC intensity or peak rainfall \citep{lavers2022era5precipitation}.

\textbf{Reproducible comparisons and future work.}
Fixed tasks, explicit scoring rules, and saved errors and provenance make comparisons repeatable. Applying the evaluation to successive model versions can track whether improvements extend across physical families and event categories. Future work can expand event and process coverage and assess uncertainty through probabilistic forecasts.

%% file: appendices/A_related_frameworks.tex
\section{Comparison with Related Evaluation Frameworks}
\label{app:related-work-comparison}

The main text discusses benchmarks that focus on extreme or severe weather. \Tabref{tab:related-work-comparison} places those tasks alongside broader weather evaluation frameworks. The comparison describes the published evaluation targets, rather than treating the absence of a particular experiment as a restriction on what a software package could support.

\input{tables/related_work_comparison}

\paragraph{General forecast evaluation.}
WeatherBench and WeatherBench~2 establish common datasets and metrics for comparing global forecast fields \citep{rasp2020weatherbench,rasp2024weatherbench2}. WeatherBench-X provides modular evaluation infrastructure \citep{weatherbenchx2025}. These resources make conventional comparisons reproducible, but a global aggregate alone cannot locate physical structure departures inside cataloged extremes. ClimateLearn includes an Extreme-ERA5 experiment in which temperature errors are evaluated on pixels selected by percentile thresholds \citep{nguyen2023climatelearn}. This conditions field accuracy on extreme temperatures; it does not organize the evaluation around the physical families used here. RealBench examines forecast performance under operational conditions and extreme weather event challenges \citep{li2026realbench}, which also makes initialization and deployment conditions part of evaluation.

\paragraph{Evaluation within extreme weather events.}
HR-Extreme provides extreme weather cases from the High-Resolution Rapid Refresh model and reports errors within event masks \citep{ran2025hrextreme}. ExtremeWeatherBench standardizes cases, observations, and metrics relevant to hazards \citep{mcgovern2026extremeweatherbench}. TCBench focuses on tropical cyclone track and intensity \citep{gomez2026tcbench}. WeatherQA uses meteorological images and expert discussions to evaluate reasoning about severe weather \citep{ma2024weatherqa}. These evaluations differ in inputs, outputs, spatial scales, and targets. ExtremePhysBench adds comparisons of physical diagnostics derived from forecast fields within an independently constructed event catalog. Its four categories do not imply broader hazard coverage than those resources.

\paragraph{Physical diagnostics.}
Studies of learned dynamics and model imbalance motivate examining physical relationships alongside field accuracy \citep{hakim2024dynamical,selz2023butterfly,bonavita2024limitations,subramaniam2025hydrostatic}. PhysMetrics.Weather brings conservation, spectral, dynamical, and thermodynamic diagnostics into a common evaluation of global forecasts \citep{kasteleyn2026physmetrics}. Its evaluation on global samples addresses a different sampling question from repeated evaluation inside extreme weather events. ExtremePhysBench selects its physical families to describe physical structures within those events and compares their forecast values with the fifth-generation reanalysis (ERA5) from the European Centre for Medium-Range Weather Forecasts. Neither the availability of diagnostic fields nor agreement with a reanalysis establishes closure of all prognostic equations from archived snapshots.

\FloatBarrier

%% file: tables/related_work_comparison.tex
\begin{table}[!htbp]
\centering
\caption{Published evaluation targets of representative frameworks. Physical structure evaluation complements conventional forecast and event verification.}
\label{tab:related-work-comparison}
\scriptsize
\setlength{\tabcolsep}{4pt}
\begin{tabularx}{\linewidth}{@{}p{0.23\linewidth}p{0.25\linewidth}Y@{}}
\toprule
\textbf{Framework} & \textbf{Evaluation setting} & \textbf{Principal comparison}\\
\midrule
HR-Extreme & Extreme weather cases & Forecast field errors within event masks.\\
ExtremeWeatherBench & Curated weather hazards & Observations and verification metrics specific to each phenomenon.\\
TCBench & Tropical cyclones & Track and intensity forecasts.\\
WeatherQA & Severe weather reasoning & Affected areas and severe convection potential from multimodal inputs.\\
\addlinespace
WeatherBench / WeatherBench2 & Global weather forecasts & Accuracy of forecast variables under shared evaluation protocols.\\
WeatherBench-X & Weather evaluation software & Modular computation of forecast verification metrics.\\
ClimateLearn & Weather and climate tasks & Includes temperature errors conditioned on extremes.\\
RealBench & Operational conditions and extremes & Performance under operational forecasting constraints.\\
PhysMetrics.Weather & Global forecast samples & Conservation, spectral, dynamical, and thermodynamic diagnostics.\\
\addlinespace
\rowcolor{gray!12}
\textbf{ExtremePhysBench} & \textbf{Cataloged extreme weather events} & \textbf{Physical structure by diagnostic, physical family, event category, and lead.}\\
\bottomrule
\end{tabularx}
\end{table}

%% file: appendices/B_dataset_details.tex
\section{Extreme Weather Events Dataset Details}
\label{app:dataset-details}

The catalog covers 2022--2025 as one evaluation period. These years are not a train, validation, or test partition for the forecast systems; model exposure dates are reported separately in \Tabref{tab:models}. Event construction used only the stated source products, and its scripts do not read forecast outputs.

\subsection{Source Data and Common Conventions}
\label{htmltoc:b-dataset-details-1}

Thermal events used ERA5-HEAT daily minimum and maximum Universal Thermal Climate Index (UTCI) fields \citep{dinapoli2021era5heat,copernicus2020era5heat}. The files contain \texttt{utci\_daily\_min} and \texttt{utci\_daily\_max} on a $0.25^\circ$ grid from $90^\circ$N to $60^\circ$S and $180^\circ$W to $179.75^\circ$E. A 1991--2020 baseline supplied the local calendar day $p_{05}$ of daily minimum UTCI and $p_{95}$ of daily maximum UTCI. Precipitation events used Version 07B Final Run daily precipitation from the Global Precipitation Measurement (GPM) mission's Integrated Multi-satellitE Retrievals for GPM (IMERG) product \citep{huffman2020imerg,huffman2023imergdaily}, in mm day$^{-1}$. The source $0.1^\circ$ product was conservatively remapped with the Climate Data Operators \texttt{remapcon} command to a $0.25^\circ$, $1440\times721$ longitude--latitude grid after selecting precipitation and replacing source missing values with zero. Its local $p_{95}$ baseline used 2001--2020.

Both baselines pooled a nominal window of 15 days centered on each calendar date across all reference years. The implementation truncated rather than wrapped the window at the start and end of the calendar, so dates near those boundaries used 8--14 days and all other dates used 15. The baselines indexed calendar days against the 366 days of year 2000; February 29 was the mean of February 28 and March 1. Event masks followed the proleptic Gregorian calendar. Event construction used complete daily source sequences, and no missing day was replaced by a neighboring day. Comparisons involving a nonfinite source value or threshold did not enter a binary event mask.

Thermal and precipitation masks used the same ERA5 land mask, aligned to each source grid by nearest grid point. A cell was land when \texttt{lsm}$\geq0.5$, and gridded detection was restricted to $60^\circ$S--$60^\circ$N. Longitude was normalized only when required to align the source and mask grids; catalog centroids and task bounds use $[0,360)$.

\subsection{Gridded Event Construction}
\label{htmltoc:b-dataset-details-2}

The heat, cold, and precipitation pipelines used the following fixed sequence.
\begin{enumerate}
  \item For each calendar day, heat cells satisfied \texttt{utci\_daily\_max}$>p_{95}$, cold cells satisfied \texttt{utci\_daily\_min}$<p_{05}$, and precipitation cells satisfied both precipitation$\geq p_{95}$ and precipitation$\geq10$ mm day$^{-1}$. All masks were intersected with the common land and latitude mask.
  \item TempestExtremes \texttt{DetectBlobs} labeled spatial components with diagonal connectivity. Monthly files were listed chronologically, and \texttt{StitchBlobs} processed the complete 2022--2025 sequence so objects could cross file, month, year, and calendar boundaries. It required at least three daily steps and at least 1,600 grid cells per thermal blob or 300 grid cells per precipitation blob. Because eligibility counts cells on a regular latitude--longitude grid rather than physical area, the implied area threshold decreases with absolute latitude within the $60^\circ$ bounds.
  \item \texttt{BlobStats} recorded the event ID, day index, date, area, centroid, and latitude--longitude bounds. The construction script applied a fixed start buffer of nine days and removed objects whose day 0 date preceded 2022-01-10; IDs were then renumbered consistently in the catalog and event label NetCDF. No further merging or post hoc deduplication was applied: one stitched label defined one catalog event.
  \item A task used 00:00 Coordinated Universal Time (UTC) on event day 1, the second retained catalog day, as its target. Because every retained object had at least three days, this was the earliest fixed index with a retained day on both sides. It supplied one common anchor without selecting a peak or consulting forecast output. Each event produced seven tasks at 24--168 h in 24 h increments. The daily catalog mask fixed the evaluation region, while the forecast represented the prescribed instantaneous target time. Thus, 00 UTC need not coincide with the subdaily UTCI extremum or precipitation episode; conditioning is on the event day and mask rather than confirmation that the instantaneous state is itself extreme.
\end{enumerate}

\subsection{Landfalling Tropical Cyclones}
\label{htmltoc:b-dataset-details-3}

Tropical cyclone (TC) construction used version 04r01 of the International Best Track Archive for Climate Stewardship (IBTrACS) \citep{knapp2010ibtracs,gahtan2024ibtracs}. We retained storms whose complete recorded lifecycle fell within 2022--2025, selected 00, 06, 12, and 18 UTC fixes with valid positions, and identified every consecutive transition from an offshore fix with \texttt{DIST2LAND}$>0$ km to a landed fix 6 h later with \texttt{DIST2LAND}$\leq0$ km. The offshore fix required \texttt{USA\_WIND}$\geq34$ kt. When a storm had multiple eligible transitions, the deterministic order was strongest wind, lowest pressure, then earliest time, all evaluated at the offshore fix. The target was 24 h before that offshore fix. The nearest best track fix to the target had to be within 3 h. Task leads ran from 6 to 120 h in 6 h increments. TC initializations were retained only from the first synoptic fix with \texttt{USA\_WIND}$\geq25$ kt onward, excluding weaker precursor states. The final TC catalog contains 125 events across six ocean basins and 1,452 forecast tasks.

\subsection{Catalog and Task Records}
\label{htmltoc:b-dataset-details-4}

Catalogs record event identifiers, times, locations, and evaluation geometry. Task tables link each event to its target time, initialization, and forecast lead. Gridded events use catalog masks; TC tasks retain the IBTrACS positions used to initialize center searches. The benchmark package contains construction scripts, derived masks, catalogs, and task tables with source versions and citations.

\subsection{Catalog Summary}
\label{htmltoc:b-dataset-details-5}

\begin{figure}[!htbp]
  \centering
  \includegraphics[width=\linewidth]{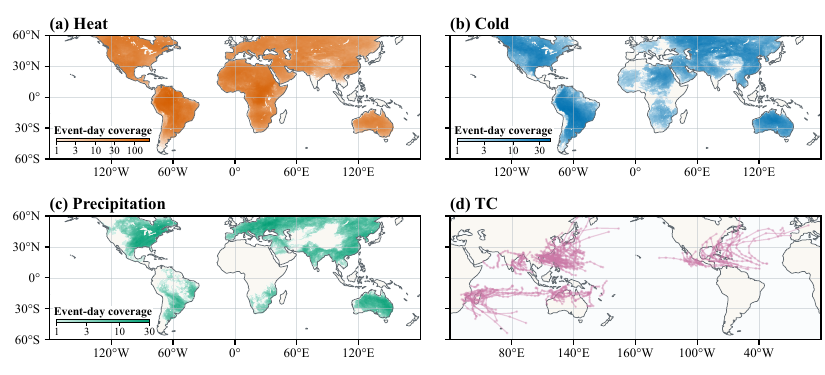}

  \caption{Global coverage of the Extreme Weather Events Dataset. Panels (a)--(c) count retained event days at each grid cell. Panel (d) shows IBTrACS record of the TC tracks during 2022--2025.}
  \label{fig:event-distribution-overview}
\end{figure}

\Tabref{tab:event-temporal-distribution} reports annual and monthly counts by target time. \Tabref{tab:event-duration-area} summarizes gridded event duration and target mask area. \Figref{fig:event-distribution-overview} shows geographic coverage.

\begin{table}[!htbp]
  \centering
  \caption{Temporal distribution of retained events by target time.}
  \label{tab:event-temporal-distribution}
  \scriptsize
  \setlength{\tabcolsep}{3.5pt}
  \begin{tabular}{@{}lrrrr@{}}
    \toprule
    \textbf{Year} & \textbf{Heat} & \textbf{Cold} & \textbf{Precipitation} & \textbf{TC} \\
    \midrule

      2022 & 52 & 38 & 70 & 32 \\
      2023 & 82 & 29 & 60 & 31 \\
      2024 & 106 & 20 & 68 & 33 \\
      2025 & 74 & 19 & 77 & 29 \\
    \bottomrule
  \end{tabular}

  \medskip
  \setlength{\tabcolsep}{2.6pt}
  \begin{tabular}{@{}lrrrrrrrrrrrr@{}}
    \toprule
    & \textbf{Jan} & \textbf{Feb} & \textbf{Mar} & \textbf{Apr} & \textbf{May} & \textbf{Jun} & \textbf{Jul} & \textbf{Aug} & \textbf{Sep} & \textbf{Oct} & \textbf{Nov} & \textbf{Dec} \\
    \midrule
      Heat & 17 & 24 & 22 & 21 & 24 & 30 & 30 & 39 & 38 & 29 & 17 & 23 \\
      Cold & 13 & 13 & 9 & 9 & 14 & 7 & 5 & 7 & 4 & 11 & 4 & 10 \\
      Precipitation & 13 & 16 & 33 & 27 & 19 & 19 & 21 & 31 & 26 & 22 & 29 & 19 \\
      TC & 6 & 7 & 7 & 3 & 6 & 6 & 13 & 15 & 22 & 22 & 11 & 7 \\
    \bottomrule
  \end{tabular}
\end{table}

\begin{table}[!htbp]
  \centering
  \caption{Duration and target region area of retained gridded events.}
  \label{tab:event-duration-area}
  \scriptsize
  \setlength{\tabcolsep}{4pt}
  \begin{tabular}{@{}lrrrrrrr@{}}
    \toprule
    & \multicolumn{4}{c}{\textbf{Duration (days)}} & \multicolumn{3}{c}{\textbf{Target area ($10^6$ km$^2$)}} \\
    \cmidrule(lr){2-5}\cmidrule(l){6-8}
    \textbf{Category} & \textbf{Min.} & \textbf{Median} & \textbf{95th pct.} & \textbf{Max.} & \textbf{5th pct.} & \textbf{Median} & \textbf{95th pct.} \\
    \midrule

      Heat & 3 & 5 & 15 & 53 & 1.06 & 1.82 & 4.11 \\
      Cold & 3 & 4 & 8 & 12 & 1.07 & 2.21 & 4.22 \\
      Precipitation & 3 & 3 & 5 & 8 & 0.23 & 0.45 & 1.14 \\
    \bottomrule
  \end{tabular}
\end{table}

%% file: appendices/C_inventory_tables.tex
\section{Benchmark Inventory, Scales, and Aggregation}
\label{app:benchmark-inventory}

\subsection{Diagnostic and Task Inventory}
\label{app:diagnostic-inventory}

\Tabref{tab:physical-structures} introduces the diagnostics in the main text. \Tabrefs{tab:universal-metrics} and~\ref{tab:event-metrics} give their units, pressure levels, and precise scored objects. \Tabref{tab:task-scope} reports event, task, and initialization counts.

\input{tables/universal_metrics}
\input{tables/event_metrics}
\input{tables/task_scope}

\subsection{Fixed Physics Score Scales}
\label{app:score-scales}

\input{tables/scale_parameters}
\input{tables/score_scales}

\begin{figure}[!htbp]
  \centering
  \includegraphics[width=0.45\linewidth]{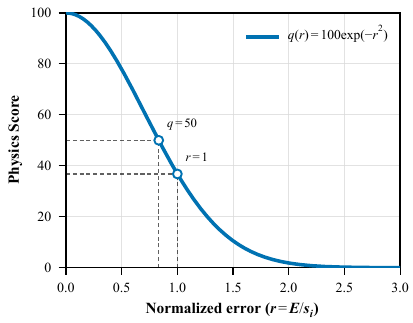}
  \caption{Fixed mapping from physical diagnostic error to task Physics Score. The mapping $q_{i,k}=100\exp(-r^2)$ uses normalized root mean squared error (RMSE) $r=E_{i,k}/(C_i\tau_i)$ and is applied before averaging. An RMSE equal to its prescribed scale maps to 36.8 points.}
  \label{fig:physics-score-mapping}
\end{figure}

\Tabref{tab:scale-parameters} lists the prescribed regime parameters used to construct $C_i$; \Tabref{tab:score-scales} gives the resulting characteristic magnitudes, tolerances, and score scales.
For every domain--diagnostic row, the benchmark declares a characteristic magnitude $C_i$ and a dimensionless tolerance $\tau_i$, with dimensional score scale
\begin{equation}
 S_i=C_i\tau_i.
 \label{eq:score-scale}
\end{equation}
This scale assigns a score of $100/e$ when the diagnostic error equals $S_i$. Heat and cold share the thermal stress scales. All scales and tolerances are fixed across systems, events, and leads and are not fitted to forecast errors or event statistics. Tolerances express scoring choices, not observational uncertainty or thresholds of physical correctness. Errors in the original diagnostic units are retained alongside scores.
\Figref{fig:physics-score-mapping} shows the bounded transformation from normalized error $r=E_{i,k}/S_i$ to a score between 0 and 100.

\subsection{Explicit Hierarchical Estimator}
\label{app:score-aggregation}

The estimator controls how much influence each event and physical family has on a summary. Mapping before aggregation is essential because the exponential map is nonlinear: mapping a mean squared error would define a different score. Let $q_{me\ell i}$ be the mapped score for model $m$, event $e$, retained task or lead $\ell$, and diagnostic $i$. Let $L_{ei}$ contain the tasks on which diagnostic $i$ is applicable according to the reference. For a complete model record, the event and diagnostic mean is
\begin{equation}
 \bar q_{mei}=\frac{1}{|L_{ei}|}\sum_{\ell\in L_{ei}}q_{me\ell i}.
 \label{eq:appendix-event-diagnostic}
\end{equation}
An empty applicability set does not receive a score. Missing or invalid model output on a reference applicable task makes the corresponding record incomplete rather than reducing its denominator. Model comparisons retain the same complete events across all compared systems within each category.

Applicability also determines which diagnostics and physical families contribute within an event. Write $I_{ef}$ for the applicable diagnostics in physical family $f$, $F_{eg}$ for the nonempty physical families in group $g\in\{U,E\}$, and $\mathcal E_c$ for the common complete events in category $c$. The hierarchy is
\begin{align}
 Q_{mef}&=\frac{1}{|I_{ef}|}\sum_{i\in I_{ef}}\bar q_{mei},\\
 Q_{meg}&=\frac{1}{|F_{eg}|}\sum_{f\in F_{eg}}Q_{mef},\\
 Q_{me}^{\mathrm{event}}&=\tfrac12(Q_{meU}+Q_{meE}),\\
 Q_{mc}^{\mathrm{category}}&=\frac{1}{|\mathcal E_c|}
   \sum_{e\in\mathcal E_c}Q_{me}^{\mathrm{event}}.
 \label{eq:appendix-category-score}
\end{align}
Thus each applicable diagnostic has equal weight within its physical family, each nonempty physical family has equal weight within its group, and the two groups have equal weight within an event. Each complete event then has equal weight within its category. Inapplicability, which is determined by the reference, is distinct from invalid forecast output. For example, an equatorial event can omit geostrophic diagnostics for every model without assigning them zero scores.

Overall retains the common universal group and equal category weights:
\begin{equation}
 Q_m^{\mathrm{overall}}=\frac14\sum_{c\in\{\mathrm{Categories}\}}
 \frac{1}{|\mathcal E_c|}\sum_{e\in\mathcal E_c}Q_{meU}.
 \label{eq:appendix-overall-score}
\end{equation}
This avoids averaging physically different event-specific diagnostics across categories. Scores at each forecast lead are averaged equally across events.

\paragraph{Physical family reports.}
For the main physical family table, let $\mathcal E_{cf}$ be the events in $\mathcal E_c$ with nonempty physical family $f$. The reported mean is
\begin{equation}
 Q_{mcf}^{\mathrm{family}}=\frac{1}{|\mathcal E_{cf}|}
   \sum_{e\in\mathcal E_{cf}}Q_{mef}.
 \label{eq:appendix-family-report}
\end{equation}
Its event count is $|\mathcal E_{cf}|$. A physical family with no supported events is reported as missing. When physical family applicability differs across events, averaging these final means does not reproduce \Eqref{eq:appendix-category-score}; reconstruction must start from the physical family scores within each event. The report retains separate universal and event-specific physical families even where their display names are both thermodynamics.

\paragraph{Weighting rationale.}
Hierarchical averaging specifies the contribution of each level rather than allowing groups with more observations or more measures to dominate automatically. The guidance from the Organisation for Economic Co-operation and Development and the European Commission Joint Research Centre on composite indicators treats normalization, weighting, and aggregation as choices that require justification and sensitivity assessment \citep{oecd2008composite}. Equal weights here express the benchmark's reporting priorities; they do not imply that the physical families have identical causal importance. Correlated diagnostics and compensation between physical families remain relevant limitations. \Appref{app:score-sensitivity} evaluates sensitivity to tolerance and weighting choices.

\paragraph{Reporting beyond a leaderboard.}
LiveBench presents category results beside an overall score, while Holistic Evaluation of Language Models evaluates multiple dimensions under shared scenarios \citep{white2024livebench,liang2023helm}. We use the corresponding reporting principle to keep the event category, physical family, and diagnostic views visible. The overall ranking is an entry point to the evaluation; the disaggregated results identify the particular physical structures responsible for it. This motivates the result layout without importing another benchmark's task definitions, units, or scoring rule.

\FloatBarrier

%% file: tables/universal_metrics.tex
\begin{table}[!htbp]
  \centering
  \begin{threeparttable}
    \caption{Universal physical structure diagnostics. Ten diagnostics provide a shared basis for evaluation across all four event categories.}
    \label{tab:universal-metrics}
    \scriptsize
    \setlength{\tabcolsep}{2.5pt}
    \begin{tabularx}{\linewidth}{@{}>{\raggedright\arraybackslash}p{0.31\linewidth}Y
      >{\raggedright\arraybackslash}p{0.14\linewidth}
      >{\raggedright\arraybackslash}p{0.07\linewidth}@{}}
      \toprule
      \textbf{Diagnostic} & \textbf{Scored physical structure} &
      \textbf{Level(s) (hPa)} & \textbf{Units} \\
      \midrule
      \DiagnosticEntry{U1}{Normalized signed hydrostatic components} & Three sublayer thickness residuals with symmetric normalization & 850, 700, 600, 500 & 1 \\
      \DiagnosticEntry{U2}{Geostrophic-wind components} & Geopotential-derived eastward and northward geostrophic wind & 500 & m s$^{-1}$ \\
      \DiagnosticEntry{U3}{Geostrophic-wind shear components} & Eastward and northward 850-minus-500 hPa geostrophic shear & 850, 500 & m s$^{-1}$ \\
      \DiagnosticEntry{U4}{Relative vorticity} & Spherical rotational structure & 850 & s$^{-1}$ \\
      \DiagnosticEntry{U5}{Horizontal divergence} & Spherical divergent structure & 850 & s$^{-1}$ \\
      \DiagnosticEntry{U6}{Relative vorticity} & Midtropospheric rotational structure & 500 & s$^{-1}$ \\
      \DiagnosticEntry{U7}{Horizontal divergence} & Midtropospheric divergent structure & 500 & s$^{-1}$ \\
      \DiagnosticEntry{U8}{Horizontal deformation components} & Signed spherical stretching and shearing components & 850 & s$^{-1}$ \\
      \DiagnosticEntry{U9}{Dry Brunt--V\"ais\"al\"a frequency squared} & Dry-potential-temperature layer stability & 850--500 & s$^{-2}$ \\
      \DiagnosticEntry{U10}{Moist static energy} & Thermodynamic energy state & 850 & J kg$^{-1}$ \\
      \bottomrule
    \end{tabularx}
    \begin{tablenotes}[flushleft]
      \scriptsize
      \item Each diagnostic is derived independently from forecast and ERA5 fields. U2 and U3 exclude $|\phi|<15^\circ$. Horizontal derivatives use spherical operators with a cyclic longitude stencil on a complete global grid. Joint-component errors preserve signed layer or vector components.
    \end{tablenotes}
  \end{threeparttable}
\end{table}

%% file: tables/event_metrics.tex
\begin{table}[!htbp]
  \centering
  \begin{threeparttable}
    \caption{Event-specific physical structure diagnostics. Each category uses five additional diagnostics for thermal-stress, moisture, or vortex structures; heat and cold share the thermal-stress block.}
    \label{tab:event-metrics}
    \scriptsize
    \setlength{\tabcolsep}{2.3pt}
    \begin{tabularx}{\linewidth}{@{}>{\raggedright\arraybackslash}p{0.13\linewidth}
      >{\raggedright\arraybackslash}p{0.34\linewidth}
      >{\raggedright\arraybackslash}p{0.14\linewidth}Y@{}}
      \toprule
      \textbf{Block} & \textbf{Diagnostic} & \textbf{Level(s) (hPa)} &
      \textbf{Scored physical structure} \\
      \midrule
      Heat/cold & \DiagnosticEntry{E1}{Temperature advection} & 850 & Horizontal dry-temperature transport \\
       & \DiagnosticEntry{E2}{Moisture advection} & 850 & Horizontal specific-humidity transport \\
       & \DiagnosticEntry{E3}{Partial-layer horizontal MSE advection} & 850--500 & Gauge-invariant $-\int \mathbf V\cdot\nabla h\,dp/g$ contribution \\
       & \DiagnosticEntry{E4}{Dry Brunt--V\"ais\"al\"a frequency squared} & 850--700 & Lower-tropospheric dry stability \\
       & \DiagnosticEntry{E5}{Horizontal kinematic temperature frontogenesis} & 850 & Horizontal change of temperature-gradient magnitude \\
      \addlinespace[2pt]
      Precipitation & \DiagnosticEntry{E1}{Free-tropospheric water vapor} & 850--300 & Layer water amount \\
       & \DiagnosticEntry{E2}{Free-tropospheric vapor-transport components} & 850--300 & Eastward and northward vapor transport \\
       & \DiagnosticEntry{E3}{Free-tropospheric vapor-flux convergence} & 850--300 & Layer-integrated vapor convergence \\
       & \DiagnosticEntry{E4}{Total moisture-flux convergence} & 850 & Low-level $-\nabla\!\cdot(q\mathbf V)$ \\
       & \DiagnosticEntry{E5}{Total MSE layer difference} & 850--500 & Signed lower-minus-midlevel thermodynamic contrast \\
      \addlinespace[2pt]
      TC & \DiagnosticEntry{E1}{Axisymmetric tangential wind} & 850; 0--150 km & Inner-core circulation profile \\
       & \DiagnosticEntry{E2}{Total Ertel potential vorticity} & 700 & Rotation, stratification, and tilting \\
       & \DiagnosticEntry{E3}{Radial relative vorticity} & 500 & Midlevel vortex profile \\
       & \DiagnosticEntry{E4}{Inner-outflow radial divergence} & 200; 0--500 km & Upper-level outflow profile \\
       & \DiagnosticEntry{E5}{Warm-core thickness anomaly} & 300--700 & Inner-core thickness relative to an ERA5-defined environment \\
      \bottomrule
    \end{tabularx}
    \begin{tablenotes}[flushleft]
      \scriptsize
      \item E1--E5 are local codes within each category block. MSE denotes moist static energy. Heat and cold share one diagnostic configuration but use distinct event masks. ERA5 defines the radial sampling support and warm-core environmental annulus; the same geometry and weights are applied around each forecast's own storm center.
    \end{tablenotes}
  \end{threeparttable}
\end{table}

%% file: tables/task_scope.tex
\begin{table}[!htbp]
  \centering
  \begin{threeparttable}
    \caption{Final event and forecast task inventory by category.}
    \label{tab:task-scope}
    \scriptsize
    \begin{tabularx}{0.6\linewidth}{@{}Y r r c r@{}}
      \toprule
      \textbf{Event category} & {\textbf{Events}} & {\textbf{Tasks}} &
      \textbf{Leads (h)} &
      {\makecell[c]{\textbf{Unique initializations}}} \\
      \midrule

      Extreme heat stress & 314 & 2198 & $24{:}24{:}168$ & 1160 \\
      Extreme cold stress & 106 & 742 & $24{:}24{:}168$ & 602 \\
      Extreme precipitation & 275 & 1925 & $24{:}24{:}168$ & 1093 \\
      Landfalling TC & 125 & 1452 & $6{:}6{:}120$ & 1296 \\
      \midrule
      \bfseries Total & {\bfseries 820} & {\bfseries 6317} & -- & {--} \\
      \bottomrule
    \end{tabularx}
  \end{threeparttable}
\end{table}

%% file: tables/scale_parameters.tex
\begin{table}[!htbp]
  \centering
  \begin{threeparttable}
    \caption{Fixed regime parameters for constructing characteristic magnitudes $C_i$.}
    \label{tab:scale-parameters}
    \scriptsize
    \setlength{\tabcolsep}{2.5pt}
    \begin{tabularx}{\linewidth}{@{}>{\raggedright\arraybackslash}p{0.17\linewidth}Y@{}}
      \toprule
      \textbf{Regime or operator} & \textbf{Pre-specified characteristic parameters} \\
      \midrule
      Synoptic universal
        & $L=250$ km, $U=15$ m s$^{-1}$, $f_0=10^{-4}$ s$^{-1}$, $\theta_0=300$ K, $H=5$ km, $\Delta\theta=20$ K, $\Delta T=10$ K, $\Delta q=0.005$ kg kg$^{-1}$, and $\Delta H=1$ km. These values apply to the universal block for stress and precipitation events. \\
      Thermal stress
        & Universal synoptic values; the 850--500-hPa MSE-advection scale additionally uses $\Delta p=350$ hPa and $\Delta h=3$ kJ kg$^{-1}$; the 850--700-hPa stability scale uses $H=1.5$ km and $\Delta\theta=10$ K. \\
      Precipitation
        & Universal synoptic values; the fixed free-tropospheric layer uses $\Delta p=550$ hPa and $\bar q=0.005$ kg kg$^{-1}$. The 850-hPa flux-convergence scale uses $q_0=0.01$ kg kg$^{-1}$ and $L=500$ km. The MSE layer-difference scale uses $\Delta T=30$ K, $\Delta q=0.008$ kg kg$^{-1}$, and $\bar T=270$ K. \\
      Tropical cyclone
        & $R=L=200$ km, $U=40$ m s$^{-1}$, $f_0=5\times10^{-5}$ s$^{-1}$, $\theta_0=300$ K, $H=5$ km, $\Delta\theta=20$ K, $\Delta T=8$ K, $\Delta q=0.008$ kg kg$^{-1}$, and $\Delta H=1.5$ km. The event block uses $U_{850}=40$, $U_{500}=30$, and $U_{200}=20$ m s$^{-1}$; potential vorticity (PV) uses $\Delta p=350$ hPa. \\
      \bottomrule
    \end{tabularx}
    \begin{tablenotes}[flushleft]
      \item Constants are $g=\SI{9.80665}{m.s^{-2}}$, $R_d=\SI{287.05}{J.kg^{-1}.K^{-1}}$, $c_p=\SI{1004.0}{J.kg^{-1}.K^{-1}}$, and $L_v=\SI{2.5e6}{J.kg^{-1}}$.
    \end{tablenotes}
  \end{threeparttable}
\end{table}

%% file: tables/score_scales.tex
\begin{table}[!htbp]
\centering
\caption{Fixed characteristic magnitudes $C_i$, tolerance fractions $\tau_i$, and score scales $S_i=C_i\tau_i$. Heat, cold, and precipitation share the universal synoptic scales; TC uses a separate regime.}
\label{tab:score-scales}
\scriptsize
\setlength{\tabcolsep}{2pt}
\begin{tabular}{@{}lp{0.055\linewidth}p{0.30\linewidth}>{\raggedright\arraybackslash}p{0.15\linewidth}rrrl@{}}
\toprule
\textbf{Regime} & \textbf{Code} & \textbf{Diagnostic} & \textbf{$C_i$ construction} & \textbf{$C_i$} & \textbf{$\tau_i$} & \textbf{$S_i$} & \textbf{Units} \\
\midrule
Synoptic universal & U1 & Hydrostatic normalized residual components & $1$ & 1 & 0.001 & 0.001 & 1 \\
 & U2 & Geostrophic-wind components, 500 & $U/\sqrt{2}$ & 10.6066 & 0.50 & 5.3033 & m s$^{-1}$ \\
 & U3 & Geostrophic-wind shear components, 850--500 & $U/\sqrt{2}$ & 10.6066 & 0.50 & 5.3033 & m s$^{-1}$ \\
 & U4/U6 & Relative vorticity, 850/500 & $U/L$ & $6.0000\times10^{-5}$ & 0.50 & $3.0000\times10^{-5}$ & s$^{-1}$ \\
 & U5/U7 & Divergence, 850/500 & $U^2/(f_0L^2)$ & $3.6000\times10^{-5}$ & 1.00 & $3.6000\times10^{-5}$ & s$^{-1}$ \\
 & U8 & Deformation components, 850 & $U/(\sqrt{2}L)$ & $4.2426\times10^{-5}$ & 1.00 & $4.2426\times10^{-5}$ & s$^{-1}$ \\
 & U9 & Dry $N^2$, 850--500 & $g\Delta\theta/(\theta_0H)$ & $1.3076\times10^{-4}$ & 0.10 & $1.3076\times10^{-5}$ & s$^{-2}$ \\
 & U10 & MSE, 850 & $c_p\Delta T+L_v\Delta q+g\Delta H$ & 32346.65 & 0.15 & 4851.9975 & J kg$^{-1}$ \\
\addlinespace
Thermal stress & E1 & Temperature advection, 850 & $U\Delta T/L$ & $6.0000\times10^{-4}$ & 0.25 & $1.5000\times10^{-4}$ & K s$^{-1}$ \\
 & E2 & Moisture advection, 850 & $U\Delta q/L$ & $3.0000\times10^{-7}$ & 0.25 & $7.5000\times10^{-8}$ & kg kg$^{-1}$ s$^{-1}$ \\
 & E3 & Partial-layer horizontal MSE advection, 850--500 & $(\Delta p/g)U\Delta h/L$ & 642.4212 & 0.50 & 321.2106 & W m$^{-2}$ \\
 & E4 & Dry $N^2$, 850--700 & $g\Delta\theta/(\theta_0H)$ & $2.1793\times10^{-4}$ & 0.25 & $5.4481\times10^{-5}$ & s$^{-2}$ \\
 & E5 & Horizontal kinematic temperature frontogenesis, 850 & $U\Delta T/L^2$ & $2.4000\times10^{-9}$ & 0.25 & $6.0000\times10^{-10}$ & K m$^{-1}$ s$^{-1}$ \\
\addlinespace
Precipitation & E1 & Free-tropospheric water vapor, 850--300 & $\Delta p\bar q/g$ & 28.0422 & 0.10 & 2.8042 & kg m$^{-2}$ \\
 & E2 & Vapor-transport components, 850--300 & $\Delta p\bar qU/g$ & 420.6329 & 0.10 & 42.0633 & kg m$^{-1}$ s$^{-1}$ \\
 & E3 & Vapor-flux convergence, 850--300 & $\Delta p\bar qU/(gL)$ & $1.6825\times10^{-3}$ & 0.25 & $4.2063\times10^{-4}$ & kg m$^{-2}$ s$^{-1}$ \\
 & E4 & Moisture-flux convergence, 850 & $q_0U/L$ & $3.0000\times10^{-7}$ & 1.00 & $3.0000\times10^{-7}$ & kg kg$^{-1}$ s$^{-1}$ \\
 & E5 & MSE layer difference, 850--500 & $|c_p\Delta T+L_v\Delta q-R_d\bar T\ln(850/500)|$ & 8994.4533 & 0.25 & 2248.6133 & J kg$^{-1}$ \\
\addlinespace
TC universal & U1 & Hydrostatic normalized residual components & $1$ & 1 & 0.001 & 0.001 & 1 \\
 & U2 & Geostrophic-wind components, 500 & $U/\sqrt{2}$ & 28.2843 & 0.50 & 14.1421 & m s$^{-1}$ \\
 & U3 & Geostrophic-wind shear components, 850--500 & $U/\sqrt{2}$ & 28.2843 & 0.50 & 14.1421 & m s$^{-1}$ \\
 & U4/U6 & Relative vorticity, 850/500 & $U/L$ & $2.0000\times10^{-4}$ & 0.50 & $1.0000\times10^{-4}$ & s$^{-1}$ \\
 & U5/U7 & Divergence, 850/500 & $U/L$ & $2.0000\times10^{-4}$ & 0.20 & $4.0000\times10^{-5}$ & s$^{-1}$ \\
 & U8 & Deformation components, 850 & $U/(\sqrt{2}L)$ & $1.4142\times10^{-4}$ & 0.50 & $7.0711\times10^{-5}$ & s$^{-1}$ \\
 & U9 & Dry $N^2$, 850--500 & $g\Delta\theta/(\theta_0H)$ & $1.3076\times10^{-4}$ & 0.10 & $1.3076\times10^{-5}$ & s$^{-2}$ \\
 & U10 & MSE, 850 & $c_p\Delta T+L_v\Delta q+g\Delta H$ & 42741.975 & 0.10 & 4274.1975 & J kg$^{-1}$ \\
\addlinespace
TC event & E1 & Tangential-wind profile, 850, 0--150 km & $U_{850}$ & 40.0 & 0.25 & 10.0 & m s$^{-1}$ \\
 & E2 & Total Ertel PV, 700 & $g(f_0+U/L)\Delta\theta/\Delta p$ & $1.4010\times10^{-6}$ & 0.25 & $3.5024\times10^{-7}$ & K m$^2$ kg$^{-1}$ s$^{-1}$ \\
 & E3 & Radial relative-vorticity profile, 500 & $U_{500}/R$ & $1.5000\times10^{-4}$ & 0.50 & $7.5000\times10^{-5}$ & s$^{-1}$ \\
 & E4 & Radial divergence profile, 200, 0--500 km & $U_{200}/R$ & $1.0000\times10^{-4}$ & 0.20 & $2.0000\times10^{-5}$ & s$^{-1}$ \\
 & E5 & Warm-core thickness anomaly, 300--700 & $R_d\Delta T\ln(700/300)/g$ & 198.4097 & 0.10 & 19.8410 & m \\
\bottomrule
\end{tabular}
\end{table}

%% file: appendices/D_physics_metrics.tex
\section{Diagnostic Operators and Error Geometry}
\label{app:physics-metrics}

This appendix specifies the observable physical structures compared by ExtremePhysBench. It does not define residuals of the complete prognostic equations. Terms requiring time tendencies, vertical motion, diabatic heating, friction, surface exchange, or subgrid tendencies are unavailable from the archived instantaneous fields on pressure levels. Every primary operator below is therefore evaluated independently on the forecast and ERA5 state, and the two resulting diagnostic fields are compared on support determined without the forecast. Classical definitions follow standard dynamic meteorology \citep{holton2013dynamic}; pressure levels, discrete geometry, support, and error aggregation are benchmark choices.

\subsection{Notation and Spherical Operators}
\label{htmltoc:d-physics-metrics-1}

Let longitude, latitude, and Earth radius be $(\lambda,\phi,a)$; let $(u,v)$ denote eastward and northward wind, $\Phi$ geopotential, $T$ dry-air temperature, $q$ specific humidity, $p$ pressure, $f=2\Omega\sin\phi$, and
\begin{equation}
  \begin{gathered}
  T_v=T(1+0.608q),\quad \theta=T(p_0/p)^{R_d/c_p},\\
  h=c_pT+\Phi+L_vq.
  \end{gathered}
  \label{eq:appendix-thermodynamic-definitions}
\end{equation}
Here $h$ denotes moist static energy (MSE). For a scalar $A$, the horizontal gradient in the local orthonormal east--north basis is
\begin{equation}
  \nabla_h A=\left(\frac{1}{a\cos\phi}\frac{\partial A}{\partial\lambda},
  \frac{1}{a}\frac{\partial A}{\partial\phi}\right).
  \label{eq:spherical-scalar-gradient}
\end{equation}
For horizontal wind, the covariant gradient components are
\begin{align}
  u_e&=\frac{1}{a\cos\phi}u_\lambda-\frac{v\tan\phi}{a}, &
  u_n&=\frac{1}{a}u_\phi,\\
  v_e&=\frac{1}{a\cos\phi}v_\lambda+\frac{u\tan\phi}{a}, &
  v_n&=\frac{1}{a}v_\phi.
  \label{eq:covariant-wind-gradient}
\end{align}
Hence divergence, vertical relative vorticity, stretching deformation, and shearing deformation are
\begin{equation}
  \begin{aligned}
  \delta&=u_e+v_n, & \zeta&=v_e-u_n,\\
  D_s&=u_e-v_n, & D_h&=v_e+u_n.
  \end{aligned}
  \label{eq:spherical-kinematics}
\end{equation}
Derivatives use centered finite differences where possible. Longitude is cyclic only when the input is a complete uniform global grid; the geographic poles are excluded from the singular eastward basis. The spherical connection terms are retained throughout the diagnostic suite.

\subsection{From Governing Relations to Observable Physical Structure}
\label{app:derivations}

The governing equations organize the diagnostic suite, but they are not themselves the scored object. For atmospheric state $X$ and diagnostic operator $\mathcal D_i$, forecast and ERA5 are processed independently,
\begin{equation}
  \begin{gathered}
  D_i^{\FC}=\mathcal D_i(X^{\FC}),\quad D_i^{\ERA}=\mathcal D_i(X^{\ERA}),\\
  \Delta D_i=D_i^{\FC}-D_i^{\ERA}.
  \end{gathered}
  \label{eq:operator-comparison-invariant}
\end{equation}
The derivations below therefore end at an observable field $D_i$ whose forecast--reference departure can be evaluated. Exact identities and coordinate transformations are written with equality signs. Geostrophic balance, source-free horizontal scalar evolution, and axisymmetric gradient-wind balance are controlled approximations; they motivate diagnostic fields or audit terms but do not assert closure of a forecast snapshot.

\subsubsection{Hydrostatic Structure, Thickness, and Stability}

Let $\rho$ be density and $z$ geometric height. The moist-air equation of state and hydrostatic balance are
\begin{equation}
  p=\rho R_dT_v,\qquad
  \frac{\partial p}{\partial z}=-\rho g,\qquad
  \mathrm d\Phi=g\,\mathrm dz.
  \label{eq:eos-hydrostatic}
\end{equation}
Eliminating density gives the pressure-coordinate differential identity
\begin{equation}
  \mathrm d\Phi=-R_dT_v\frac{\mathrm dp}{p}
  =-R_dT_v\,\mathrm d\ln p.
  \label{eq:hypsometric-differential}
\end{equation}
For lower pressure surface $p_l$ and upper surface $p_u<p_l$, integration yields the hypsometric relation
\begin{equation}
  \Phi_u-\Phi_l=R_d\int_{p_u}^{p_l}T_v\,\mathrm d\ln p.
  \label{eq:hypsometric-integral}
\end{equation}
Endpoint trapezoidal quadrature of \Eqref{eq:hypsometric-integral} produces the two independently represented layer-thickness terms $A_{lu}$ and $B_{lu}$ in U1. Their normalized signed difference is retained because the benchmark compares the forecast residual with the ERA5 residual; it does not require either residual to vanish.

Potential temperature follows from dry adiabatic compression or expansion,
\begin{equation}
  \theta=T\left(\frac{p_0}{p}\right)^\kappa,\qquad
  \kappa=\frac{R_d}{c_p}.
  \label{eq:potential-temperature-derivation}
\end{equation}
The dry squared Brunt--V\"ais\"al\"a frequency is the restoring-acceleration coefficient for a small vertical parcel displacement,
\begin{equation}
  N^2=\frac{g}{\theta}\frac{\partial\theta}{\partial z}.
  \label{eq:continuous-n2}
\end{equation}
Replacing the derivative by a finite layer difference and using $\Delta z=(\Phi_u-\Phi_l)/g$ gives \Eqref{eq:dry-n2}. Thus U9 and thermal-stress E4 are finite-layer dry-stability structures, not moist parcel-instability or convective available potential energy diagnostics.

\Eqref{eq:hypsometric-integral} also explains the TC warm-core operator. The thickness
\begin{equation}
  Z_{300:700}=\frac{\Phi_{300}-\Phi_{700}}{g}
  =\frac{R_d}{g}\int_{300}^{700}T_v\,\mathrm d\ln p
  \label{eq:warm-core-hypsometric}
\end{equation}
increases with layer-mean virtual temperature. Subtracting the ERA5-defined outer environment therefore isolates a warm-core thickness anomaly while removing the regional background.

\subsubsection{Rotating Momentum, Geostrophic Wind, and Thermal Wind}

On a pressure surface, horizontal momentum may be written
\begin{equation}
  \frac{D\bm V}{Dt}+f\bm k\times\bm V
  =-\nabla_h\Phi+\bm F,
  \label{eq:pressure-surface-momentum}
\end{equation}
where $\bm k$ is the local vertical unit vector and $\bm F$ collects frictional and unresolved momentum tendencies. Away from the equator, neglecting acceleration and $\bm F$ relative to the pressure-gradient and Coriolis terms gives the geostrophic approximation
\begin{equation}
  f\bm k\times\bm V_g\approx-\nabla_h\Phi,\qquad
  (u_g,v_g)=\left(-\frac{\Phi_y}{f},\frac{\Phi_x}{f}\right).
  \label{eq:geostrophic-derivation}
\end{equation}
The current U2 operator scores these two geopotential-derived components at 500 hPa. U3 scores their 850-minus-500-hPa endpoint difference, so neither U2 nor U3 mixes forecast wind with reference geopotential. The approximation is used to define an observable balanced-flow structure rather than to label ageostrophic motion as forecast error \citep{charney1948scale,holton2013dynamic}.

Differentiating \Eqref{eq:geostrophic-derivation} with respect to $\ln p$ and substituting \Eqref{eq:hypsometric-differential} gives
\begin{equation}
  \frac{\partial u_g}{\partial\ln p}
  =\frac{R_d}{f}\frac{\partial T_v}{\partial y},\qquad
  \frac{\partial v_g}{\partial\ln p}
  =-\frac{R_d}{f}\frac{\partial T_v}{\partial x}.
  \label{eq:thermal-wind-differential}
\end{equation}
Integrating from 500 to 850 hPa yields
\begin{equation}
  \bm V_{g,850}-\bm V_{g,500}
  =\frac{R_d}{f}
  \left(\partial_y,-\partial_x\right)
  \int_{500}^{850}T_v\,\mathrm d\ln p.
  \label{eq:thermal-wind-integral}
\end{equation}
The implementation evaluates the log-pressure integral by trapezoidal quadrature over 500, 600, 700, and 850 hPa. Its difference from the geopotential-derived endpoint shear is exported as a signed audit residual. It is not the primary U3 score, and both sides retain the $|\phi|\ge15^\circ$ applicability restriction.

\subsubsection{Horizontal Budgets, Layer Mass, and Moist Transport}

Let $\omega=Dp/Dt$ be pressure vertical velocity, $\dot Q$ diabatic heating per unit mass, and $S_q$ the net moisture source. The temperature and moisture equations in pressure coordinates contain
\begin{align}
  \frac{\partial T}{\partial t}+\bm V\cdot\nabla_hT
  +\omega\frac{\partial T}{\partial p}
  -\frac{\kappa T}{p}\omega&=\frac{\dot Q}{c_p},
  \label{eq:temperature-budget}\\
  \frac{\partial q}{\partial t}+\bm V\cdot\nabla_hq
  +\omega\frac{\partial q}{\partial p}&=S_q.
  \label{eq:moisture-budget}
\end{align}
Instantaneous archived fields do not identify every term in these equations. The benchmark therefore extracts only the observable horizontal contributions $-\bm V\cdot\nabla_hT$ and $-\bm V\cdot\nabla_hq$ for thermal-stress E1 and E2. These operators diagnose resolved redistribution at one time; they are not estimates of the complete temperature or moisture tendency.

Hydrostatic balance also supplies the exact layer-mass conversion
\begin{equation}
  \rho\,\mathrm dz=-\frac{\mathrm dp}{g}.
  \label{eq:pressure-layer-mass}
\end{equation}
Consequently, for an upper bound $p_u$ and lower bound $p_l>p_u$,
\begin{equation}
  \int_{z_l}^{z_u}\rho A\,\mathrm dz
  =\frac{1}{g}\int_{p_u}^{p_l}A\,\mathrm dp.
  \label{eq:pressure-integral-conversion}
\end{equation}
Choosing $A=q$ gives partial-layer water vapor; choosing $A=q\bm V$ gives partial-layer horizontal vapor transport. Applying negative horizontal divergence gives
\begin{equation}
  -\nabla_h\cdot\left(\frac{1}{g}
  \int_{300}^{850}q\bm V\,\mathrm dp\right),
  \label{eq:integrated-vapor-convergence-derivation}
\end{equation}
which is precipitation E3. The fixed 850--300-hPa limits make these free-tropospheric rather than total-column quantities.

For the instantaneous MSE $h=c_pT+\Phi+L_vq$, horizontal differentiation gives the exact decomposition
\begin{equation}
  \bm V\cdot\nabla_hh
  =c_p\bm V\cdot\nabla_hT
  +\bm V\cdot\nabla_h\Phi
  +L_v\bm V\cdot\nabla_hq.
  \label{eq:mse-advection-decomposition}
\end{equation}
Pressure integration of its negative over 500--850 hPa gives thermal-stress E3. An additive constant in $h$ disappears under $\nabla_h$, making this advective form gauge invariant. The conservative identity
\begin{equation}
  -\nabla_h\cdot(h\bm V)
  =-\bm V\cdot\nabla_hh-h\nabla_h\cdot\bm V
  \label{eq:mse-product-rule}
\end{equation}
provides audit terms but not a closed energy budget. Likewise, precipitation E4 follows from the moisture-flux product rule
\begin{equation}
  -\nabla_h\cdot(q\bm V)
  =-\bm V\cdot\nabla_hq-q\nabla_h\cdot\bm V.
  \label{eq:moisture-product-rule}
\end{equation}

\subsubsection{Kinematic Frontogenesis}

Let $\bm g_T=\nabla_hT$. Under the explicitly restricted, source-free horizontal evolution $\partial_tT+\bm V\cdot\nabla_hT=0$, covariant differentiation gives
\begin{equation}
  \frac{D_h\bm g_T}{Dt}
  =-(\nabla_h\bm V)^{\mathsf T}\bm g_T.
  \label{eq:temperature-gradient-tendency}
\end{equation}
Projecting this tendency onto $\bm g_T/|\bm g_T|$ yields
\begin{equation}
  \frac{D_h|\bm g_T|}{Dt}
  =-\frac{\bm g_T^{\mathsf T}(\nabla_h\bm V)\bm g_T}
  {|\bm g_T|},
  \label{eq:frontogenesis-derivation}
\end{equation}
which is the thermal-stress E5 field \citep{petterssen1936frontogenesis}. The derivation explains its kinematic meaning while making clear that vertical motion, tilting, diabatic heating, and turbulent mixing are omitted.

\subsubsection{Ertel Potential Vorticity and Storm-Relative Balance}

Ertel potential vorticity (PV) begins with the coordinate-independent scalar
\begin{equation}
  P=\rho^{-1}\bm\omega_a\cdot\nabla\theta.
  \label{eq:ertel-coordinate-free}
\end{equation}
Using pressure as the vertical coordinate, hydrostatic balance converts this identity to
\begin{equation}
  P=-g\left[(f+\zeta)\theta_p-v_p\theta_x+u_p\theta_y\right].
  \label{eq:ertel-pressure-derivation}
\end{equation}
The first term couples absolute vertical rotation to static stability; the remaining two are tilting contributions. TC E2 evaluates all terms at 700 hPa, with $p$ derivatives obtained from the Lagrange polynomial through 850, 700, and 500 hPa. This is an instantaneous PV-structure diagnostic, not a test of material PV conservation, which would additionally require adiabatic and frictionless evolution \citep{ertel1942vorticity,hoskins1985pv}.

For TC radial structure, let $\bm e_r$ be the local outward great-circle direction and $\bm e_\theta$ its counterclockwise tangent. Then
\begin{equation}
  V_\theta=\bm V\cdot\bm e_\theta
  \label{eq:tangential-projection}
\end{equation}
defines the primitive quantity whose complete-ring mean is scored by E1. Under the additional audit assumptions of axisymmetry, quasi-steady tangential flow, and negligible radial acceleration and friction, the radial momentum equation reduces to
\begin{equation}
  \frac{\partial\Phi}{\partial r}
  \approx\frac{V_\theta^2}{r}+fV_\theta,\qquad
  R_{\mathrm{GW}}=\frac{\partial\Phi}{\partial r}
  -\frac{V_\theta^2}{r}-fV_\theta.
  \label{eq:gradient-wind-audit}
\end{equation}
The signed and normalized $R_{\mathrm{GW}}$ profiles are audit outputs rather than the E1 score \citep{willoughby1990gradient}. TC E3 and E4 instead apply the exact spherical covariant vorticity and divergence definitions in \Eqref{eq:spherical-kinematics} before complete-ring averaging. Together with the hypsometric warm-core construction in \Eqref{eq:warm-core-hypsometric}, these chains connect the TC block to rotation, outflow, and thermal structure without claiming a closed cyclone model.

\subsection{Reference-Only Support}
\label{htmltoc:d-physics-metrics-2}

For metric $i$, let $A_e(x)$ be the catalog event mask, $p_s^{\ERA}(x)$ ERA5 surface pressure, $p_i^{\max}$ the deepest required pressure level, and $H_i$ the operator-stencil validity mask. Before reference-diagnostic finiteness is checked, the requested support is
\begin{equation}
  \widetilde\Omega_{ie}=
  A_e\cap\{p_s^{\ERA}\ge p_i^{\max}\}\cap H_i.
  \label{eq:requested-support}
\end{equation}
The pressure inequality has no added terrain margin. U2 and U3 additionally require $|\phi|\ge15^\circ$. The final support is
\begin{equation}
  \begin{aligned}
  \Omega_{ie}=\widetilde\Omega_{ie}\cap\{&\text{all primary ERA5 diagnostic components are finite}\}.
  \end{aligned}
  \label{eq:truth-diagnostic-support}
\end{equation}
There is no arbitrary minimum event-area or valid-fraction cutoff. A joint-component sample is retained only if all reference components are finite. Forecast values do not alter $\Omega_{ie}$; a nonfinite forecast on this support is recorded as invalid model output. Area fields use $\cos\phi$ weights on the common $0.25^\circ$ grid.

\subsection{Universal Diagnostics U1--U10}
\label{htmltoc:d-physics-metrics-3}

The same ten universal diagnostics are used for all four event categories.

\paragraph{U1: normalized signed hydrostatic sublayer residuals.}
For each adjacent pair $(p_l,p_u)$ in $(850,700)$, $(700,600)$, and $(600,500)$ hPa,
\begin{align}
  A_{lu}&=\Phi_u-\Phi_l,\\
  B_{lu}&=\frac{R_d}{2}(T_{v,l}+T_{v,u})\ln(p_l/p_u),\\
  R^{H}_{lu}&=\frac{A_{lu}-B_{lu}}
  {\tfrac12(|A_{lu}|+|B_{lu}|)}.
  \label{eq:hydrostatic-components}
\end{align}
The three signed $R^H_{lu}$ fields are scored jointly with pressure-thickness component weights. The comparison is forecast residual versus ERA5 residual, not residual versus zero.

\paragraph{U2--U3: geopotential-derived balanced-flow structure.}
At pressure $p$,
\begin{equation}
  (u_g,v_g)_p=\left(-\frac{\Phi_y}{f},\frac{\Phi_x}{f}\right)_p.
  \label{eq:geostrophic-wind}
\end{equation}
U2 jointly scores the eastward and northward components at 500 hPa. U3 jointly scores the endpoint shear $(u_{g,850}-u_{g,500},v_{g,850}-v_{g,500})$. A log-pressure virtual-temperature thermal-wind residual is exported for audit, but it is not the primary U3 field. The $15^\circ$ latitude exclusion avoids division by small $f$.

\paragraph{U4--U8: spherical kinematics.}
U4 and U6 are $\zeta$ at 850 and 500 hPa; U5 and U7 are $\delta$ at those levels. U8 jointly scores the signed $(D_s,D_h)$ components at 850 hPa. The unsigned deformation magnitude is retained only as an audit summary.

\paragraph{U9: dry layer stability.}
For lower and upper levels $(l,u)=(850,500)$ hPa,
\begin{equation}
  \begin{gathered}
  N^2_{lu}=g\frac{\theta_u-\theta_l}
  {\bar\theta_{lu}\,\Delta z_{lu}},\\
  \bar\theta_{lu}=\tfrac12(\theta_l+\theta_u),\quad
  \Delta z_{lu}=\frac{\Phi_u-\Phi_l}{g}.
  \end{gathered}
  \label{eq:dry-n2}
\end{equation}
Samples with nonpositive $\bar\theta_{lu}$ or $\Delta z_{lu}$ are not diagnostically valid in ERA5 and therefore do not enter reference support.

\paragraph{U10: moist static energy.}
U10 is $h_{850}$ from \Eqref{eq:appendix-thermodynamic-definitions}. It compares the observable thermodynamic state and is not an MSE budget-closure test \citep{neelin1987mse}.

\subsection{Thermal-Stress Diagnostics E1--E5}
\label{htmltoc:d-physics-metrics-4}

Heat and cold events share the same thermal stress operator block.

\paragraph{E1--E2: horizontal scalar advection.}
At 850 hPa,
\begin{equation}
  A_T=-\bm V\cdot\nabla_hT,\qquad
  A_q=-\bm V\cdot\nabla_hq.
  \label{eq:scalar-advection}
\end{equation}

\paragraph{E3: partial-layer horizontal MSE advection.}
The primary field is the gauge-invariant advective contribution
\begin{equation}
  A_h=-\frac{1}{g}\int_{500\,\mathrm{hPa}}^{850\,\mathrm{hPa}}
  \bm V\cdot\nabla_hh\,dp,
  \label{eq:mse-horizontal-advection}
\end{equation}
evaluated by trapezoidal pressure integration over 500, 600, 700, and 850 hPa. The conservative $-\nabla_h\cdot(h\bm V)$ form, its $-h\nabla_h\cdot\bm V$ term, and their discrete product-rule residual are retained for audit only. E3 excludes vertical transport, storage, sources, sinks, and the atmosphere outside 850--500 hPa; it is not a column budget.

\paragraph{E4: lower-tropospheric dry stability.}
E4 uses \Eqref{eq:dry-n2} for 850--700 hPa.

\paragraph{E5: horizontal kinematic temperature frontogenesis.}
With $\bm g_T=\nabla_hT$ and the covariant wind-gradient tensor $\nabla_h\bm V$,
\begin{equation}
  F_T=-\frac{\bm g_T^{\mathsf T}(\nabla_h\bm V)\bm g_T}
  {|\bm g_T|},
  \label{eq:frontogenesis}
\end{equation}
with zero returned where $|\bm g_T|=0$. This is the two-dimensional kinematic contribution to the material change of temperature-gradient magnitude \citep{petterssen1936frontogenesis}; vertical-motion, tilting, and diabatic terms are outside its scope.

\subsection{Precipitation Diagnostics E1--E5}
\label{htmltoc:d-physics-metrics-5}

The free-tropospheric operators use the fixed levels 850, 700, 600, 500, 400, and 300 hPa. Defining $\mathcal I[A]=-g^{-1}\int_{850}^{300}A\,dp$,
\begin{align}
  W_{850:300}&=\mathcal I[q],\\
  \bm Q_{850:300}&=\mathcal I[q\bm V],\\
  C_{850:300}&=-\nabla_h\cdot\bm Q_{850:300}.
  \label{eq:free-tropospheric-transport}
\end{align}
E1 scores $W$, E2 jointly scores the signed eastward and northward components of $\bm Q$, and E3 scores $C$. The fixed 850--300-hPa bounds mean that these are neither total-column water nor total-column transport quantities \citep{benton1954watervapor}.

E4 is the conservative flux-form 850-hPa moisture-flux convergence
\begin{equation}
  C_q=-\nabla_h\cdot(q\bm V).
  \label{eq:moisture-flux-convergence}
\end{equation}
The advective and divergent product-rule components and their discrete residual are exported for audit \citep{banacos2005mfc}. E5 is the signed layer difference
\begin{equation}
  \begin{aligned}
  \Delta h_{850:500}&=h_{850}-h_{500}\\
  &=c_p(T_{850}-T_{500})\\
  &\quad +(\Phi_{850}-\Phi_{500})+L_v(q_{850}-q_{500}).
  \end{aligned}
  \label{eq:mse-layer-difference}
\end{equation}

\subsection{Tropical-Cyclone Diagnostics E1--E5}
\label{htmltoc:d-physics-metrics-6}

\begin{figure}[!htbp]
  \centering
  \input{figures/tc_geometry}
  \caption{Storm-relative evaluation geometry. Panel (a) shows circular sampling for radial diagnostics; (b) shows spatial fields and the E5 environmental annulus. Forecast and reference centers are diagnosed independently. ERA5 fixes the radial support, E5 boundary $r_0$, and annular weights for both fields; environmental values come from each field itself. Circles and grid spacing are schematic.}
  \label{fig:tcgeometry}
\end{figure}
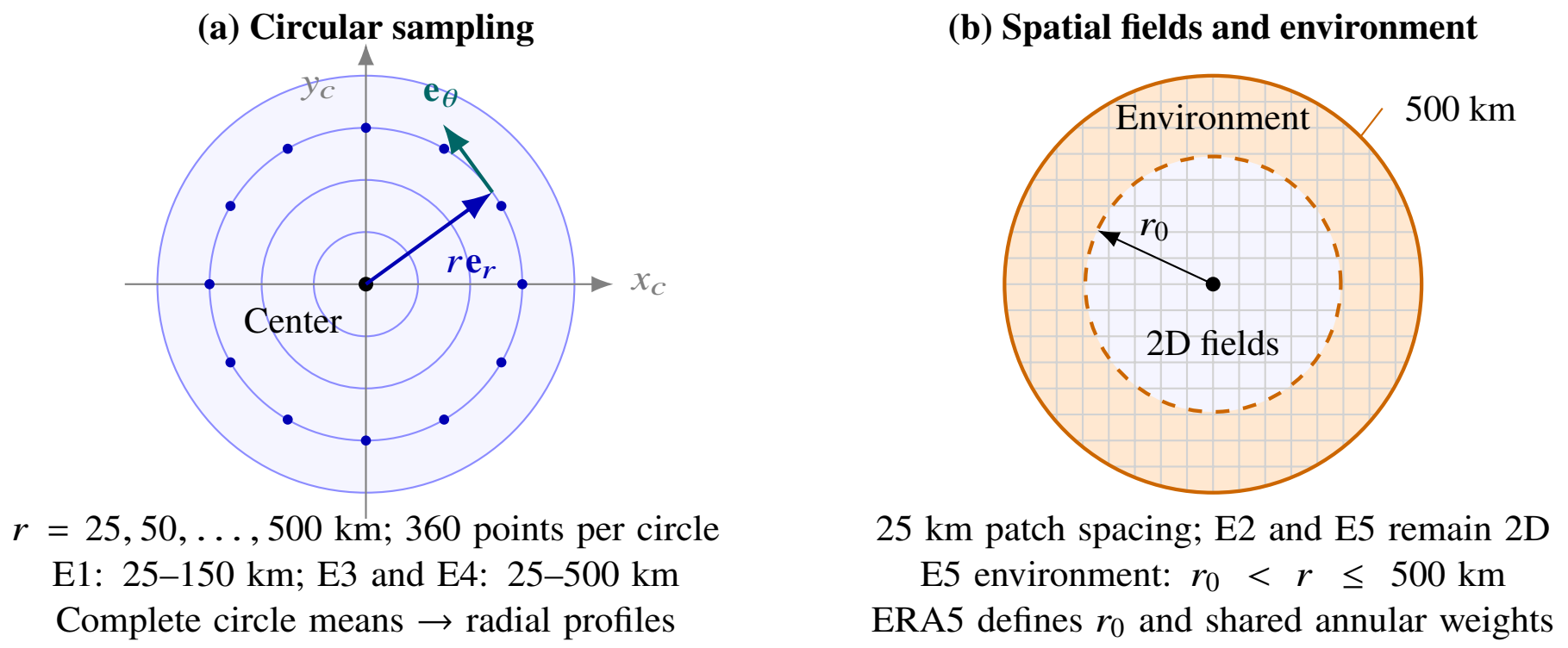

At every task valid time, the IBTrACS position seeds separate searches for the ERA5 and forecast minimum mean-sea-level pressure within $\pm8^\circ$. Each state is then sampled relative to its own center. The spherical azimuthal-equidistant patch extends to 500 km at 25-km spacing. Exact-radius great-circle rings contain 360 azimuths and a radial value is valid only when the complete ring is available. This construction isolates storm-relative organization; it deliberately does not score center displacement. \Figref{fig:tcgeometry} summarizes the sampling and scoring geometry.

\paragraph{E1: inner-core tangential wind.}
At each ring, wind is projected onto the local great-circle tangential direction and azimuthally averaged. E1 scores the complete-ring profile $\overline{V_\theta}(r)$ for $0<r\le150$ km at 850 hPa. The associated gradient-wind terms are audit outputs, not the primary score.

\paragraph{E2: total Ertel potential vorticity.}
At 700 hPa,
\begin{equation}
  P=-g\left[(f+\zeta)\theta_p-v_p\theta_x+u_p\theta_y\right].
  \label{eq:ertel-pv}
\end{equation}
The vertical derivatives use three-level Lagrange differentiation at 850, 700, and 500 hPa. The stretching term and both tilting terms are exported separately, while their total is scored \citep{ertel1942vorticity,hoskins1985pv}.

\paragraph{E3--E4: radial rotation and outflow structure.}
E3 is the complete-ring azimuthal-mean profile of full covariant $\zeta_{500}$ over $0<r\le500$ km. E4 is the corresponding $\delta_{200}$ profile over the same inner-outflow radius. Their profile mean squared error gives equal weight to each retained radius bin.

\paragraph{E5: warm-core thickness anomaly.}
Let $Z_{300:700}=(\Phi_{300}-\Phi_{700})/g$. ERA5 first defines a storm outer radius as the first zero crossing after the positive maximum of the hemisphere-signed, azimuthal-mean $\zeta_{500}$ profile. The environmental mean is the spherical-annulus-weighted mean of complete $Z_{300:700}$ rings outside that radius and within 500 km. That ERA5-derived annulus and its weights are frozen and reused for the forecast. E5 scores
\begin{equation}
  Z'_{300:700}(x)=Z_{300:700}(x)-\langle Z_{300:700}\rangle_{\mathrm{env}}.
  \label{eq:warm-core-thickness}
\end{equation}
If ERA5 has no valid outer zero crossing or complete environment, the diagnostic is inapplicable for every model on that task rather than receiving a substituted climatological radius.

\subsection{Diagnostic Error Geometries}
\label{app:diagnostic-error-geometries}

A diagnostic field must be reduced to one task error before the score map can be applied. For a scalar field, grid cell area weights $w_x$ keep a regular longitude and latitude grid from overrepresenting small cells at higher latitudes. On the frozen diagnostic support $\Omega_{i,k}$, the mean squared error is
\begin{equation}
 \begin{gathered}
 M_{i,k}=\frac{\sum_{x\in\Omega_{i,k}}w_x
   \left(D^{\FC}_{i,k}(x)-D^{\ERA}_{i,k}(x)\right)^2}
   {\sum_{x\in\Omega_{i,k}}w_x},
 \\ E_{i,k}=\sqrt{M_{i,k}}.
 \end{gathered}
 \label{eq:diagnostic-mse}
\end{equation}
Squaring follows differencing, so signs and directions are retained in the comparison rather than discarded by comparing component magnitudes. To give multiple components the declared relative influence, joint component diagnostics use
\begin{equation}
  M_i=\frac{\sum_j\gamma_j\sum_{x\in\Omega_i}w_x
  (D^{\FC}_{ij}(x)-D^{\ERA}_{ij}(x))^2}
  {\sum_j\gamma_j\sum_{x\in\Omega_i}w_x},
  \label{eq:component-mse}
\end{equation}
where $j$ indexes components and $\gamma_j$ is equal for vector/deformation components and proportional to sublayer pressure thickness for U1. Storm-patch fields use spherical patch-cell areas, while E1, E3, and E4 use equal radius-bin weights after complete-ring averaging. Exact spherical-annulus weights are used for the E5 environmental mean, not for radial-profile mean squared error. These choices preserve signed direction, component, layer, and radial structure until after forecast--ERA5 differencing. The resulting RMSE is normalized and mapped as specified in \Secref{sec:scoring}, then aggregated according to \Appref{app:score-aggregation}.

\FloatBarrier

%% file: figures/tc_geometry.tex
\begin{tikzpicture}[x=1cm,y=1cm,>=Latex,font=\footnotesize]
  \node[font=\footnotesize\bfseries] at (3.1,4.75) {(a) Circular sampling};
  \node[font=\footnotesize\bfseries] at (9.6,4.75) {(b) Spatial fields and environment};

  \begin{scope}[shift={(3.1,2.8)}]
    \fill[blue!4] (0,0) circle (1.6);
    \foreach \r in {0.4,0.8,1.2,1.6} {\draw[blue!45] (0,0) circle (\r);}
    \draw[->,gray] (-1.85,0)--(1.9,0) node[right] {$x_c$};
    \draw[->,gray] (0,-1.8)--(0,1.85);
    \node[gray,anchor=east] at (-0.08,1.5) {$y_c$};
    \fill (0,0) circle (1.6pt);
    \node[anchor=north east] at (-0.06,-0.06) {Center};
    \foreach \a in {0,30,...,330} {\fill[blue!70!black] (\a:1.2) circle (1.1pt);}
    \coordinate (P) at (36:1.2);
    \draw[->,thick,blue!70!black] (0,0)--(P) node[midway,below right] {$r\mathbf e_r$};
    \draw[->,thick,teal!80!black] (P)--++(126:0.65) node[above] {$\mathbf e_\theta$};
  \end{scope}
  \node[align=center,text width=6cm] at (3.1,0.55)
    {$r=25,50,\ldots,500$ km; 360 points per circle\\
     E1: 25--150 km; E3 and E4: 25--500 km\\
     Complete circle means $\rightarrow$ radial profiles};

  \begin{scope}[shift={(9.6,2.8)}]
    \fill[orange!18] (0,0) circle (1.6);
    \fill[blue!4] (0,0) circle (0.98);
    \begin{scope}
      \clip (0,0) circle (1.6);
      \draw[gray!35,step=0.2] (-1.6,-1.6) grid (1.6,1.6);
    \end{scope}
    \draw[thick,orange!80!black] (0,0) circle (1.6);
    \draw[thick,dashed,orange!80!black] (0,0) circle (0.98);
    \fill (0,0) circle (1.6pt);
    \draw[->] (0,0)--(155:0.98) node[midway,above] {$r_0$};
    \node[align=center] at (0,-0.45) {2D fields};
    \node[anchor=west] at (1.35,1.35) {500 km};
    \draw[orange!80!black] (45:1.6)--(1.3,1.35);
    \node[align=center,text width=2.4cm] at (0,1.28) {Environment};
  \end{scope}
  \node[align=center,text width=6cm] at (9.6,0.55)
    {25 km patch spacing; E2 and E5 remain 2D\\
     E5 environment: $r_0<r\le500$ km\\
     ERA5 defines $r_0$ and shared annular weights};
\end{tikzpicture}

%% file: appendices/E_forecast_systems.tex
\section{Forecast Systems and Inference Details}
\label{app:forecast-systems}

\subsection{Earth2Studio Execution Layer}
\label{htmltoc:e-forecast-systems-1}

All six AI models were run through NVIDIA Earth2Studio 0.14.0 \citep{earth2studio2024}, using the released model weights and implementations. Initial conditions for the AI models came from the Analysis-Ready Cloud-Optimized ERA5 (ARCO-ERA5) dataset, while operational Global Forecast System (GFS) forecasts were retrieved from the National Oceanic and Atmospheric Administration's National Centers for Environmental Prediction (NOAA/NCEP). Forecast outputs were standardized to common coordinates and selected at the prescribed lead times, while GFS retained its operational initial conditions.

\Tabrefs{tab:models} and~\ref{tab:model-variables} record the components and default artifact identifiers selected by the benchmark code together with the complete variable interfaces resolved from the versioned Earth2Studio documentation and implementation \citep{earth2studio014docs}. Architecture and original model descriptions follow the corresponding publications \citep{bi2023pangu,bodnar2025aurora,bonev2023sfno,chen2023fengwu,bonev2025fcn3,lam2023graphcast}, whereas supplementary training data provenance follows the released Spherical Fourier Neural Operator (SFNO) and FourCastNet~3 (FCN3) model cards and the GraphCast repository documentation \citep{nvidia2024sfnocheckpoint,nvidia2025fcn3checkpoint,deepmind2023graphcastrepo}. We distinguish the training protocols reported in papers from metadata specific to the selected checkpoints; differences between these sources are recorded below. \Tabref{tab:model-variables} reports the full state variable coordinates exposed by each pinned wrapper and separately identifies internally supplied static fields and derived forcings where applicable. Its principal comparison quantity is the common output pool $\mathcal{C}=\bigcap_s\mathcal{O}_s=\mathcal{A}_5\cup\{\texttt{msl},\texttt{u10m},\texttt{v10m},\texttt{t2m}\}$, which contains 69 variables across all seven systems. Every raw forecast field used by a diagnostic was selected from $\mathcal{C}$; the benchmark did not score every field in $\mathcal{C}$, and no system received credit for an output unavailable from another system. Every AI model was advanced autoregressively in 6-h increments and supplied global fields on the common nominal $0.25^\circ$ grid used by the diagnostics.

\input{tables/forecast_systems}

\paragraph{Training dates and model variants.}
Pangu's paper reports ERA5 training in 1979--2017 and validation in 2019. GraphCast's paper uses 2016--2017 for development validation, while the selected Operational checkpoint is documented as pretrained on ERA5 in 1979--2017 and fine-tuned on HRES in 2016--2021; the development validation split is not a separately documented validation split for that checkpoint. SFNO's paper (Section 5.2) reports training in 1979--2015, validation in 2016--2017, and two-step autoregressive fine-tuning; its NVIDIA NGC package does not independently specify these dates. FengWu's Supplementary Notes 2 and 4 report ERA5 training and fine-tuning in 1979--2015 and validation in 2016--2017; the main text describes the broader ERA5 period as 1979--2017. The original repository distinguishes \texttt{fengwu\_v1.onnx}, intended for ERA5 evaluation without operational-analysis transfer learning, from \texttt{fengwu\_v2.onnx}, which was fine-tuned on operational analysis data through 2021 \citep{openearthlabFengWuRepo}. The benchmark uses the default FengWu v1 checkpoint. The 2021 cutoff therefore describes v2 rather than the selected checkpoint. Aurora's Supplementary Table C3 and Section C.4 give the source-dependent 1950--2020 pretraining coverage and HRES 2020 validation; the selected pretrained checkpoint does not use the subsequent task-specific fine-tuning \citep{bi2023pangu,lam2023graphcast,deepmind2023graphcastrepo,bonev2023sfno,nvidia2024sfnocheckpoint,chen2023fengwu,bodnar2025aurora}.

\paragraph{FCN3 source differences.}
\Tabref{tab:models} reports the FCN3 paper's training split from its Table 3 and Appendix E.4 (1980--2016), fine-tuning split from its Appendix E.3 (2012--2016), and out-of-sample validation period (2018--2021, with reported metrics in 2020). The same paper's Appendix E.2 instead starts pretraining in 1979, an internal difference in that source. The pinned artifact model card lists training in 1980--2015, testing in 2016--2017, and evaluation in 2018--2019; its configuration contains a run-directory label indicating fine-tuning in 2013--2016. These records differ, so the paper dates describe its reported experiments and do not establish the exact training history of the selected weights \citep{bonev2025fcn3,nvidia2025fcn3checkpoint}.

\input{tables/model_variables}

\Tabref{tab:models} identifies the evaluated checkpoints and reports the paper training protocols with explicit source labels. For FCN3, the selected v1 checkpoint metadata and the paper report different splits, as detailed above.

\subsection{Initial Conditions and Forecast Outputs}
\label{htmltoc:e-forecast-systems-2}

For the AI models, the local ARCO-ERA5 adapter served 6-hourly ERA5 fields in the variable vocabulary and coordinate order requested by the corresponding Earth2Studio wrapper. Models with one initial state consumed the analysis at $t_0$; Aurora, FengWu, and GraphCast additionally consumed $t_0-6$ h. Static fields required by Aurora and GraphCast were loaded from their released packages. Although the benchmark stored only variables needed by the downstream diagnostics, the inference call always received the complete state required by the checkpoint. Every scored lead came from the corresponding model integration output.

Inference uses the released implementations: ONNX for Pangu and FengWu, JAX for GraphCast, and PyTorch for the remaining AI models.
For FCN3, the random number generator was reset to seed 333 when the model was loaded, and only one realization was evaluated.

GFS was treated differently because it is an operational numerical forecast, not a checkpoint initialized from ERA5. The benchmark used Earth2Studio's \texttt{GFS\_FX} source to retrieve the NOAA/NCEP forecast initialized at the task cycle. Consequently, GFS retained its operational data assimilation state while sharing the same valid time, requested variables, and verification operators as the AI models.

\FloatBarrier

%% file: tables/forecast_systems.tex
\begin{table}[!htbp]
  \centering
  \begin{threeparttable}
    \caption{Forecast systems and the provenance of the checkpoints used in ExtremePhysBench. Metadata describe the evaluated checkpoints rather than all versions of each model.}
    \label{tab:models}
    \scriptsize
    \setlength{\tabcolsep}{2pt}
    \begin{tabularx}{\linewidth}{@{}>{\raggedright\arraybackslash}p{0.08\linewidth}>{\raggedright\arraybackslash}p{0.14\linewidth}>{\raggedright\arraybackslash}p{0.16\linewidth}>{\raggedright\arraybackslash}p{0.27\linewidth}Y@{}}
      \toprule
      \textbf{System} & \textbf{Architecture} & \textbf{Earth2Studio component} &
      \textbf{Checkpoint} & \textbf{Training data} \\
      \midrule
      GFS & NWP & \texttt{GFS\_FX} & Operational forecast archive & N/A \\
      \addlinespace[2pt]
      Pangu & 3D transformer & \texttt{Pangu6} & Original-publication ONNX checkpoints: \texttt{pangu\_weather\_6.onnx} and \texttt{pangu\_weather\_24.onnx} & \textit{Train:} ERA5, 1979--2017\newline\textit{Val.:} 2019\newline\textit{FT.:} none reported \\
      \addlinespace[2pt]
      GraphCast & graph network & \texttt{GraphCast}\newline\texttt{Operational} & Operational checkpoint at 0.25$^\circ$ with 13 pressure levels; ERA5 pretraining and HRES fine tuning & \textit{Train:} ERA5, 1979--2017\newline\textit{Val.:} 2016--2017\newline\textit{FT.:} HRES, 2016--2021 \\
      \addlinespace[2pt]
      FengWu & multimodal transformer & \texttt{FengWu} & ERA5-evaluation ONNX checkpoint without operational-analysis transfer learning: \texttt{fengwu\_v1.onnx}; 69 variables & \textit{Train:} ERA5, 1979--2015\newline\textit{Val.:} 2016--2017\newline\textit{FT.:} ERA5, 1979--2015 \\
      \addlinespace[2pt]
      SFNO & spherical neural operator & \texttt{SFNO} & Modulus-Makani checkpoint: \texttt{sfno\_73ch\_small}, package version 0.1.0 & \textit{Train:} ERA5, 1979--2015\newline\textit{Val.:} 2016--2017\newline\textit{FT.:} two-step autoregressive training; no separate date split reported \\
      \addlinespace[2pt]
      Aurora & foundation transformer & \texttt{Aurora} & Original-publication checkpoint; default loader selects \path{aurora-0.25-pretrained.ckpt} without LoRA & \textit{Pretrain:} mixture of ten sources, 1950--2020; coverage varies by source\newline\textit{Val.:} IFS HRES 0.25$^\circ$, 2020\newline\textit{FT.:} none; pretrained checkpoint without LoRA \\
      \addlinespace[2pt]
      FCN3 & geometric neural operator & \texttt{FCN3} & NVIDIA FourCastNet 3 v1 checkpoint & \textit{Train:} ERA5, 1980--2016\newline\textit{Val.:} 2018--2021; reported metrics use 2020\newline\textit{FT.:} ERA5, 2012--2016 \\
      \bottomrule
    \end{tabularx}
    \begin{tablenotes}[flushleft]
      \scriptsize
      \item Checkpoint descriptions follow the Earth2Studio 0.14.0 documentation \citep{earth2studio014docs}; the selected checkpoints correspond to the default models in that release. Training dates and protocols follow the sources discussed below. GraphCast validation dates refer to the development split. Paper training protocols are not independent verification of the released weights; source differences are detailed below. N/A: not applicable; Val.: validation; FT.: fine tuning. GFS uses an operational forecast archive rather than a checkpoint.
    \end{tablenotes}
  \end{threeparttable}
\end{table}

%% file: tables/model_variables.tex
\begin{table}[!htbp]
  \centering
  \begin{threeparttable}
    \caption{Earth2Studio model interfaces and forecast settings. The common output intersection contains 69 variables, defining the admissible pool of diagnostic inputs.}
    \label{tab:model-variables}
    \scriptsize
    \setlength{\tabcolsep}{2pt}
    \begin{tabular}{@{}>{\raggedright\arraybackslash}p{0.11\linewidth}>{\raggedright\arraybackslash}p{0.39\linewidth}>{\raggedright\arraybackslash}p{0.27\linewidth}>{\raggedright\arraybackslash}p{0.18\linewidth}@{}}
      \toprule
      \textbf{System} & \textbf{Complete input interface} &
      \textbf{Complete output interface} & \textbf{Context / forecast setting} \\
      \midrule
      GFS & N/A: operational source; model inputs are not exposed & 246 requestable fields: $\{u,v,w,z,t,r,q\}\!\times\!\mathcal{L}_{\mathrm{GFS}}$ and $\mathcal{S}_{\mathrm{GFS}}$ & Operational cycle; deterministic archived forecast \\
      \addlinespace[2pt]
      Pangu & $\mathcal{A}_5\cup\{\texttt{msl},\texttt{u10m},\texttt{v10m},\texttt{t2m}\}$ (69) & Same 69 state variables & $t_0$; 6-h deterministic rollout \\
      \addlinespace[2pt]
      GraphCast & $\mathcal{A}_6\cup\{\texttt{t2m},\texttt{msl},\texttt{u10m},\texttt{v10m}\}$ (82); derived $\mathcal{F}$; static \path{land_sea_mask} and \path{geopotential_at_surface} & Same 82 state variables plus \texttt{tp06} (83) & $t_0-6$ h, $t_0$; 6-h deterministic rollout \\
      \addlinespace[2pt]
      FengWu & $\mathcal{A}_5\cup\{\texttt{u10m},\texttt{v10m},\texttt{t2m},\texttt{msl}\}$ (69) & Same 69 state variables & $t_0-6$ h, $t_0$; 6-h deterministic rollout \\
      \addlinespace[2pt]
      SFNO & $\mathcal{A}_5$ plus \texttt{u10m}, \texttt{v10m}, \texttt{u100m}, \texttt{v100m}, \texttt{t2m}, \texttt{sp}, \texttt{msl}, and \texttt{tcwv} (73) & Same 73 state variables & $t_0$; 6-h deterministic rollout \\
      \addlinespace[2pt]
      Aurora & Same 69 state variables as Pangu; packaged static \texttt{z}, \texttt{slt}, and \texttt{lsm} & Same 69 state variables & $t_0-6$ h, $t_0$; 6-h deterministic rollout \\
      \addlinespace[2pt]
      FCN3 & $\mathcal{A}_5$ plus \texttt{u10m}, \texttt{v10m}, \texttt{u100m}, \texttt{v100m}, \texttt{t2m}, \texttt{msl}, and \texttt{tcwv} (72) & Same 72 state variables & $t_0$; one realization, seed 333 \\
      \midrule
      \textbf{Common intersection} & N/A & $\mathcal{C}=\mathcal{A}_5\cup\{\texttt{msl},\texttt{u10m},\texttt{v10m},\texttt{t2m}\}$ (69) & Admissible pool of evaluation variables; outputs specific to individual systems excluded \\
      \bottomrule
    \end{tabular}
    \begin{tablenotes}[flushleft]
      \scriptsize
      \item Variable codes follow Earth2Studio 0.14.0. The set $\mathcal{L}_{13}$ of 13 levels contains 50, 100, 150, 200, 250, 300, 400, 500, 600, 700, 850, 925, and 1000 hPa; $\mathcal{A}_5=\{z,q,t,u,v\}\times\mathcal{L}_{13}$ and $\mathcal{A}_6=\{z,q,t,u,v,w\}\times\mathcal{L}_{13}$. The GraphCast forcing set $\mathcal{F}$ contains \texttt{toa\_incident\_solar\_radiation}, \texttt{year\_progress\_sin}, \texttt{year\_progress\_cos}, \texttt{day\_progress\_sin}, and \texttt{day\_progress\_cos}. The set $\mathcal{L}_{\mathrm{GFS}}$ of GFS pressure levels contains 1, 2, 3, 5, 7, 10, 15, 20, 30, 40, 50, 70, 100, 150, 200, 250, 300, 350, 400, 450, 500, 550, 600, 650, 700, 750, 800, 850, 900, 925, 950, 975, and 1000 hPa. Its set $\mathcal{S}_{\mathrm{GFS}}$ of surface and single level variables contains \texttt{u10m}, \texttt{v10m}, \texttt{u100m}, \texttt{v100m}, \texttt{t2m}, \texttt{d2m}, \texttt{r2m}, \texttt{q2m}, \texttt{sp}, \texttt{msl}, \texttt{tcwv}, \texttt{tp}, \texttt{2d}, \texttt{fg10m}, and \texttt{refc}; \texttt{2d} is a second Earth2Studio alias for 2-m dew point temperature, and some GFS fields are available only beyond lead zero. Here $u$ and $v$ are horizontal wind, $w$ is pressure vertical velocity, $z$ is geopotential, $t$ is temperature, $q$ is specific humidity, and $r$ is relative humidity; Aurora's static \texttt{z}, \texttt{slt}, and \texttt{lsm} denote surface geopotential, soil type, and land--sea mask. Parenthetical counts include state variables only.
    \end{tablenotes}
  \end{threeparttable}
\end{table}

%% file: appendices/F_additional_results.tex
\section{Additional Experimental Results}
\label{app:additional-results}

\subsection{Conventional RMSE Comparison}
\label{app:rmse-results}

The conventional comparison uses RMSE for the 15 shared fields on pressure levels $\{z,t,q,u,v\}\times\{850,700,500\}$ on the same event support defined by the reference. For each event category and variable, task RMSE is averaged within event and then log transformed and standardized across the model--event panel. The RMSE composite skill for each event is the negative mean of these standardized log RMSE values, so larger values mean lower field error. This composite is an internal summary for comparison with the Physics Score; it is not an externally calibrated accuracy scale.

For heat, cold, and precipitation, the conventional and physical structure evaluations use the catalog mask. For TC, primitive fields are sampled on the common grid with 25 km spacing relative to each storm center after the forecast and ERA5 centers are diagnosed independently, matching the centering convention of the TC physical structure diagnostics.

\begin{figure}[!htbp]
  \centering
  \includegraphics[width=\linewidth]{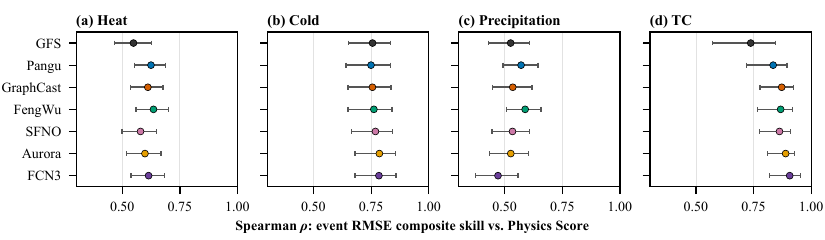}

  \caption{Within-model associations between RMSE composite skill and Event Physics Score. Panels (a--d) show heat, cold, precipitation, and TC events. Points show within-model Spearman correlations across events; bars show nominal percentile bootstrap 95\% intervals, all excluding zero. All models are evaluated on the same events within each category. RMSE composite skill and Event Physics Score are positively associated in all 28 model--category pairs.}
  \label{fig:main-rmse-correlation}
\end{figure}

\begin{figure}[!htbp]
  \centering
  \includegraphics[width=0.45\linewidth]{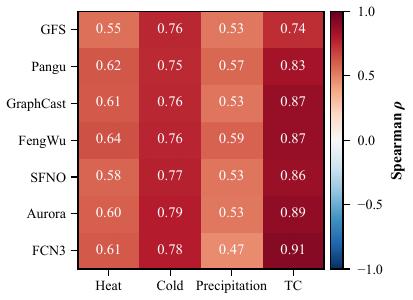}

  \caption{Correlations between RMSE composite skill and Event Physics Score by model and event category. Cells show within-model Spearman correlations between RMSE composite skill and Event Physics Score, using the same events for all seven systems within each category. TC has the strongest median within-model association between RMSE composite skill and Event Physics Score ($\rho=0.867$), despite its larger disagreement in model ranks.}
  \label{fig:rmse-correlation-heatmap}
\end{figure}

Across seven systems and four event categories, all 28 correlations between RMSE composite skill and Event Physics Score are positive, and every nominal percentile bootstrap 95\% interval excludes zero. The median correlations across systems are 0.611 for heat, 0.762 for cold, 0.534 for precipitation, and 0.867 for TC. \Figref{fig:main-rmse-correlation} shows the coefficients and their intervals; \Figref{fig:rmse-correlation-heatmap} summarizes the coefficients in a heatmap. These associations support relatedness, not equivalence: a monotone relationship across events does not imply that RMSE identifies which balance, kinematic, thermodynamic, transport, or vortex structure differs.

\begin{figure}[!htbp]
  \centering
  \includegraphics[width=\linewidth]{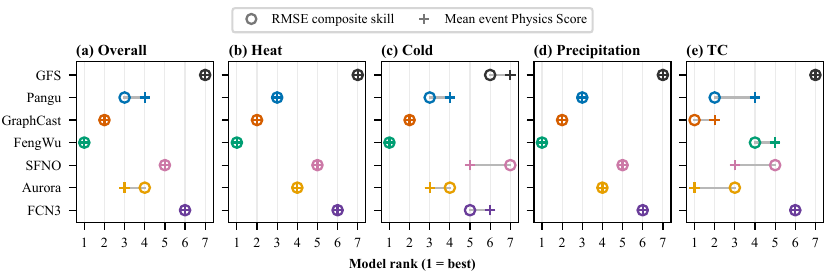}

  \caption{Model ranks from conventional RMSE composite skill and mean Event Physics Score. Panels (a--e) show Overall, heat, cold, precipitation, and TC comparisons. TC shows the largest rank disagreement between conventional RMSE composite skill and mean Event Physics Score.}
  \label{fig:main-rmse-ranks}
\end{figure}

\begin{figure}[!htbp]
  \centering
  \includegraphics[width=0.8\linewidth]{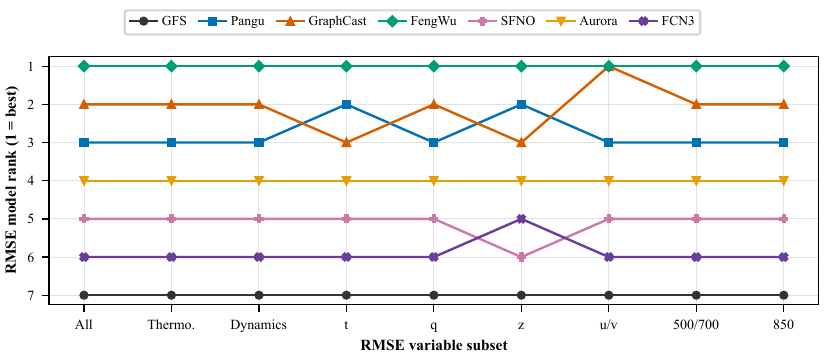}

  \caption{Sensitivity of conventional RMSE rankings to nine variable and pressure-level subsets. Within each subset, variable-wise mean RMSE ranks are averaged over variables and then equally over event categories. Rank 1 is best. Conventional RMSE rankings depend on the selected variables and pressure levels, although FengWu remains first across all nine subsets.}
  \label{fig:rmse-variable-sensitivity}
\end{figure}

\Figref{fig:rmse-variable-sensitivity} shows that the conventional model order also depends on which primitive variables are summarized. The main comparison uses all 15 fields; the subsets are sensitivity analyses, not alternative benchmark leaderboards.

The formal leaderboard uses the complete declared support for each summary: its correlation between model ranks is 0.893 Overall, 1.000 for heat, 0.857 for cold, 1.000 for precipitation, and 0.703 for TC. The view of Event Physics Scores on matched support drawn in \Figref{fig:main-rmse-ranks} is slightly different because its plotted Overall coordinate averages Event Physics Scores on common support; its correlations are 0.964 Overall and 0.750 for TC. In either construction, TC exhibits the largest rank disagreement. This estimator distinction is why the main text does not treat a rank plot as another definition of the formal Overall Physics Score.

\subsection{Aggregate Model Comparisons}
\label{app:aggregate-model-comparisons}

\begin{figure}[!htbp]
  \centering
  \includegraphics[width=\linewidth]{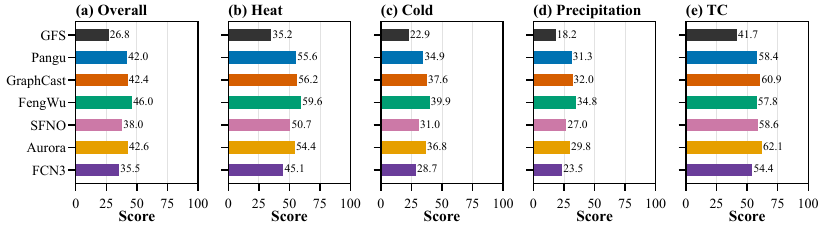}
  \caption{Overall and Category Physics Scores. Panels (a--e) show Overall, heat, cold, precipitation, and TC scores. Overall Physics Score averages universal scores equally across the four categories; Category Physics Scores combine universal and applicable event-specific groups. Higher scores indicate closer physical structure agreement with ERA5 under fixed scales and tolerances. FengWu leads Overall, heat, cold, and precipitation, while Aurora leads TC.}
  \label{fig:model-scores-event-category}
\end{figure}

\begin{figure}[!htbp]
  \centering
  \includegraphics[width=0.45\linewidth]{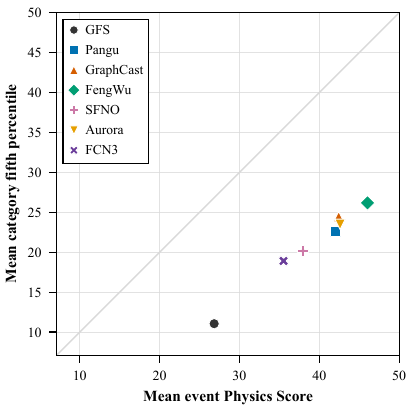}
  \caption{Mean and lower-tail performance of universal diagnostics on common finite event support across seven systems. The horizontal axis averages category event means; the vertical axis averages category fifth percentiles, with equal category weights. Both axes use universal scores. The diagonal marks equality; distance below it is not an uncertainty interval. FengWu achieves the highest mean and fifth-percentile universal scores, indicating stronger physical structure agreement with ERA5 both on average and in the lower tail of event performance.}
  \label{fig:model-accuracy-robustness}
\end{figure}

\begin{figure}[!htbp]
  \centering
  \includegraphics[width=\linewidth]{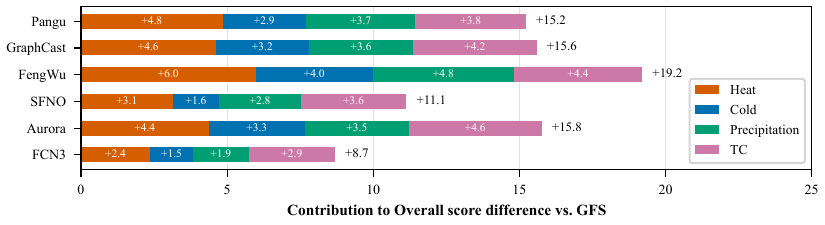}
  \caption{Category contributions to Overall Physics Score differences from GFS. Each segment is one quarter of the category mean difference in universal scores; the total is their sum. Event-specific diagnostics are excluded. All six AI models improve on GFS in every category contribution to Overall Physics Score, with FengWu gaining the most overall.}
  \label{fig:overall-contributions-gfs}
\end{figure}

\Figrefs{fig:model-scores-event-category}--\ref{fig:overall-contributions-gfs} complement the formal scoreboard with category, contribution, and lower tail views. FengWu has the highest Overall Physics Score and leads heat, cold, and precipitation, while Aurora has the highest TC Category Physics Score. All six AI models have positive contributions from the universal group relative to GFS in all four event categories. The lower tail view in \Figref{fig:model-accuracy-robustness} also places FengWu highest in the mean of the category fifth percentiles. These are descriptive summaries of physical structure fidelity, not additional hypothesis tests or measures of hazard skill.

\subsection{Event Distributions and Paired Comparisons}
\label{app:event-comparisons}

\begin{figure}[!htbp]
  \centering
  \includegraphics[width=\linewidth]{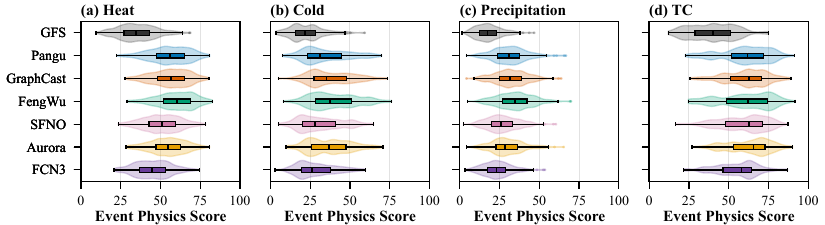}
  \caption{Distributions of Event Physics Scores. Panels (a--d) show heat, cold, precipitation, and TC events. Within each event, mapped task scores are averaged within diagnostic, diagnostics within family, and families within group; the universal and event-specific groups receive equal weight.}
  \label{fig:event-score-distributions}
\end{figure}

\begin{figure}[!htbp]
  \centering
  \includegraphics[width=0.5\linewidth]{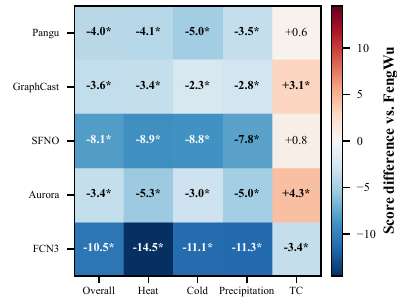}
  \caption{Paired Physics Score differences from FengWu for the other five AI models. Positive differences indicate higher Physics Scores than FengWu. Category intervals resample paired events; Overall equally weights category differences in universal scores. Asterisks mark nominal paired bootstrap 95\% intervals excluding zero, without multiplicity adjustment. FengWu leads the other AI models Overall and in heat, cold, and precipitation; Aurora and GraphCast lead FengWu for TC.}
  \label{fig:paired-bootstrap}
\end{figure}

\Figref{fig:event-score-distributions} shows the distributions across events behind the category means. Their overlap makes clear that an ordering of model averages is not a statement about every event. \Figref{fig:paired-bootstrap} therefore compares each AI model with FengWu on paired events. All five Overall differences are negative, with nominal 95\% intervals excluding zero. By category, every comparison system is lower for heat, cold, and precipitation. TC differs: Aurora is $+4.33$ points $[1.94,6.88]$ and GraphCast is $+3.11$ $[0.70,5.82]$ relative to FengWu; intervals for Pangu and SFNO include zero; FCN3 is lower by $-3.43$ $[-6.23,-0.73]$. These are nominal intervals without multiplicity adjustment. For Overall, paired events are resampled within each category before averaging universal scores with equal category weights.

\begin{figure}[!htbp]
  \centering
  \includegraphics[width=\linewidth]{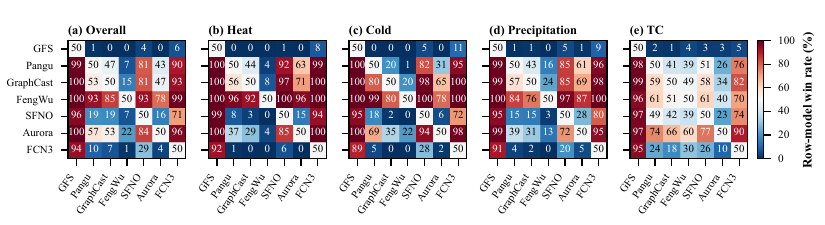}

  \caption{Pairwise Event Physics Score win rates. Panels (a--e) show Overall, heat, cold, precipitation, and TC comparisons. Cells report how often the row system has a higher Event Physics Score than the column system, with ties counting as half a win. The diagonal is 50\%; Overall averages the four category matrices equally. Win rates measure frequency, not score-gap magnitude or uncertainty. Pairwise advantages vary by event category, especially for TCs.}
  \label{fig:pairwise-win-rates}
\end{figure}

Pairwise win rates in \Figref{fig:pairwise-win-rates} answer a different question from mean differences. They count the fraction of paired events on which the row system has the larger score, with ties contributing one half. Consequently, they need not reproduce the order of average score differences.

\subsection{Physical Structure Profiles by Diagnostic}
\label{app:diagnostic-results}

\begin{figure}[!htbp]
  \centering
  \includegraphics[width=\linewidth]{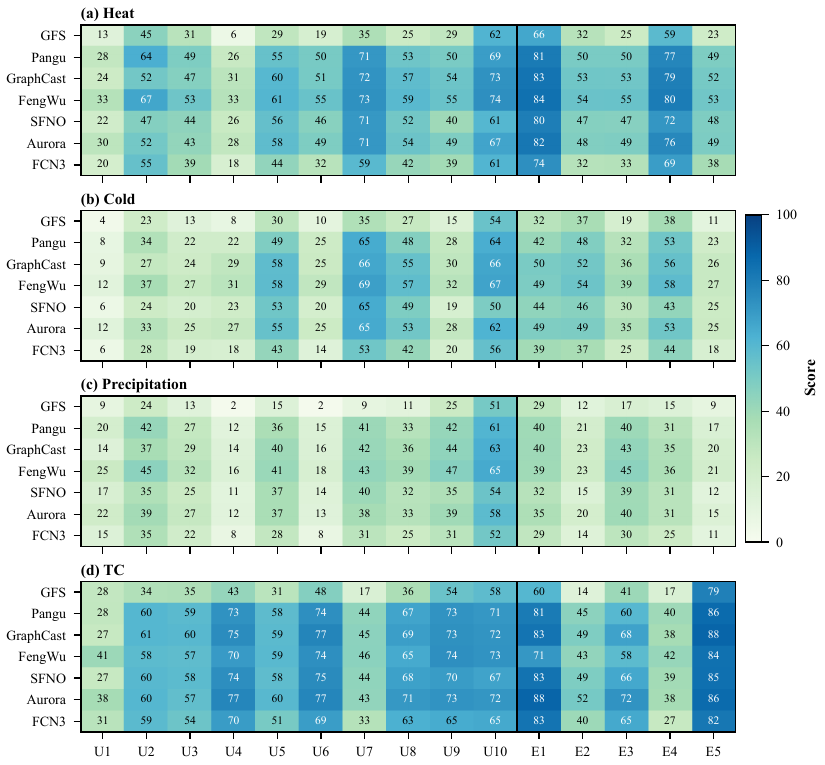}
  \caption{Mean Diagnostic Physics Scores by system. Panels (a--d) show heat, cold, precipitation, and TC events. Rows show systems; codes follow \Tabrefs{tab:universal-metrics} and~\ref{tab:event-metrics}. Mapped scores are averaged over leads within event, then over applicable events. Support differs from the common complete events in \Tabref{tab:family-scores}. Diagnostic leaders differ within the same event category: for TC, Aurora leads circulation (E1), while GraphCast leads warm-core structure (E5).}
  \label{fig:metric-heatmap}
\end{figure}

\begin{figure}[!htbp]
  \centering
  \includegraphics[width=\linewidth]{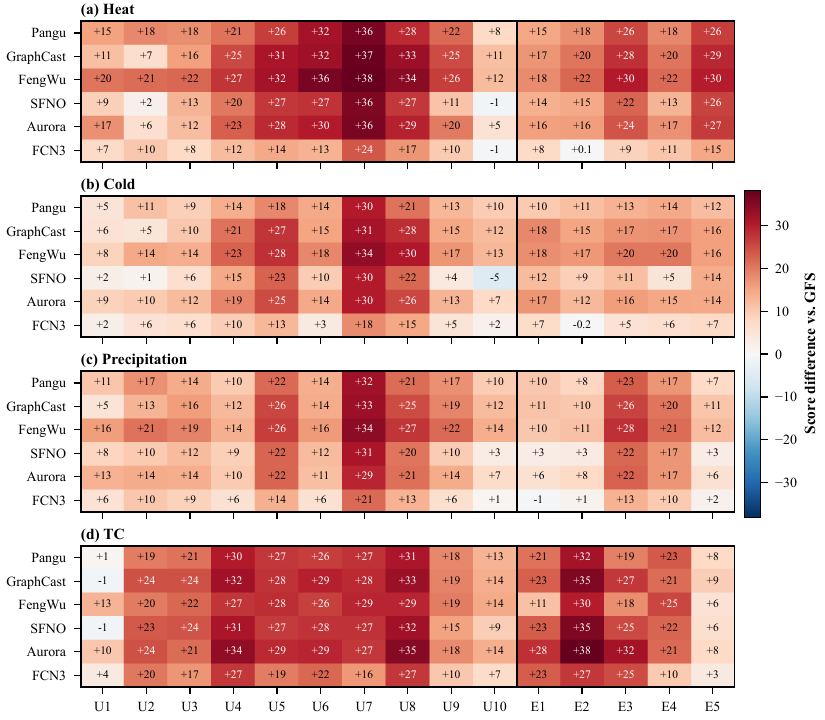}
  \caption{Diagnostic Physics Score differences from GFS. Panels (a--d) show heat, cold, precipitation, and TC events. Each system is paired with GFS on common available events; positive Diagnostic Physics Score differences favor the named system. Values are point estimates without individual diagnostic uncertainty intervals. Gains over GFS vary across diagnostics within each event category.}
  \label{fig:metric-delta-gfs}
\end{figure}

\begin{figure}[!htbp]
  \centering
  \includegraphics[width=\linewidth]{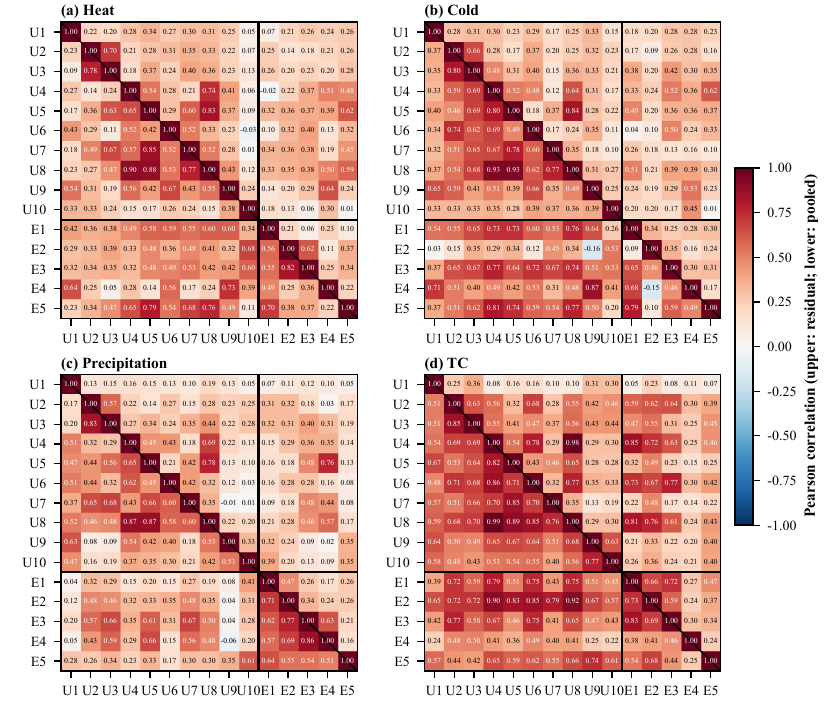}

  \caption{Correlations among Diagnostic Physics Scores. Panels (a--d) show heat, cold, precipitation, and TC events. Lower triangles show Pearson correlations of Diagnostic Physics Scores pooled over model--event observations; upper triangles correlate residuals after removing additive model and event effects. Heavy rules separate universal (U1--U10) from event-specific (E1--E5) diagnostics. Many diagnostic correlations weaken after removing additive model and event effects.}
  \label{fig:metric-correlations}
\end{figure}

\Figref{fig:metric-heatmap} reports the Diagnostic Physics Scores from which the physical family and group summaries are constructed. Every column is mapped using its own fixed dimensional $C_i\tau_i$ scale before averaging, so color differences between columns are score differences under the declared tolerances rather than comparisons of raw units. \Figref{fig:metric-delta-gfs} provides paired descriptive differences from GFS, and \Figref{fig:metric-correlations} separates pooled correlations from correlations after removing additive model and event effects. None of these auxiliary views changes the aggregation weights in the formal scoreboard.

\textbf{Diagnostic contrasts within an extreme weather event category.}
The heatmap in \Figref{fig:metric-heatmap} shows that the leading system can change even between closely related diagnostics. For extreme precipitation events, GraphCast has the highest water vapor score (E1), while FengWu leads vapor-flux convergence (E3), low-level moisture-flux convergence (E4), and the MSE layer difference (E5). Thus, agreement in layer water amount does not by itself establish agreement in its transport or vertical thermodynamic context. For TC events, Aurora leads axisymmetric tangential wind (E1) and the two event-specific rotation diagnostics (E2--E3), whereas GraphCast leads warm-core thickness (E5) and FengWu leads upper-level outflow (E4). These diagnostic comparisons complement the physical family comparisons on common event support in the main text.

\Figref{fig:metric-delta-gfs} makes the GFS comparison local to each diagnostic. Broad improvements can coexist with small or negative differences in individual columns; for example, the TC hydrostatic scores (U1) of Pangu, GraphCast, and SFNO are close to the GFS value in the absolute heatmap. The paired differences are the appropriate view for assessing those contrasts because they restrict each comparison to shared available events. A large positive cell identifies a diagnostic contribution worth examining, but its size alone does not establish statistical significance or determine the category ranking, which also depends on the physical family and group weights.

\textbf{Shared variation among diagnostics.}
In \Figref{fig:metric-correlations}, the pooled lower triangles contain widespread positive associations: model--event combinations with high scores for one physical structure often also score well for others. Many associations weaken in the upper triangles after additive model and event effects are removed. This reduction indicates that part of the pooled association reflects differences in average model performance and event difficulty. Some links remain strong, including the relationship between geostrophic wind and geostrophic shear (U2--U3), consistent with their shared dependence on geopotential gradients. The residual correlations describe co-variation beyond the fitted additive effects; they do not isolate causal relationships or demonstrate that diagnostics are interchangeable.

\begin{figure}[!htbp]
  \centering
  \includegraphics[width=0.9\linewidth]{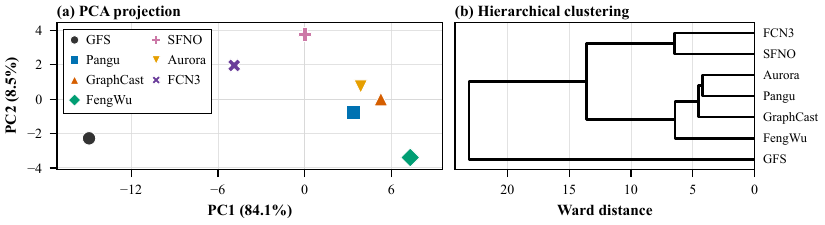}
  \caption{Similarity among systems in Diagnostic Physics Score profiles. Features comprise 15 diagnostics in each of four categories, standardized across seven systems. Panel (a) shows the two-component projection; (b) applies Ward clustering to all 60 features. Proximity describes score profiles, not shared architecture or statistically distinct model families. The first two principal components explain 92.6\% of variation in standardized diagnostic scores across the seven systems.}
  \label{fig:model-metric-similarity}
\end{figure}

\begin{figure}[!htbp]
  \centering
  \includegraphics[width=0.8\linewidth]{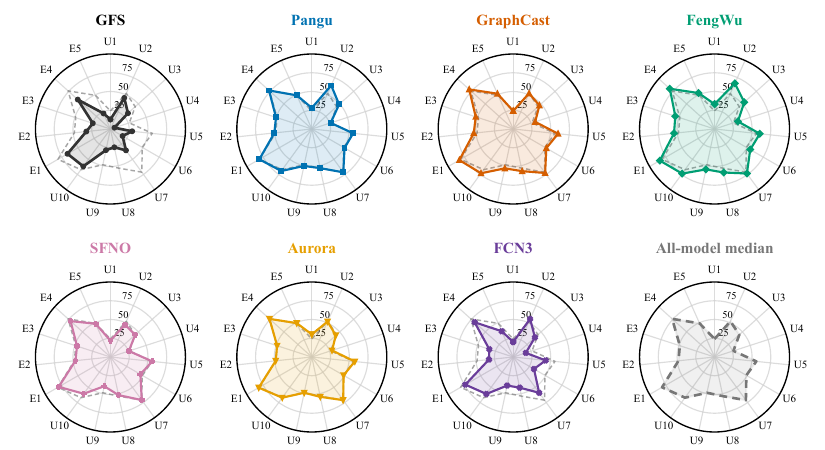}
  \caption{Diagnostic Physics Score profiles for heat events. Panels show individual systems and the seven-system median. Each system is compared with the seven-system median for each diagnostic. Codes follow \Tabrefs{tab:universal-metrics} and~\ref{tab:event-metrics}. Radial values use fixed scales and tolerances; enclosed area is not an aggregate score.}
  \label{fig:metric-radar-heat}
\end{figure}

\begin{figure}[!htbp]
  \centering
  \includegraphics[width=0.8\linewidth]{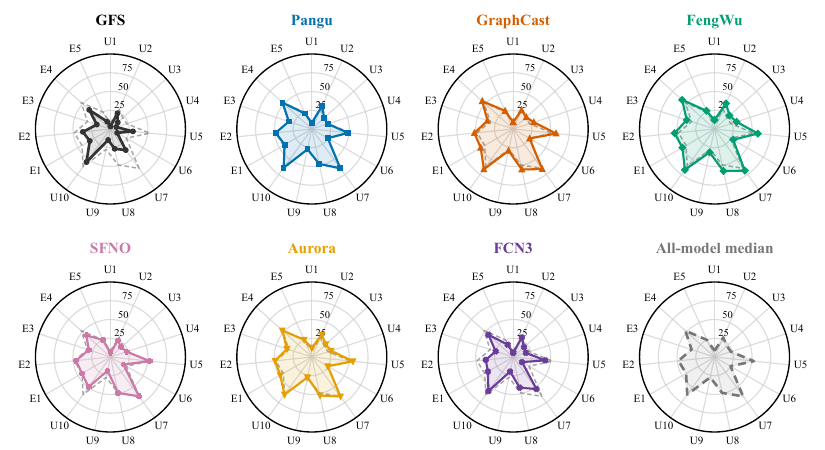}
  \caption{Like \Figref{fig:metric-radar-heat}, but for cold events.}
  \label{fig:metric-radar-cold}
\end{figure}

\begin{figure}[!htbp]
  \centering
  \includegraphics[width=0.8\linewidth]{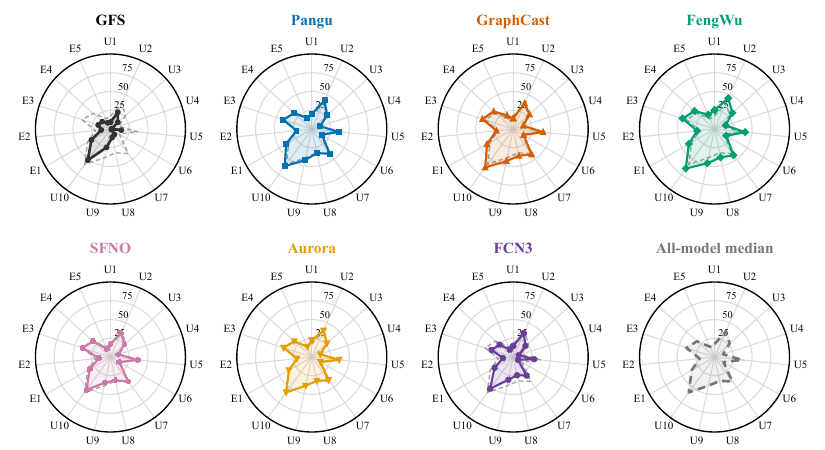}
  \caption{Like \Figref{fig:metric-radar-heat}, but for precipitation events.}
  \label{fig:metric-radar-precipitation}
\end{figure}

\begin{figure}[!htbp]
  \centering
  \includegraphics[width=0.8\linewidth]{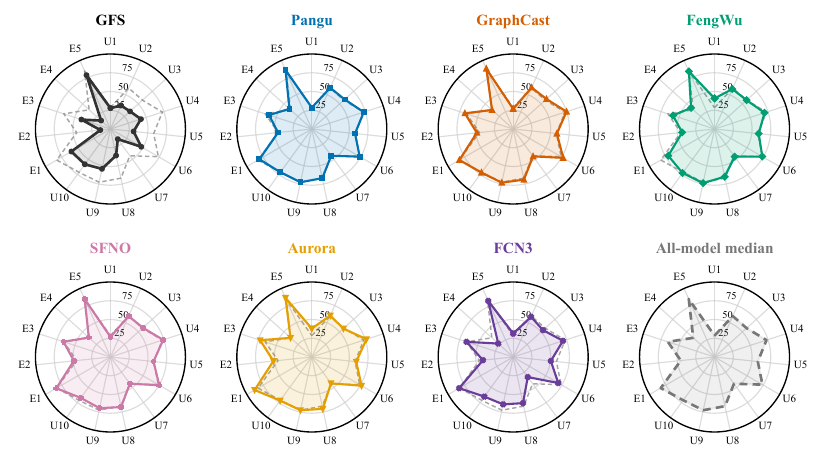}
  \caption{Like \Figref{fig:metric-radar-heat}, but for TC events.}
  \label{fig:metric-radar-tc}
\end{figure}

\Figref{fig:model-metric-similarity} summarizes similarities among the diagnostic profiles, while \Figrefs{fig:metric-radar-heat}--\ref{fig:metric-radar-tc} retain individual diagnostic identities within each event category. The radar views compare each system with the median across the seven systems, helping locate relative strengths and weaknesses that a category mean can hide. Neither geometric proximity nor radar area defines an additional Physics Score.

\textbf{Similarity of complete score profiles.}
The first two principal components in \Figref{fig:model-metric-similarity} explain 84.1\% and 8.5\% of the variance, respectively, retaining 92.6\% of the variance in this standardized representation. The projection therefore captures much of the variation across the seven systems, while the Ward tree provides a view using all retained features. Each category--diagnostic feature is standardized across systems, so a diagnostic with a narrow absolute score range can still influence the geometry. These equal feature weights differ from the benchmark's physical family and group aggregation; proximity in this figure should be read as similarity of relative diagnostic profiles.

\textbf{Category-specific strengths in the radar views.}
For extreme heat stress events (\Figref{fig:metric-radar-heat}), FengWu's profile is broadly strong across both the universal and event-specific blocks, consistent with its advantage across all seven applicable physical families. The profiles for extreme cold stress events (\Figref{fig:metric-radar-cold}) are less uniform: Aurora slightly exceeds FengWu in hydrostatic structure (U1), and GraphCast leads temperature advection (E1), while FengWu retains advantages in moisture and MSE advection (E2--E3) and lower-layer stability (E4). These local departures explain why the best category mean need not imply the best value on every axis.

For extreme precipitation events (\Figref{fig:metric-radar-precipitation}), the contrast between water amount (E1), vapor supply and convergence (E2--E4), and vertical energy contrast (E5) is particularly useful: the axes separate processes that are combined in the category summary. The profiles for TC events (\Figref{fig:metric-radar-tc}) instead separate circulation, rotation, outflow, and warm-core structure. Aurora's strong circulation profile and GraphCast's strong warm-core score therefore represent different physical structures within TC events. In all four panels, the median is an empirical reference across the seven evaluated systems, not a physical acceptance threshold.

\subsection{Detailed Views by Forecast Lead}
\label{app:lead-results}

\begin{figure}[!htbp]
  \centering
  \includegraphics[width=\linewidth]{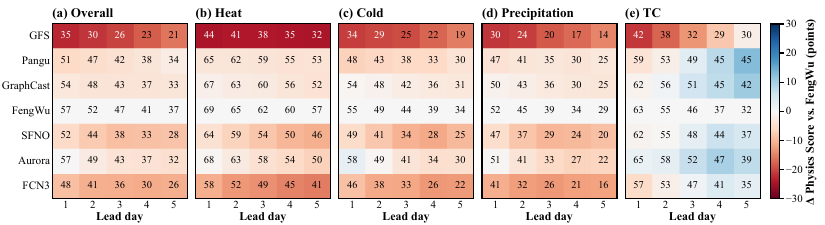}

  \caption{Physics Score scorecard over 24--120 h. Panels (a--e) show Overall, heat, cold, precipitation, and TC scores. Labels show absolute Physics Scores; shading shows differences from FengWu. Overall equally weights universal scores across categories; category panels combine universal and event-specific groups. Aurora leads Overall at 24 h; FengWu leads at subsequent daily leads through 120 h.}
  \label{fig:lead-scorecard}
\end{figure}

\begin{figure}[!htbp]
  \centering
  \includegraphics[width=\linewidth]{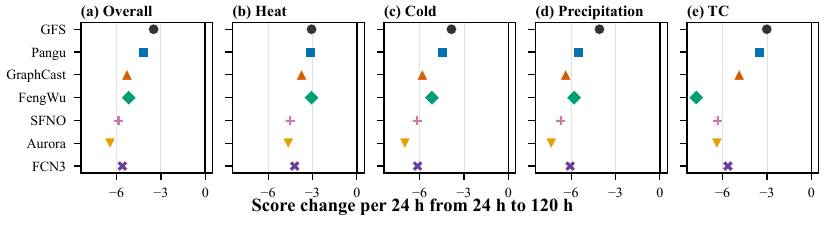}

  \caption{Endpoint changes in Physics Scores from 24 to 120 h. Panels (a--e) show Overall, heat, cold, precipitation, and TC changes. Markers show the endpoint change $[S(120)-S(24)]/4$ in score points per 24 h, not fitted rates. A smaller decline alone does not imply a higher score. All systems lose Physics Score between 24 and 120 h, with smaller declines for heat than for cold or precipitation.}
  \label{fig:lead-degradation}
\end{figure}

\Figref{fig:main-lead-scores} summarizes score evolution. The common Overall window is 24--120 h because the gridded events use daily leads and TC uses 6-hourly leads. \Figref{fig:lead-scorecard} annotates the absolute values. Aurora is 0.15 points above FengWu Overall at 24 h; FengWu leads from 48 h onward. Between 24 and 120 h, all displayed model--scope series decline. \Tabref{tab:lead-endpoint-changes} and \Figref{fig:lead-degradation} report the endpoint changes per 24 h. The Overall change ranges from $-6.45$ for Aurora to $-3.50$ for GFS. The smaller GFS decline occurs from a substantially lower starting score and is not evidence of better absolute structure fidelity.

\input{tables/lead_endpoint_changes}

Heat scores decline less rapidly than cold or precipitation for most systems. The bounded score map compresses large errors, so score changes must be read with their starting values and are not dimensional error growth rates.

\begin{figure}[!htbp]
  \centering
  \includegraphics[width=\linewidth]{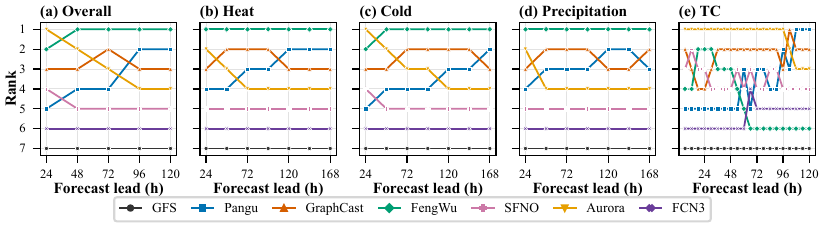}

  \caption{System rankings by forecast lead. Panels (a--e) show Overall, heat, cold, precipitation, and TC ranks. Rank 1 denotes the highest Physics Score at each plotted lead. FengWu ranks first for heat, cold, and precipitation at every plotted lead, while TC rankings change more often.}
  \label{fig:lead-rank}
\end{figure}

\begin{figure}[!htbp]
  \centering
  \includegraphics[width=\linewidth]{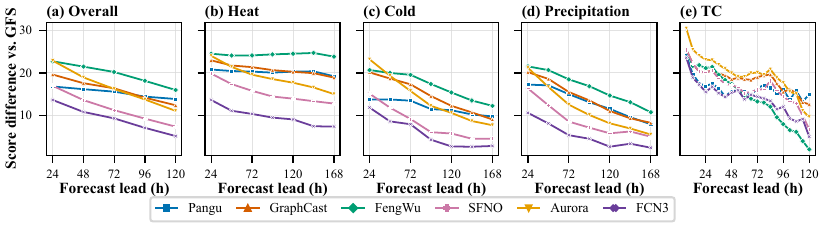}

  \caption{Physics Score differences from GFS by lead. Panels (a--e) show Overall, heat, cold, precipitation, and TC comparisons. Positive differences favor the named system. Overall uses universal diagnostics with equal category weights; category scores combine universal and event-specific groups. Physics Score advantages over GFS generally narrow with increasing lead.}
  \label{fig:lead-delta-gfs}
\end{figure}

\begin{figure}[!htbp]
  \centering
  \includegraphics[width=\linewidth]{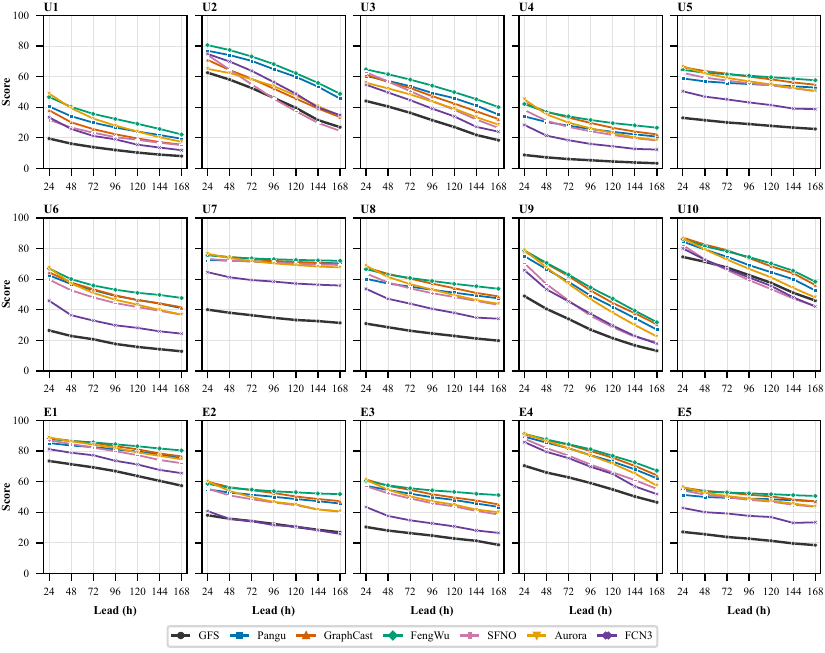}
  \caption{Diagnostic Physics Scores by forecast lead for heat events. Each panel shows one diagnostic (U1--U10 or E1--E5) across systems. Codes follow \Tabrefs{tab:universal-metrics} and~\ref{tab:event-metrics}. For heat events, geostrophic wind and shear scores (U2--U3) decline more strongly with lead than temperature advection (E1) or frontogenesis (E5).}
  \label{fig:metric-lead-heat}
\end{figure}

\begin{figure}[!htbp]
  \centering
  \includegraphics[width=\linewidth]{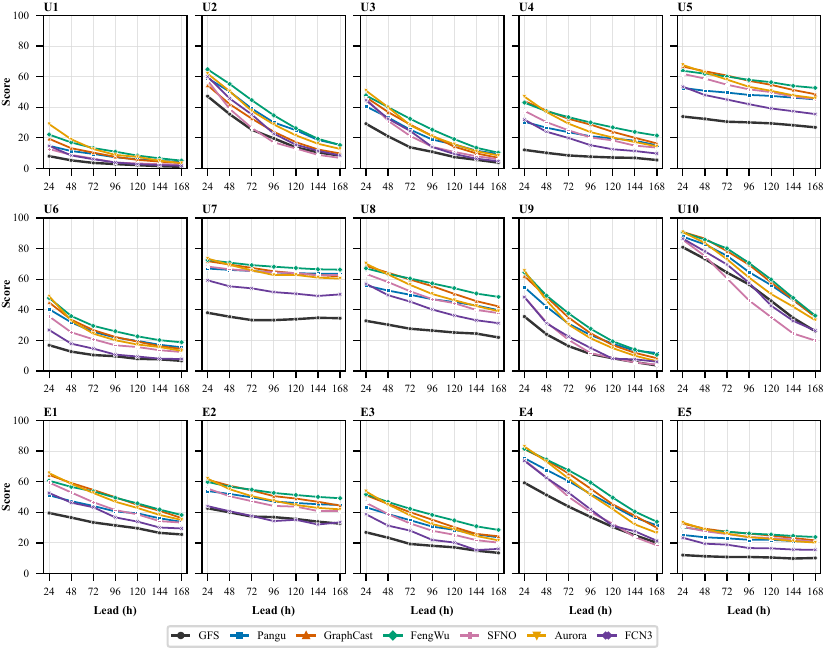}
  \caption{Like \Figref{fig:metric-lead-heat}, but for cold events. Lead-dependent losses vary by diagnostic, with larger changes in geostrophic wind and shear (U2--U3) than in divergence (U5, U7).}
  \label{fig:metric-lead-cold}
\end{figure}

\begin{figure}[!htbp]
  \centering
  \includegraphics[width=\linewidth]{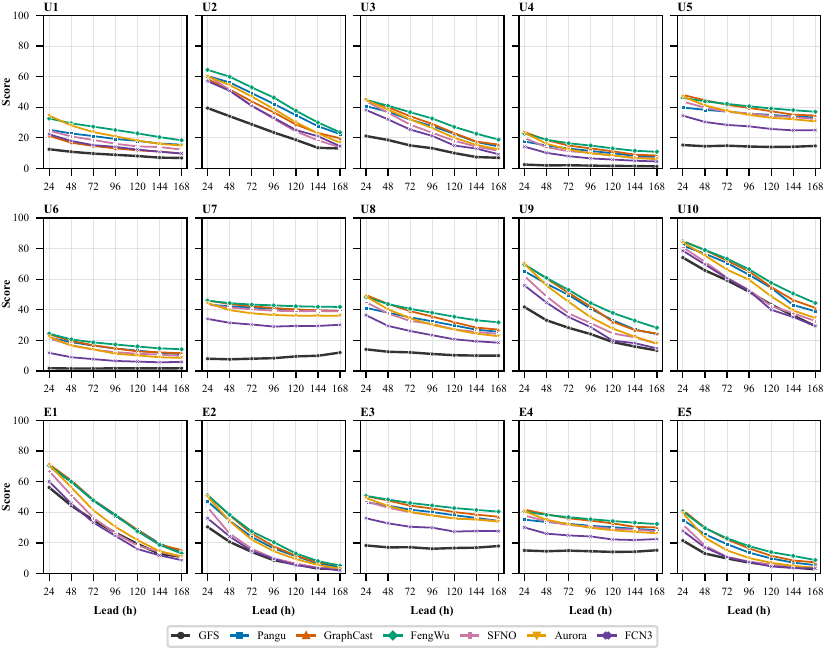}
  \caption{Like \Figref{fig:metric-lead-heat}, but for precipitation events. Water-content and vapor-transport scores (E1--E2) decline more strongly with lead than moisture-convergence scores (E3--E4).}
  \label{fig:metric-lead-precipitation}
\end{figure}

\begin{figure}[!htbp]
  \centering
  \includegraphics[width=\linewidth]{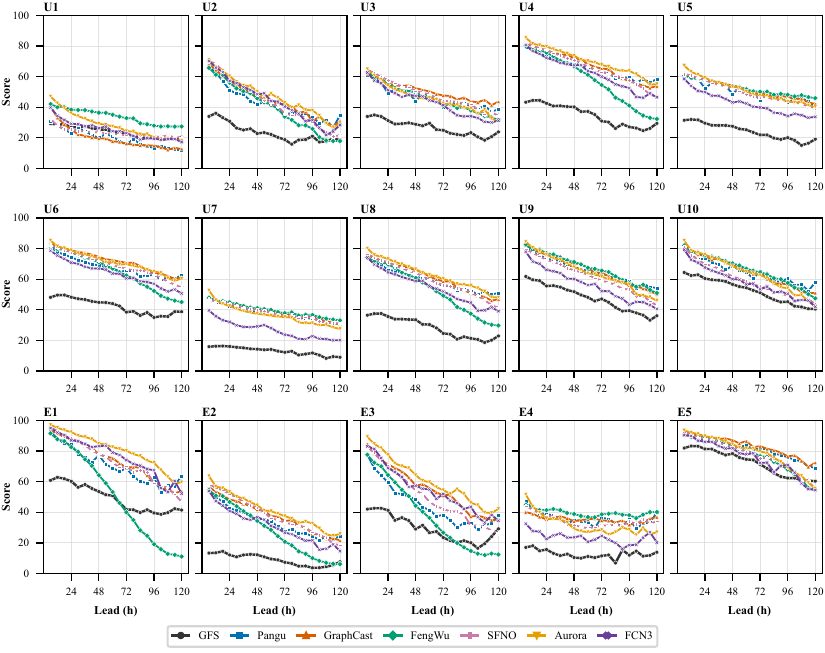}
  \caption{Like \Figref{fig:metric-lead-heat}, but for TC events. FengWu's circulation score (E1) falls below the other AI models at longer leads, while its upper-level outflow score (E4) remains comparatively strong.}
  \label{fig:metric-lead-tc}
\end{figure}

\Figref{fig:lead-rank} gives the corresponding rank paths, and \Figref{fig:lead-delta-gfs} shows absolute score differences from GFS. \Figrefs{fig:metric-lead-heat}--\ref{fig:metric-lead-tc} reproduce the diagnostic lead curves for all four event categories, showing how individual physical structures evolve rather than reducing them to one category mean.

\textbf{Rank changes and absolute margins.}
\Figref{fig:lead-rank} emphasizes crossings between systems, whereas \Figref{fig:lead-delta-gfs} retains the score margins relative to a common baseline. Reading them together distinguishes changes in ordering from changes in the size of an advantage: a rank crossing can occur with a small score difference, and an unchanged rank can accompany a substantial decline.

\textbf{Heat events: differences in score decline across physical structures.}
\Figref{fig:metric-lead-heat} shows diagnostic scores for extreme heat stress events through 168 h. Geostrophic wind and shear (U2--U3), dry stability (U9), and MSE (U10) show pronounced score declines. By comparison, horizontal divergence (U5 and U7), temperature advection (E1), and frontogenesis (E5) retain flatter score trajectories. The Category Physics Score for heat events therefore summarizes agreement in physical structures with different lead dependence. FengWu remains strong across many diagnostics, but the separation among systems and the amount of score loss vary by axis.

\textbf{Cold events: declining scores for stability and transport.}
For extreme cold stress events (\Figref{fig:metric-lead-cold}), dry stability scores fall markedly in both the universal diagnostic for the 850--500 hPa layer (U9) and the event-specific diagnostic for the 850--700 hPa layer (E4). Scores for geostrophic wind and shear (U2--U3) and temperature and MSE advection (E1 and E3) also decline, while divergence scores (U5 and U7) change more gradually. Several low-scoring diagnostics become more closely grouped at long leads. Such convergence near the lower part of the bounded score scale should not be interpreted as convergence of the underlying dimensional errors.

\textbf{Precipitation events: different score trends for water supply and convergence.}
\Figref{fig:metric-lead-precipitation} shows steep score declines for layer water amount (E1), vapor-transport components (E2), and vertical MSE contrast (E5). The vapor-flux and low-level moisture-flux convergence scores (E3--E4) vary more gradually, despite their connection to the same moisture field. The universal diagnostics likewise distinguish declining geostrophic and thermodynamic scores from relatively flatter divergence scores. These curves concern the physical structures within the selected extreme precipitation events; they do not directly measure accumulated precipitation error.

\textbf{TC events: different score trends for circulation and outflow.}
In \Figref{fig:metric-lead-tc}, FengWu's tangential-wind score (E1) declines especially strongly and falls below the other AI models at later leads. Its event-specific PV and midlevel-vorticity scores (E2--E3) also lose ground, whereas its outflow score (E4) remains comparatively strong. Scores for warm-core thickness (E5) decline across systems but remain relatively high.

\subsection{Sensitivity to the GFS Verifying Analysis}
\label{app:gfs-reference}

\begin{figure}[!htbp]
  \centering
  \includegraphics[width=\linewidth]{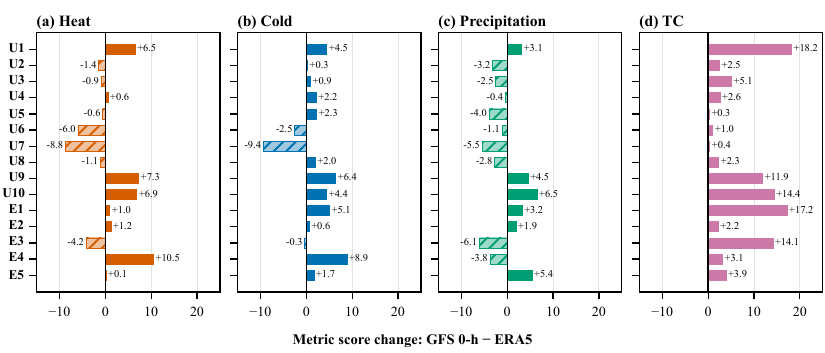}

  \caption{Sensitivity of GFS Diagnostic Physics Scores to the verifying analysis. Panels (a--d) show changes from ERA5 to GFS 0-h verification for heat, cold, precipitation, and TC. Hatched bars mark decreases. Each formal diagnostic aggregate uses the named reference\textquotesingle s own admissible support. Switching GFS verification from ERA5 to GFS 0-h raises all TC diagnostic scores but has mixed effects in other categories.}
  \label{fig:gfs-metric-sensitivity}
\end{figure}

\begin{figure}[!htbp]
  \centering
  \includegraphics[width=\linewidth]{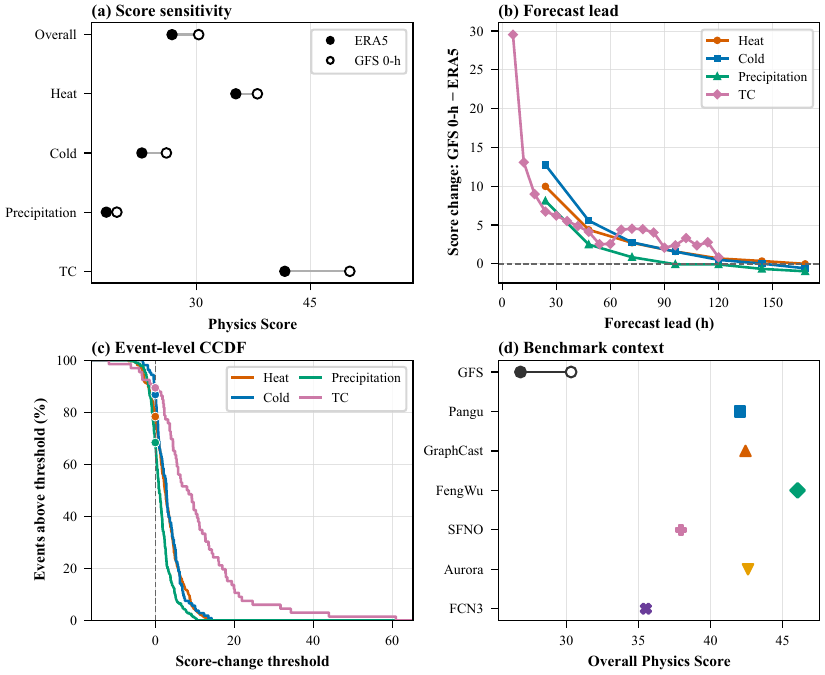}

  \caption{Sensitivity of GFS Physics Scores to the verifying analysis. Panel (a) compares aggregate scores under ERA5 and GFS 0-h verification; (b) shows their differences (GFS 0-h minus ERA5) by lead; (c) shows the percentage of paired events whose score change exceeds each threshold. Formal scores in (a--b) use each reference's own support. Panel (d) places the alternative GFS endpoint beside the ERA5-referenced benchmark for context, not as a ranking across references. GFS scores increase under GFS 0-h verification, most strongly for TC.}
  \label{fig:gfs-reference}
\end{figure}

Replacing ERA5 with archived GFS 0-h analyses changes both the reference diagnostic and the support defined by that reference. The formal GFS Overall score rises from 26.82 to 30.33, a $+3.51$ point change. The changes are $+2.83$ for heat, $+3.23$ for cold, $+1.39$ for precipitation, and $+8.55$ for TC. \Tabref{tab:gfs-reference} reports the endpoints. These are differences between formal aggregates, each computed on its reference's own support; they therefore do not receive paired confidence intervals. \Figref{fig:gfs-metric-sensitivity} resolves the reference sensitivity into individual diagnostics. \Figref{fig:gfs-reference} summarizes the aggregate and lead comparisons, paired event-score changes, and benchmark context; only its event-distribution panel uses explicitly paired events.

\input{tables/gfs_reference}

\textbf{The reference effect is diagnostic dependent.}
\Figref{fig:gfs-metric-sensitivity} shows that the aggregate increase is not a uniform upward shift. For extreme heat and cold stress events, the lower-layer stability score (E4) increases by about 10.5 and 8.9 points, respectively, while the midlevel divergence score (U7) decreases by about 8.8 and 9.4 points. Changes for extreme precipitation events also have mixed signs: the MSE score (U10) increases by about 6.5 points, whereas the layer vapor-flux convergence score (E3) decreases by about 6.1 points. All displayed TC diagnostic changes are positive, with particularly large increases in hydrostatic structure (U1; 18.2 points), tangential wind (E1; 17.2 points), and midlevel radial vorticity (E3; 14.1 points). The sensitivity therefore depends on which physical structure is evaluated, in addition to the event category.

\textbf{Aggregate, lead, and event comparisons.}
\Figref{fig:gfs-reference} compares aggregate scores, lead dependence, and paired event changes under the two references. The alternative Overall value is shown beside the main benchmark for context only: different verifying analyses do not define a common cross-system ranking.

\subsection{Sensitivity to Score Mapping and Aggregation}
\label{app:score-sensitivity}

We test tolerances at $0.5$, $1$, and $2$ times their default values, using either equal weights for physical families or equal weights for diagnostics within each group. Scores are recomputed from the original errors using \Eqref{eq:physics-score}, with $\tau_i$ replaced by $a\tau_i$. Errors, tasks, reference fields, and common eligible events are fixed. Scores are mapped before averaging, and all other aggregation rules remain unchanged.

\begin{figure}[!htbp]
 \centering
 \includegraphics[width=\linewidth]{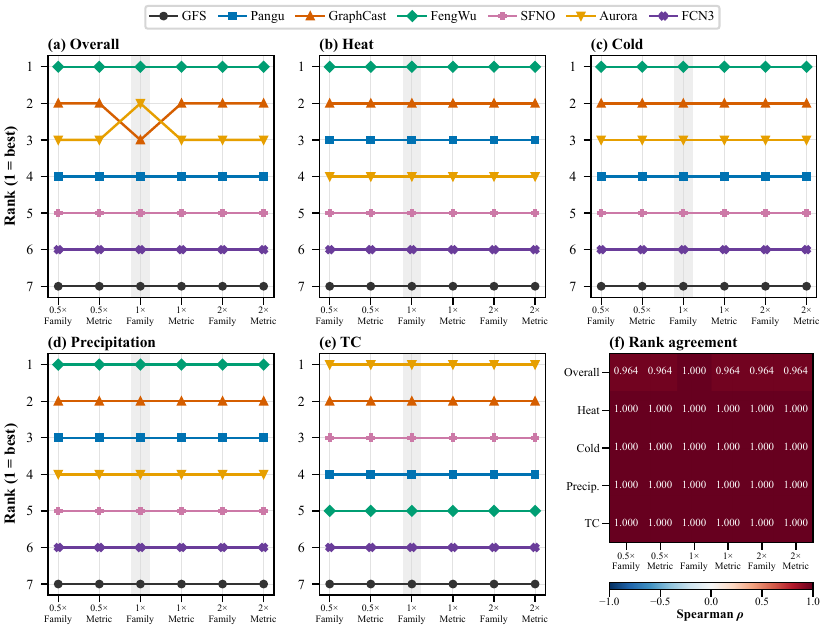}
 \caption{Sensitivity of Physics Score rankings to tolerance and aggregation. Panels (a--e) show Overall and category ranks; (f) shows Spearman correlation with the baseline ($1\times$ tolerance, equal-family weights). All category rankings persist across six tolerance and aggregation settings; only Aurora and GraphCast swap Overall positions.}
 \label{fig:score-mapping-robustness}
\end{figure}

Rankings in all four categories remain unchanged (\Figref{fig:score-mapping-robustness}). FengWu leads Overall, heat, cold, and precipitation; Aurora leads TC. In the five alternative settings, GraphCast and Aurora exchange second and third place Overall, with all other ranks unchanged ($\rho\geq0.964$ relative to the default scheme). All six AI models remain above GFS.

\begin{figure}[!htbp]
 \centering
 \includegraphics[width=\linewidth]{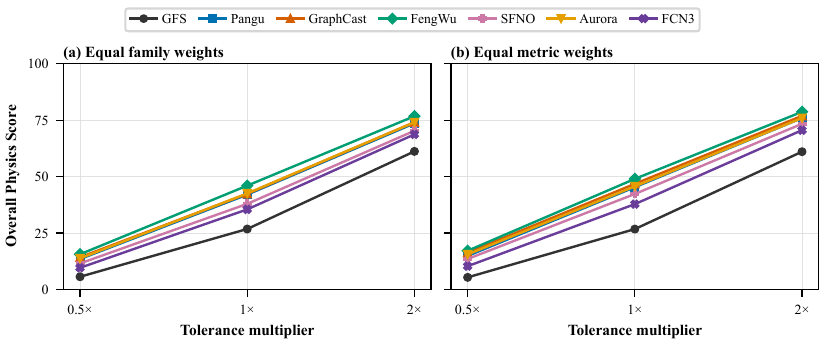}
 \caption{Overall Physics Scores under alternative tolerance and aggregation settings. The left and right panels use equal-family and equal-diagnostic weights, respectively. Larger tolerances raise Overall Physics Scores for the same forecast errors; these changes reflect the mapping, not improved forecasts.}
 \label{fig:score-mapping-absolute}
\end{figure}

Larger tolerances raise absolute scores without changing the forecasts (\Figref{fig:score-mapping-absolute}). Rank stability here concerns the aggregate results across these six settings.

\subsection{Case Studies: Field Errors and Physical Structure}
\label{app:case-studies}

We select two cases per event category from model pairs evaluated on common eligible events (\Tabref{tab:selected-rmse-physics-cases}). Let $r_j$ denote the RMSE ratio A/B for each of the 15 fields $\{z,t,q,u,v\}\times\{850,700,500\}$, where A has the higher Event Physics Score under the default scheme. Ratios compare the same variable and event, avoiding averages of errors with different units.

Pairs satisfy one of three rules: (i) all ratios are within a factor of 1.10 and the score gap is at least five points; (ii) $\exp[\frac{1}{15}\sum_j|\log r_j|]-1\leq0.10$, at least 12 ratios are within a factor of 1.20, and the gap is at least eight points; or (iii) the geometric mean ratio is at least 1.10, at least ten ratios exceed one, and the gap is at least eight points. We prioritize comparisons between AI models and select distinct events with larger Physics Score differences. Each pair must keep the same score ordering across all six scoring settings. We include comparisons with GFS when no eligible AI model pair is available for an unused event. These comparisons retain differences in initialization and verification reference.

\input{tables/selected_cases}

The cases show that similar or lower field RMSE need not imply better physical structure agreement. TC MEARI illustrates similar field errors but different Physics Scores; heat event 198 shows higher physical scores despite larger field errors. Physical family scores and diagnostic fields reveal differences in frontogenesis, moisture convergence, circulation, and outflow. These selected examples do not estimate prevalence or imply improvement in every diagnostic.
All eight figures compare field RMSE ratios in panel (a), physical family scores in panel (b), and diagnostic maps or profiles in the remaining panels.

\clearpage
\begin{figure}[!htbp]
  \centering
  \includegraphics[width=\linewidth,height=0.73\textheight,keepaspectratio]{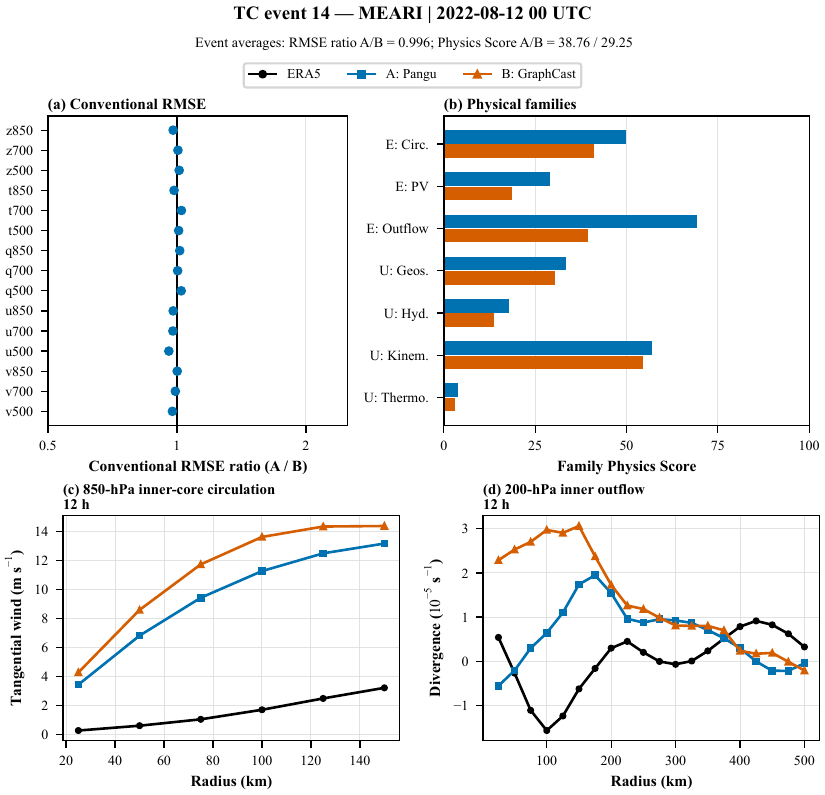}
  \caption{TC MEARI: Pangu (A) and GraphCast (B). (a) Field RMSE ratios A/B. (b) Physical family comparison. (c) Tangential wind at 850 hPa and (d) outflow at 200 hPa in ERA5 and forecasts, shown at 12 h relative to each storm's own center. Pangu has an Event Physics Score 9.51 points higher than GraphCast despite nearly equal conventional errors (geometric mean RMSE ratio 0.996).}
  \label{fig:case-tropical-cyclone-0014-Pangu-vs-GraphCast}
\end{figure}

\clearpage
\begin{figure}[!htbp]
  \centering
  \includegraphics[width=\linewidth,height=0.73\textheight,keepaspectratio]{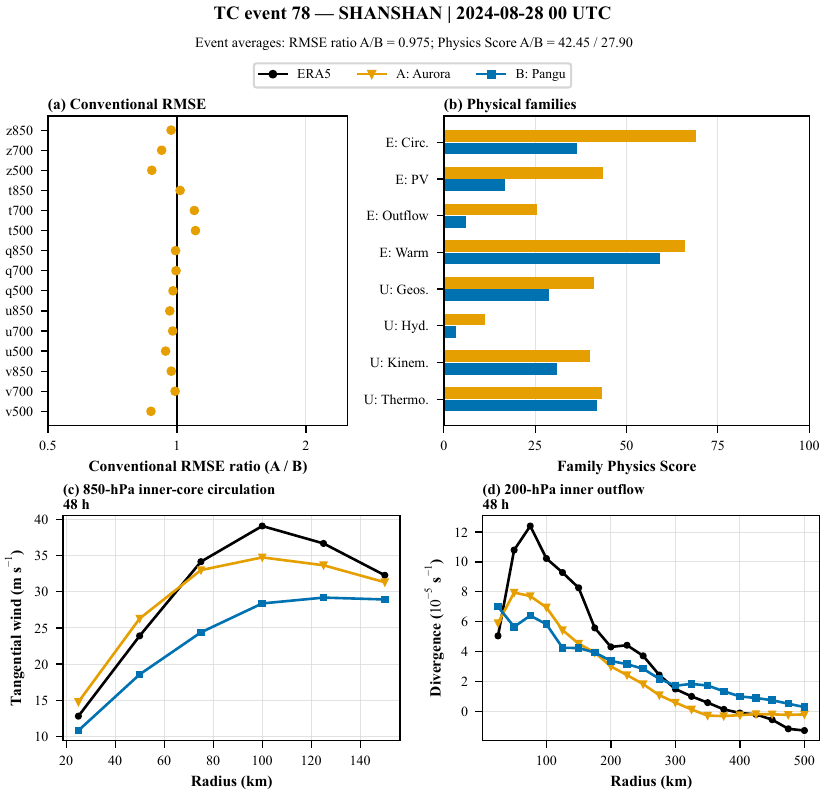}
  \caption{TC SHANSHAN: Aurora (A) and Pangu (B). (a) Field RMSE ratios A/B. (b) Physical family comparison. (c) Tangential wind at 850 hPa and (d) outflow at 200 hPa in ERA5 and forecasts, shown at 48 h relative to each storm's own center. Aurora has an Event Physics Score 14.55 points higher than Pangu with similar conventional errors (geometric mean RMSE ratio 0.975).}
  \label{fig:case-tropical-cyclone-0078-Aurora-vs-Pangu}
\end{figure}

\clearpage
\begin{figure}[!htbp]
  \centering
  \includegraphics[width=\linewidth,height=0.73\textheight,keepaspectratio]{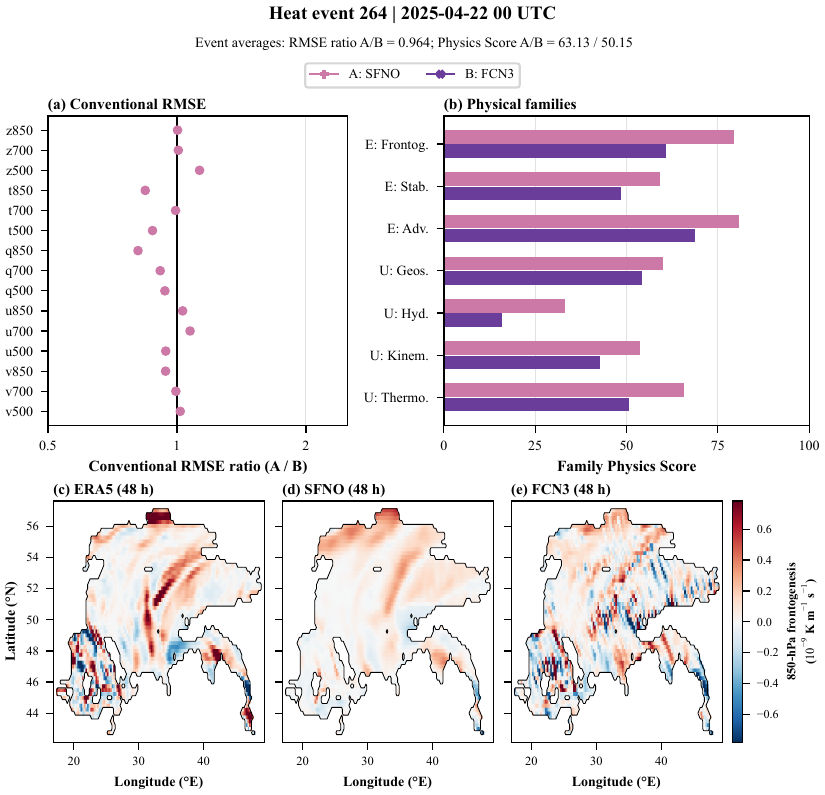}
  \caption{Heat event 264: SFNO (A) and FCN3 (B). (a) Field RMSE ratios A/B. (b) Physical family comparison. (c--e) Temperature frontogenesis at 850 hPa in ERA5, SFNO, and FCN3 at 48 h. Contours mark the event region. SFNO has an Event Physics Score 12.98 points higher than FCN3 with similar conventional errors (geometric mean RMSE ratio 0.964).}
  \label{fig:case-heat-0264-SFNO-vs-FCN3}
\end{figure}

\clearpage
\begin{figure}[!htbp]
  \centering
  \includegraphics[width=\linewidth,height=0.73\textheight,keepaspectratio]{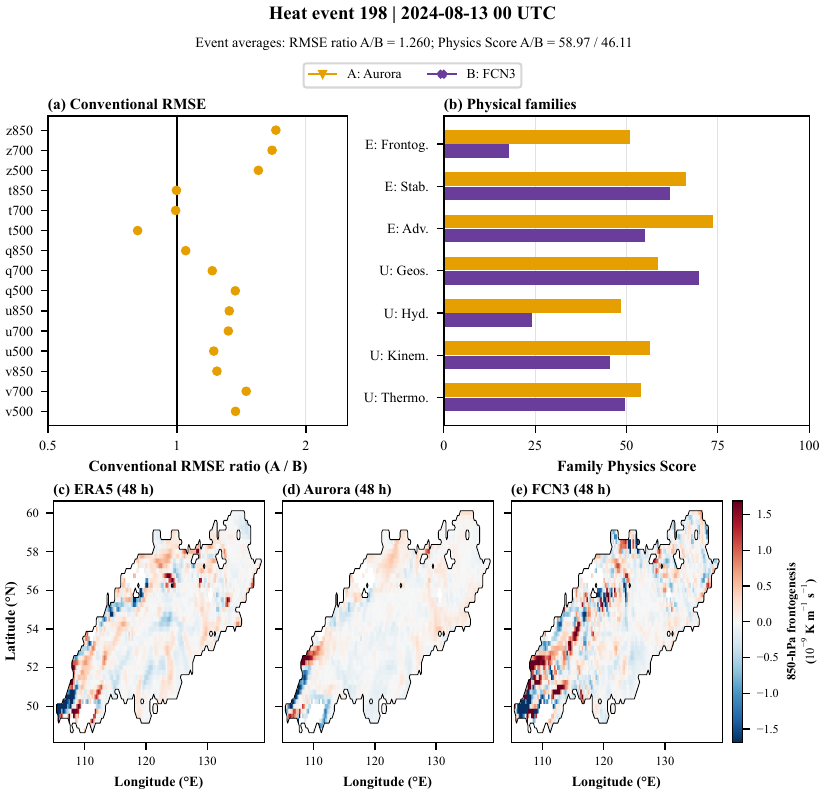}
  \caption{Heat event 198: Aurora (A) and FCN3 (B). (a) Field RMSE ratios A/B. (b) Physical family comparison. (c--e) Temperature frontogenesis at 850 hPa in ERA5, Aurora, and FCN3 at 48 h. Contours mark the event region. Aurora has an Event Physics Score 12.86 points higher than FCN3 despite larger conventional errors (geometric mean RMSE ratio 1.260).}
  \label{fig:case-heat-0198-Aurora-vs-FCN3}
\end{figure}

\clearpage
\begin{figure}[!htbp]
  \centering
  \includegraphics[width=\linewidth,height=0.73\textheight,keepaspectratio]{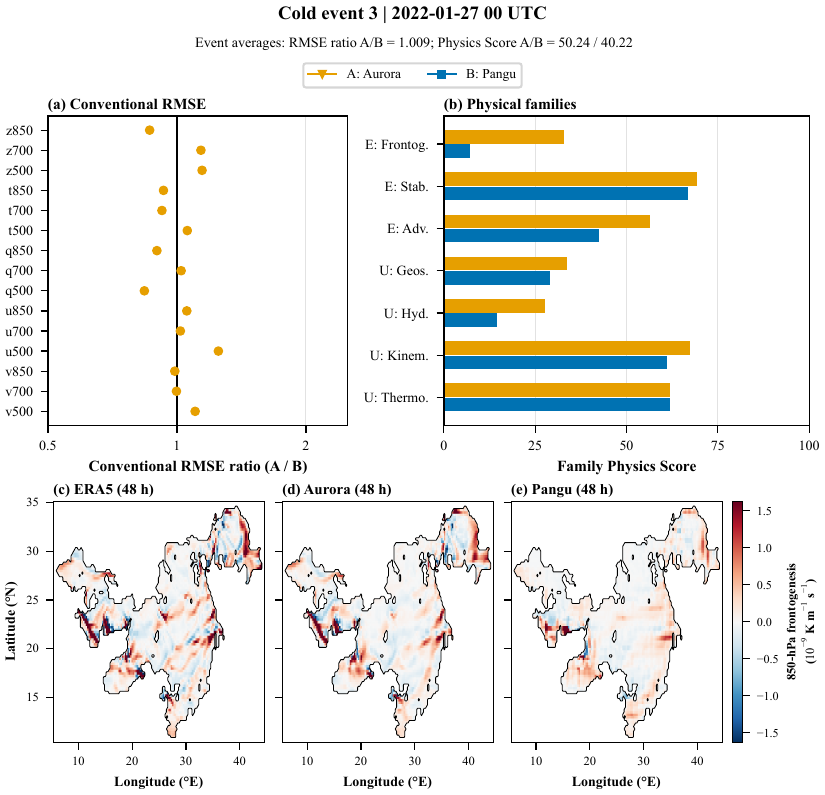}
  \caption{Cold event 3: Aurora (A) and Pangu (B). (a) Field RMSE ratios A/B. (b) Physical family comparison. (c--e) Temperature frontogenesis at 850 hPa in ERA5, Aurora, and Pangu at 48 h. Contours mark the event region. Aurora has an Event Physics Score 10.02 points higher than Pangu despite nearly equal conventional errors (geometric mean RMSE ratio 1.009).}
  \label{fig:case-cold-0003-Aurora-vs-Pangu}
\end{figure}

\clearpage
\begin{figure}[!htbp]
  \centering
  \includegraphics[width=\linewidth,height=0.73\textheight,keepaspectratio]{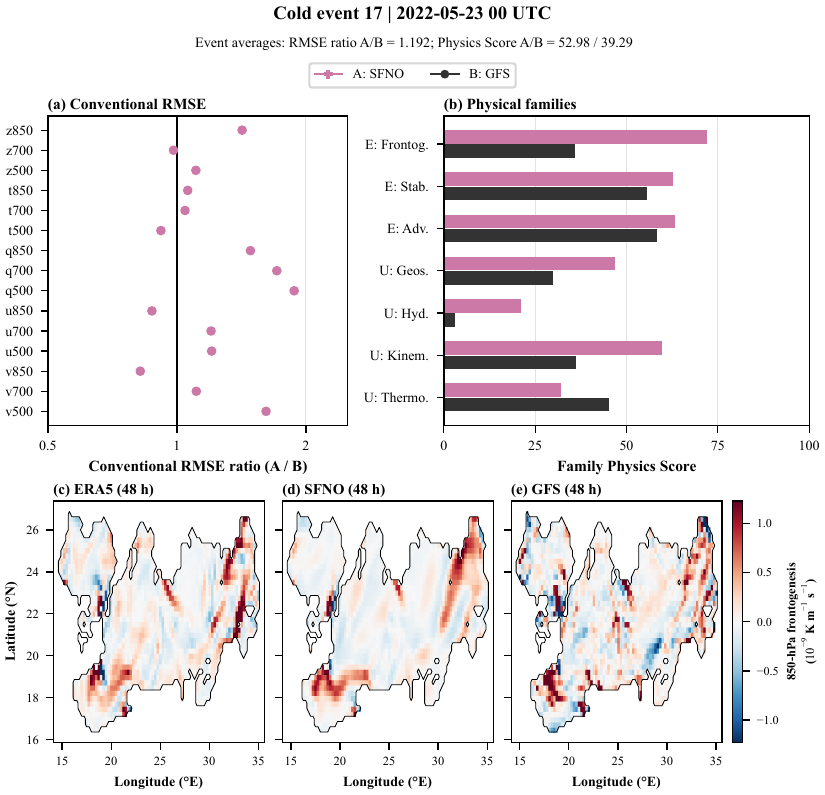}
  \caption{Cold event 17: SFNO (A) and GFS (B). (a) Field RMSE ratios A/B. (b) Physical family comparison. (c--e) Temperature frontogenesis at 850 hPa in ERA5, SFNO, and GFS at 48 h. Contours mark the event region. SFNO has an Event Physics Score 13.69 points higher than GFS despite larger conventional errors (geometric mean RMSE ratio 1.192).}
  \label{fig:case-cold-0017-SFNO-vs-GFS}
\end{figure}

\clearpage
\begin{figure}[!htbp]
  \centering
  \includegraphics[width=\linewidth,height=0.73\textheight,keepaspectratio]{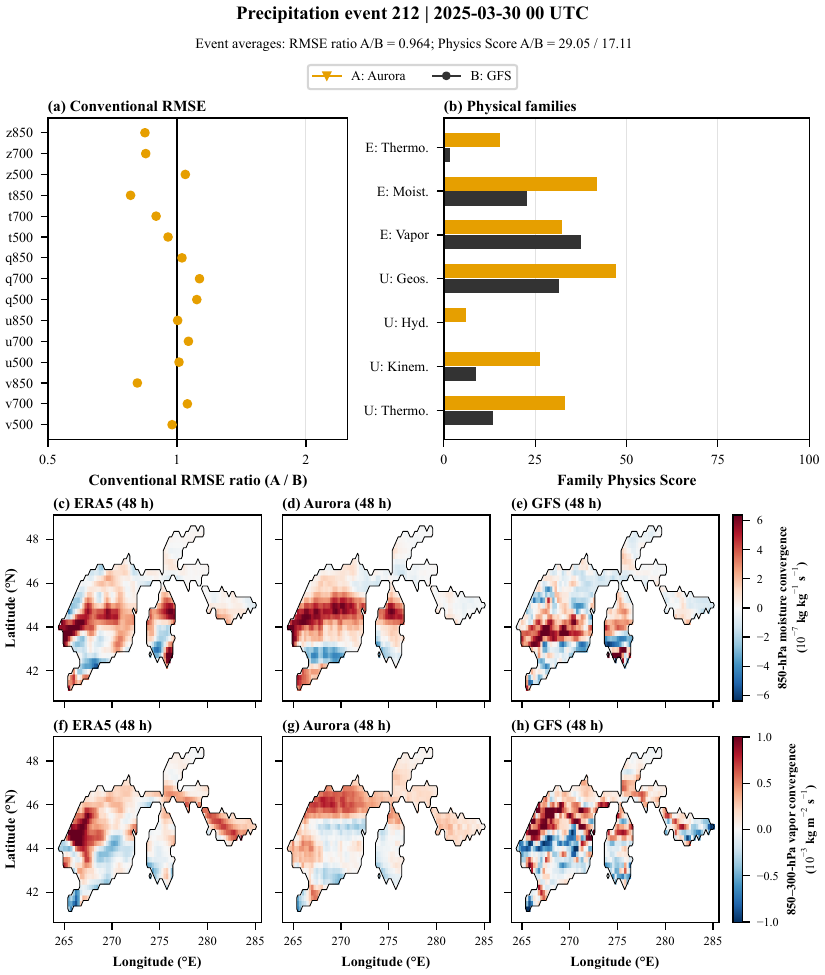}
  \caption{Precipitation event 212: Aurora (A) and GFS (B). (a) Field RMSE ratios A/B. (b) Physical family comparison. (c--e) Moisture flux convergence at 850 hPa and (f--h) vapor flux convergence over 850--300 hPa, comparing ERA5, Aurora, and GFS at 48 h. Contours mark the event region. Aurora has an Event Physics Score 11.94 points higher than GFS with similar conventional errors (geometric mean RMSE ratio 0.964).}
  \label{fig:case-precipitation-0212-Aurora-vs-GFS}
\end{figure}

\clearpage
\begin{figure}[!htbp]
  \centering
  \includegraphics[width=\linewidth,height=0.73\textheight,keepaspectratio]{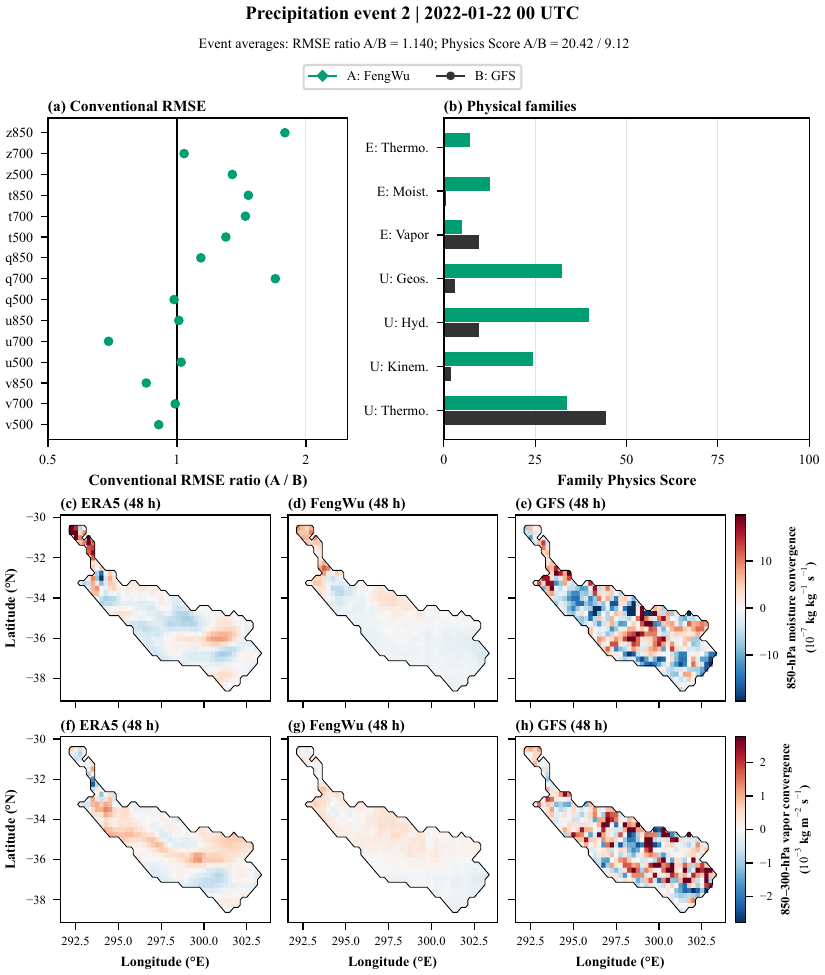}
  \caption{Precipitation event 2: FengWu (A) and GFS (B). (a) Field RMSE ratios A/B. (b) Physical family comparison. (c--e) Moisture flux convergence at 850 hPa and (f--h) vapor flux convergence over 850--300 hPa, comparing ERA5, FengWu, and GFS at 48 h. Contours mark the event region. FengWu has an Event Physics Score 11.30 points higher than GFS despite larger conventional errors (geometric mean RMSE ratio 1.140).}
  \label{fig:case-precipitation-0002-FengWu-vs-GFS}
\end{figure}

%% file: tables/lead_endpoint_changes.tex
\begin{table}[!htbp]
  \centering
  \begin{threeparttable}
    \caption{Physics Score changes from 24 to 120 h by system and evaluation scope. Values report endpoint changes per 24 h, not fitted rates or absolute forecast skill. All systems lose score over this interval.}
    \label{tab:lead-endpoint-changes}
    \scriptsize
    \begin{tabular*}{0.9\linewidth}{@{\extracolsep{\fill}}l *{5}{r}@{}}
      \toprule
      \textbf{Model} & {\textbf{Overall}} & {\textbf{Heat}} & {\textbf{Cold}} &
      {\textbf{Precipitation}} & {\textbf{TC}} \\
      \midrule

      FengWu & \TableScore{-5.18}{0} & \TableScore{-3.06}{0} & \TableScore{-5.17}{0} & \TableScore{-5.80}{0} & \TableScore{-7.79}{0} \\
      Aurora & \TableScore{-6.45}{0} & \TableScore{-4.64}{0} & \TableScore{-7.00}{0} & \TableScore{-7.35}{0} & \TableScore{-6.39}{0} \\
      GraphCast & \TableScore{-5.30}{0} & \TableScore{-3.74}{0} & \TableScore{-5.82}{0} & \TableScore{-6.37}{0} & \TableScore{-4.89}{0} \\
      Pangu & \TableScore{-4.23}{0} & \TableScore{-3.18}{0} & \TableScore{-4.49}{0} & \TableScore{-5.51}{0} & \TableScore{-3.51}{0} \\
      SFNO & \TableScore{-5.89}{0} & \TableScore{-4.54}{0} & \TableScore{-6.19}{0} & \TableScore{-6.72}{0} & \TableScore{-6.35}{0} \\
      FCN3 & \TableScore{-5.61}{0} & \TableScore{-4.20}{0} & \TableScore{-6.13}{0} & \TableScore{-6.07}{0} & \TableScore{-5.64}{0} \\
      GFS & \TableScore{-3.50}{1} & \TableScore{-3.06}{1} & \TableScore{-3.85}{1} & \TableScore{-4.09}{1} & \TableScore{-3.01}{1} \\
      \bottomrule
    \end{tabular*}
    \begin{tablenotes}[flushleft]
      \TableNoteFont
      \item Each entry is $[S(120)-S(24)]/4$; bold marks the largest (least negative) value in each column. Values are descriptive endpoint changes, not fitted slopes.
    \end{tablenotes}
  \end{threeparttable}
\end{table}

%% file: tables/gfs_reference.tex
\begin{table}[!htbp]
  \centering
  \begin{threeparttable}
    \caption{Sensitivity of formal GFS Physics Scores to the verifying analysis. Scores use each reference's own admissible support. Replacing ERA5 with GFS 0-h raises scores.}
    \label{tab:gfs-reference}
    \scriptsize
    \begin{tabular*}{0.78\linewidth}{@{\extracolsep{\fill}}l *{3}{r}@{}}
      \toprule
      \textbf{Scope} & {\textbf{ERA5}} & {\textbf{GFS 0-h}} & {\textbf{Difference}} \\
      \midrule
      Overall & {\TableNum{26.82}} & {\TableNum{30.33}} & {\TableSignedNum{3.51}} \\
      Heat & {\TableNum{35.22}} & {\TableNum{38.05}} & {\TableSignedNum{2.83}} \\
      Cold & {\TableNum{22.86}} & {\TableNum{26.08}} & {\TableSignedNum{3.23}} \\
      Precipitation & {\TableNum{18.18}} & {\TableNum{19.56}} & {\TableSignedNum{1.39}} \\
      TC & {\TableNum{41.66}} & {\TableNum{50.21}} & {\TableSignedNum{8.55}} \\
      \bottomrule
    \end{tabular*}
  \end{threeparttable}
\end{table}

%% file: tables/selected_cases.tex
\begin{table}[!htbp]
  \centering
  \begin{threeparttable}
    \caption{Selected event-level comparisons of conventional RMSE and Event Physics Scores. Columns report model pairs, geometric mean RMSE ratios, baseline scores, and minimum score gaps across six tolerance and aggregation settings. Model A retains higher scores in all settings despite similar or larger conventional errors.}
    \label{tab:selected-rmse-physics-cases}
    \scriptsize
    \setlength{\tabcolsep}{3pt}
    \begin{tabular*}{\linewidth}{@{\extracolsep{\fill}}llrrr@{}}
      \toprule
      \textbf{Case} & \textbf{Models A / B} &
      \shortstack{\textbf{RMSE ratio (A / B)}} &
      \shortstack{\textbf{Physics Score (A / B)}} &
      \shortstack{\textbf{Min. gap (six settings)}} \\
      \midrule
      TC 14 & Pangu / GraphCast & 0.996 & 38.76 / 29.25 & $+3.71$ \\
      TC 78 & Aurora / Pangu & 0.975 & 42.45 / 27.90 & $+9.15$ \\
      \addlinespace[2pt]
      Heat 264 & SFNO / FCN3 & 0.964 & 63.13 / 50.15 & $+5.73$ \\
      Heat 198 & Aurora / FCN3 & 1.260 & 58.97 / 46.11 & $+5.31$ \\
      \addlinespace[2pt]
      Cold 3 & Aurora / Pangu & 1.009 & 50.24 / 40.22 & $+4.53$ \\
      Cold 17 & SFNO / GFS & 1.192 & 52.98 / 39.29 & $+6.28$ \\
      \addlinespace[2pt]
      Precipitation 212 & Aurora / GFS & 0.964 & 29.05 / 17.11 & $+4.69$ \\
      Precipitation 2 & FengWu / GFS & 1.140 & 20.42 / 9.12 & $+0.99$ \\
      \bottomrule
    \end{tabular*}
    \begin{tablenotes}[flushleft]
      \scriptsize
      \item A denotes the model with the higher baseline Event Physics Score. The RMSE ratio is the geometric mean of 15 dimensionless, within-event variable-wise RMSE ratios; values above one indicate larger conventional errors for A. Scores use the baseline tolerance and equal-family aggregation. The minimum gap is the smallest score difference (A minus B) across the six tolerance/aggregation settings. In these selected cases, model A scores higher than model B under all six scoring settings.
    \end{tablenotes}
  \end{threeparttable}
\end{table}

%% file: ExtremePhysBench.bib
@misc{ipcc2021extremes,
  author={{IPCC}},
  title={Weather and Climate Extreme Events in a Changing Climate},
  year={2021},
  howpublished={Climate Change 2021: The Physical Science Basis, Chapter 11},
  url={https://www.ipcc.ch/report/ar6/wg1/chapter/chapter-11/}
}

@article{bi2023pangu,
  title={Accurate medium-range global weather forecasting with {3D} neural networks},
  author={Bi, Kaifeng and Xie, Lingxi and Zhang, Hengheng and Chen, Xin and Gu, Xiaotao and Tian, Qi},
  journal={Nature},
  volume={619},
  pages={533--538},
  year={2023},
  doi={10.1038/s41586-023-06185-3}
}

@article{lam2023graphcast,
  title={Learning skillful medium-range global weather forecasting},
  author={Lam, Remi and Sanchez-Gonzalez, Alvaro and Willson, Matthew and Wirnsberger, Peter and Fortunato, Meire and Alet, Ferran and Ravuri, Suman and Ewalds, Timo and Eaton-Rosen, Zach and Hu, Weihua and others},
  journal={Science},
  volume={382},
  number={6677},
  pages={1416--1421},
  year={2023},
  doi={10.1126/science.adi2336}
}

@article{bodnar2025aurora,
  title={A foundation model for the {Earth} system},
  author={Bodnar, Cristian and Bruinsma, Wessel P. and Lucic, Ana and Stanley, Megan and Allen, Anna and Brandstetter, Johannes and Garvan, Patrick and Riechert, Maik and Weyn, Jonathan A. and Dong, Haiyu and others},
  journal={Nature},
  volume={641},
  pages={1180--1187},
  year={2025},
  doi={10.1038/s41586-025-09005-y}
}

@inproceedings{bonev2023sfno,
  title={Spherical {Fourier} Neural Operators: Learning Stable Dynamics on the Sphere},
  author={Bonev, Boris and Kurth, Thorsten and Hundt, Christian and Pathak, Jaideep and Baust, Maximilian and Kashinath, Karthik and Anandkumar, Anima},
  booktitle={Proceedings of the 40th International Conference on Machine Learning},
  series={Proceedings of Machine Learning Research},
  volume={202},
  pages={2806--2823},
  publisher={PMLR},
  url={https://proceedings.mlr.press/v202/bonev23a.html},
  year={2023}
}

@misc{nvidia2024sfnocheckpoint,
  author={{NVIDIA}},
  title={Modulus Checkpoints: Spherical {Fourier} Neural Operator},
  howpublished={NVIDIA NGC model card, version 0.1.0},
  url={https://catalog.ngc.nvidia.com/orgs/nvidia/modulus/models/sfno_73ch_small},
  year={2024},
  note={Accessed 2026-08-10}
}

@misc{nvidia2025fcn3checkpoint,
  author={{NVIDIA}},
  title={{Earth-2} Checkpoints: {FourCastNet 3}},
  howpublished={Hugging Face model card},
  url={https://huggingface.co/nvidia/fourcastnet3/tree/76ef0c60237e458b33196ba027134e27f3fc4538},
  year={2025},
  note={Accessed 2026-08-10}
}

@misc{deepmind2023graphcastrepo,
  author={{Google DeepMind}},
  title={{GraphCast}: Learning Skillful Medium-Range Global Weather Forecasting},
  howpublished={GitHub repository, release v0.1},
  url={https://github.com/google-deepmind/weathernext/tree/858301cde5de5c728f8172f782dafba1ea07ac2e},
  year={2023},
  note={Version 0.1, released 2023-10-31; accessed 2026-08-10}
}

@article{chen2023fengwu,
  title={The Operational Medium-Range Deterministic Weather Forecasting Can Be Extended beyond a 10-Day Lead Time},
  author={Chen, Kang and Han, Tao and Ling, Fenghua and Gong, Junchao and Bai, Lei and Wang, Xinyu and Luo, Jing-Jia and Fei, Ben and Zhang, Wenlong and Chen, Xi and Ma, Leiming and Zhang, Tianning and Su, Rui and Ci, Yuanzheng and Li, Bin and Yang, Xiaokang and Ouyang, Wanli},
  journal={Communications Earth \& Environment},
  volume={6},
  number={1},
  pages={518},
  year={2025},
  doi={10.1038/s43247-025-02502-y}
}

@misc{bonev2025fcn3,
  title={{FourCastNet 3}: A Geometric Approach to Probabilistic Machine-Learning Weather Forecasting at Scale},
  author={Bonev, Boris and Kurth, Thorsten and Mahesh, Ankur and Bisson, Mauro and Kossaifi, Jean and Kashinath, Karthik and Anandkumar, Anima and Collins, William D. and Pritchard, Michael S. and Keller, Alexander},
  howpublished={arXiv preprint arXiv:2507.12144},
  url={https://arxiv.org/abs/2507.12144},
  year={2025}
}

@article{rasp2020weatherbench,
  title={{WeatherBench}: A Benchmark Data Set for Data-Driven Weather Forecasting},
  author={Rasp, Stephan and Dueben, Peter D. and Scher, Sebastian and Weyn, Jonathan A. and Mouatadid, Soukayna and Thuerey, Nils},
  journal={Journal of Advances in Modeling Earth Systems},
  volume={12},
  number={11},
  pages={e2020MS002203},
  year={2020},
  doi={10.1029/2020MS002203}
}

@article{rasp2024weatherbench2,
  title={{WeatherBench 2}: A Benchmark for the Next Generation of Data-Driven Global Weather Models},
  author={Rasp, Stephan and Hoyer, Stephan and Merose, Alexander and Langmore, Ian and Battaglia, Peter and Russell, Tyler and Sanchez-Gonzalez, Alvaro and Yang, Vivian and Carver, Rob and Agrawal, Shreya and Chantry, Matthew and Ben Bouallegue, Zied and Dueben, Peter and Bromberg, Carla and Sisk, Jared and Barrington, Luke and Bell, Aaron and Sha, Fei},
  journal={Journal of Advances in Modeling Earth Systems},
  volume={16},
  number={6},
  pages={e2023MS004019},
  year={2024},
  doi={10.1029/2023MS004019}
}

@misc{weatherbenchx2025,
  author={{Google LLC}},
  title={{WeatherBench-X}: A Modular Framework for Evaluating Weather Forecasts},
  howpublished={GitHub software release},
  url={https://github.com/google-research/weatherbenchX/releases/tag/v2025.02.1},
  year={2025},
  note={Version v2025.02.1; successor to the WeatherBench 2 evaluation code; accessed 2026-08-18}
}

@inproceedings{nguyen2023climatelearn,
  title={{ClimateLearn}: Benchmarking Machine Learning for Weather and Climate Modeling},
  author={Nguyen, Tung and Jewik, Jason and Bansal, Hritik and Sharma, Prakhar and Grover, Aditya},
  booktitle={Advances in Neural Information Processing Systems},
  volume={36},
  pages={75009--75025},
  publisher={Curran Associates, Inc.},
  year={2023},
  doi={10.52202/075280-3279},
  url={https://proceedings.neurips.cc/paper_files/paper/2023/file/ed73c36e771881b232ef35fa3a1dec14-Paper-Datasets_and_Benchmarks.pdf}
}

@article{pasche2025extremes,
  title={Validating Deep Learning Weather Forecast Models on Recent High-Impact Extreme Events},
  author={Pasche, Olivier C. and Wider, Jonathan and Zhang, Zhongwei and Zscheischler, Jakob and Engelke, Sebastian},
  journal={Artificial Intelligence for the Earth Systems},
  volume={4},
  number={1},
  pages={e240033},
  year={2025},
  doi={10.1175/AIES-D-24-0033.1}
}

@misc{mcgovern2026extremeweatherbench,
  title={Extreme Weather Bench: A Framework and Benchmark for Evaluation of High-Impact Weather},
  author={McGovern, Amy and Mandelbaum, Taylor and Rothenberg, Daniel and Loveday, Nicholas and Potvin, Corey and Flora, Montgomery and Magnusson, Linus and Gilleland, Eric and Allen, John},
  howpublished={arXiv preprint arXiv:2605.01126},
  url={https://arxiv.org/abs/2605.01126},
  year={2026}
}

@misc{li2026realbench,
  title={{RealBench}: Benchmarking Data-Driven Numerical Weather Forecasting Under Operational Conditions and Extreme Event Challenges},
  author={Li, Ruize and Wen, Zhibin and Han, Tao and Chen, Hao and Ling, Fenghua and Zhang, Wei and Guo, Song and Bai, Lei},
  howpublished={arXiv preprint arXiv:2605.24945},
  url={https://arxiv.org/abs/2605.24945},
  year={2026}
}

@misc{gomez2026tcbench,
  title={{TCBench}: A Benchmark for Tropical Cyclone Track and Intensity Forecasting at the Global Scale},
  author={Gomez, Milton and McGraw, Marie and Ganesh S., Saranya and Tam, Frederick Iat-Hin and Azizi, Ilia and Darmon, Samuel and Feldmann, Monika and Bourdin, Stella and Poulain--Auz{\'e}au, Louis and Camargo, Suzana J. and Lin, Jonathan and Chavas, Dan and Lee, Chia-Ying and Gupta, Ritwik and Jenney, Andrea and Beucler, Tom},
  howpublished={arXiv preprint arXiv:2601.23268},
  url={https://arxiv.org/abs/2601.23268},
  year={2026}
}

@article{hakim2024dynamical,
  title={Dynamical Tests of a Deep Learning Weather Prediction Model},
  author={Hakim, Gregory J. and Masanam, Sanjit},
  journal={Artificial Intelligence for the Earth Systems},
  volume={3},
  number={3},
  pages={e230090},
  year={2024},
  doi={10.1175/AIES-D-23-0090.1}
}

@article{selz2023butterfly,
  title={Can Artificial Intelligence-Based Weather Prediction Models Simulate the Butterfly Effect?},
  author={Selz, Tobias and Craig, George C.},
  journal={Geophysical Research Letters},
  volume={50},
  number={20},
  pages={e2023GL105747},
  year={2023},
  doi={10.1029/2023GL105747}
}

@article{bonavita2024limitations,
  title={On Some Limitations of Current Machine Learning Weather Prediction Models},
  author={Bonavita, Massimo},
  journal={Geophysical Research Letters},
  volume={51},
  number={12},
  pages={e2023GL107377},
  year={2024},
  doi={10.1029/2023GL107377}
}

@misc{subramaniam2025hydrostatic,
  title={Imposing the Fundamental Dynamical Constraint of Hydrostatic Balance to Improve Global {ML} Weather Prediction},
  author={Subramaniam, Akshay and Durran, Dale and Pruitt, David and Cresswell-Clay, Nathaniel and Yik, William},
  howpublished={arXiv preprint arXiv:2506.08285},
  url={https://arxiv.org/abs/2506.08285},
  year={2025}
}

@misc{kasteleyn2026physmetrics,
  title={{PhysMetrics.Weather}: An Evaluation Framework for Physical Consistency in {ML} Weather Models},
  author={Kasteleyn, Emma and Maier, Timo and Lauer, Axel and Eyring, Veronika and Gentine, Pierre and Lucic, Ana},
  howpublished={arXiv preprint arXiv:2606.10642},
  url={https://arxiv.org/abs/2606.10642},
  year={2026}
}

@article{hersbach2020era5,
  title={The {ERA5} global reanalysis},
  author={Hersbach, Hans and Bell, Bill and Berrisford, Paul and Hirahara, Shoji and Hor{\'a}nyi, Andr{\'a}s and Mu{\~n}oz-Sabater, Joaqu{\'i}n and Nicolas, Julien and Peubey, Carole and Radu, Raluca and Schepers, Dinand and others},
  journal={Quarterly Journal of the Royal Meteorological Society},
  volume={146},
  number={730},
  pages={1999--2049},
  year={2020},
  doi={10.1002/qj.3803}
}

@misc{ncepGFS,
  author={{NCEP, NOAA}},
  title={The Global Forecast System ({GFS})},
  howpublished={\url{https://www.emc.ncep.noaa.gov/emc/pages/numerical_forecast_systems/gfs.php}},
  note={Accessed 2026-08-11},
  year={{n.d.}}
}

@article{dinapoli2021era5heat,
  title={{ERA5-HEAT}: A global gridded historical dataset of human thermal comfort indices from climate reanalysis},
  author={Di Napoli, Claudia and Barnard, Christopher and Prudhomme, Christel and Cloke, Hannah L. and Pappenberger, Florian},
  journal={Geoscience Data Journal},
  volume={8},
  number={1},
  pages={2--10},
  year={2021},
  doi={10.1002/gdj3.102}
}

@incollection{huffman2020imerg,
  title={Integrated Multi-satellite Retrievals for the Global Precipitation Measurement ({GPM}) Mission ({IMERG})},
  author={Huffman, George J. and Bolvin, David T. and Braithwaite, Dan and Hsu, Kuo-Lin and Joyce, Robert J. and Kidd, Christopher and Nelkin, Eric J. and Sorooshian, Soroosh and Stocker, Erich F. and Tan, Jackson and Wolff, David B. and Xie, Pingping},
  booktitle={Satellite Precipitation Measurement},
  editor={Levizzani, Vincenzo and Kidd, Christopher and Kirschbaum, Dalia B. and Kummerow, Christian D. and Nakamura, Kenji and Turk, F. Joseph},
  series={Advances in Global Change Research},
  volume={67},
  pages={343--353},
  publisher={Springer International Publishing},
  year={2020},
  doi={10.1007/978-3-030-24568-9_19}
}

@misc{huffman2023imergdaily,
  author={Huffman, George J. and Stocker, Erich F. and Bolvin, David T. and Nelkin, Eric J. and Tan, Jackson},
  title={{GPM IMERG} Final Precipitation {L3} 1 Day 0.1 Degree x 0.1 Degree {V07}},
  year={2023},
  publisher={Goddard Earth Sciences Data and Information Services Center},
  doi={10.5067/GPM/IMERGDF/DAY/07},
  note={Dataset}
}

@article{knapp2010ibtracs,
  title={The International Best Track Archive for Climate Stewardship ({IBTrACS}): Unifying tropical cyclone data},
  author={Knapp, Kenneth R. and Kruk, Michael C. and Levinson, David H. and Diamond, Howard J. and Neumann, Charles J.},
  journal={Bulletin of the American Meteorological Society},
  volume={91},
  number={3},
  pages={363--376},
  year={2010},
  doi={10.1175/2009BAMS2755.1}
}

@misc{gahtan2024ibtracs,
  author={Gahtan, Jennifer and Knapp, Kenneth R. and Schreck, III, Carl J. and Diamond, Howard J. and Kossin, James P. and Kruk, Michael C.},
  title={International Best Track Archive for Climate Stewardship ({IBTrACS}) Project, Version 4r01},
  year={2024},
  publisher={NOAA National Centers for Environmental Information},
  doi={10.25921/82ty-9e16},
  note={Dataset}
}

@misc{earth2studio2024,
  author={{Earth2Studio Contributors}},
  title={{NVIDIA Earth2Studio}},
  howpublished={\url{https://github.com/NVIDIA/earth2studio}},
  year={2024},
  note={Open-source AI weather and climate workflow toolkit; repository tag 0.14.0 used in this work}
}

@misc{earth2studio014docs,
  author={{NVIDIA}},
  title={{earth2studio.models.px}: Prognostics --- {Earth2Studio}},
  howpublished={Versioned software documentation, version 0.14.0},
  url={https://nvidia.github.io/earth2studio/v/0.14.0/modules/models_px.html},
  year={2026},
  note={Accessed 2026-08-10}
}

@article{ullrich2021tempestextremes,
  title={{TempestExtremes} v2.1: A Community Framework for Feature Detection, Tracking, and Analysis in Large Datasets},
  author={Ullrich, Paul A. and Zarzycki, Colin M. and McClenny, Elizabeth E. and Pinheiro, Marielle C. and Stansfield, Alyssa M. and Reed, Kevin A.},
  journal={Geoscientific Model Development},
  volume={14},
  number={8},
  pages={5023--5048},
  year={2021},
  doi={10.5194/gmd-14-5023-2021}
}

@misc{carver2023arcoera5,
  title={{ARCO-ERA5}: An Analysis-Ready Cloud-Optimized Reanalysis Dataset},
  author={Carver, Robert W. and Merose, Alex},
  howpublished={Conference abstract and oral presentation at the 22nd Conference on Artificial Intelligence for Environmental Science, 103rd AMS Annual Meeting},
  year={2023},
  note={Denver, Colorado, 2023-01-09, Paper 4A.1},
  url={https://ams.confex.com/ams/103ANNUAL/meetingapp.cgi/Paper/415842}
}

@misc{copernicus2020era5heat,
  author={Di Napoli, Claudia},
  title={Thermal Comfort Indices Derived from {ERA5} Reanalysis},
  year={2020},
  publisher={ECMWF},
  doi={10.24381/cds.553b7518},
  note={Dataset, version 1.1; accessed 2026-08-24}
}

@book{holton2013dynamic,
  title={An Introduction to Dynamic Meteorology},
  author={Holton, James R. and Hakim, Gregory J.},
  edition={5},
  series={International Geophysics},
  volume={88},
  publisher={Academic Press},
  year={2013},
  isbn={978-0-12-384866-6},
  doi={10.1016/C2009-0-63394-8}
}

@article{hoskins1985pv,
  title={On the Use and Significance of Isentropic Potential Vorticity Maps},
  author={Hoskins, Brian J. and McIntyre, Michael E. and Robertson, Andrew W.},
  journal={Quarterly Journal of the Royal Meteorological Society},
  volume={111},
  number={470},
  pages={877--946},
  year={1985},
  doi={10.1002/qj.49711147002}
}

@article{petterssen1936frontogenesis,
  title={Contribution to the Theory of Frontogenesis},
  author={Petterssen, Sverre},
  journal={Geofysiske Publikasjoner},
  volume={11},
  number={6},
  pages={5--27},
  year={1936},
  url={https://geofysikk.org/NGF/GeoPub/NGF_GP_Vol11_no6.pdf}
}

@article{ertel1942vorticity,
  title={Ein neuer hydrodynamischer Wirbelsatz},
  author={Ertel, Hans},
  journal={Meteorologische Zeitschrift},
  volume={59},
  number={9},
  pages={277--281},
  year={1942},
  url={https://empslocal.ex.ac.uk/people/staff/gv219/classics.d/Schubert-Ertel04.pdf},
  note={English translation reproduced as paper 1 in Schubert et al. (2004)}
}

@article{charney1948scale,
  title={On the Scale of Atmospheric Motions},
  author={Charney, Jule G.},
  journal={Geofysiske Publikasjoner},
  volume={17},
  number={2},
  pages={3--17},
  year={1948},
  url={https://geofysikk.org/NGF/GeoPub/NGF_GP_Vol17_no2.pdf}
}

@article{benton1954watervapor,
  title={Water-Vapor Transfer over the North American Continent},
  author={Benton, George S. and Estoque, Mariano A.},
  journal={Journal of Meteorology},
  volume={11},
  number={6},
  pages={462--477},
  year={1954},
  doi={10.1175/1520-0469(1954)011<0462:WVTOTN>2.0.CO;2}
}

@article{neelin1987mse,
  title={Modeling Tropical Convergence Based on the Moist Static Energy Budget},
  author={Neelin, J. David and Held, Isaac M.},
  journal={Monthly Weather Review},
  volume={115},
  number={1},
  pages={3--12},
  year={1987},
  doi={10.1175/1520-0493(1987)115<0003:MTCBOT>2.0.CO;2}
}

@article{willoughby1990gradient,
  title={Gradient Balance in Tropical Cyclones},
  author={Willoughby, Hugh E.},
  journal={Journal of the Atmospheric Sciences},
  volume={47},
  number={2},
  pages={265--274},
  year={1990},
  doi={10.1175/1520-0469(1990)047<0265:GBITC>2.0.CO;2}
}

@article{banacos2005mfc,
  title={The Use of Moisture Flux Convergence in Forecasting Convective Initiation: Historical and Operational Perspectives},
  author={Banacos, Peter C. and Schultz, David M.},
  journal={Weather and Forecasting},
  volume={20},
  number={3},
  pages={351--366},
  year={2005},
  doi={10.1175/WAF858.1}
}

@inproceedings{ran2025hrextreme,
  title={{HR-Extreme}: A High-Resolution Dataset for Extreme Weather Forecasting},
  author={Ran, Nian and Xiao, Peng and Wang, Yue and Shi, Wesley and Lin, Jianxin and Meng, Qi and Allmendinger, Richard},
  booktitle={International Conference on Learning Representations},
  year={2025},
  url={https://arxiv.org/abs/2409.18885v2}
}

@misc{ma2024weatherqa,
  title={{WeatherQA}: Can Multimodal Language Models Reason about Severe Weather?},
  author={Ma, Chengqian and Hua, Zhanxiang and Anderson-Frey, Alexandra and Iyer, Vikram and Liu, Xin and Qin, Lianhui},
  howpublished={arXiv preprint arXiv:2406.11217},
  year={2024},
  url={https://arxiv.org/abs/2406.11217v2}
}

@misc{white2024livebench,
  title={{LiveBench}: A Challenging, Contamination-Limited {LLM} Benchmark},
  author={White, Colin and Dooley, Samuel and Roberts, Manley and Pal, Arka and Feuer, Benjamin and Jain, Siddhartha and Shwartz-Ziv, Ravid and Jain, Neel and Saifullah, Khalid and Dey, Sreemanti and {Shubh-Agrawal} and Sandha, Sandeep Singh and Naidu, Siddartha and Hegde, Chinmay and LeCun, Yann and Goldstein, Tom and Neiswanger, Willie and Goldblum, Micah},
  howpublished={arXiv preprint arXiv:2406.19314},
  year={2024},
  note={Version 2, revised April 2025},
  url={https://arxiv.org/abs/2406.19314v2}
}

@book{oecd2008composite,
  title={Handbook on Constructing Composite Indicators: Methodology and User Guide},
  author={{OECD} and {European Union} and {European Commission Joint Research Centre}},
  publisher={OECD Publishing},
  address={Paris},
  year={2008},
  doi={10.1787/9789264043466-en}
}

@article{liang2023helm,
  title={Holistic Evaluation of Language Models},
  author={Liang, Percy and Bommasani, Rishi and Lee, Tony and Tsipras, Dimitris and Soylu, Dilara and Yasunaga, Michihiro and Zhang, Yian and Narayanan, Deepak and Wu, Yuhuai and Kumar, Ananya and others},
  journal={Transactions on Machine Learning Research},
  year={2023},
  url={https://arxiv.org/abs/2211.09110v2}
}

@article{hodson2021mse,
  title={Mean Squared Error, Deconstructed},
  author={Hodson, Timothy O. and Over, Thomas M. and Foks, Sydney S.},
  journal={Journal of Advances in Modeling Earth Systems},
  volume={13},
  number={12},
  pages={e2021MS002681},
  year={2021},
  doi={10.1029/2021MS002681}
}

@misc{openearthlabFengWuRepo,
  author={{OpenEarthLab}},
  title={{FengWu}: Released Models and Inference Code},
  howpublished={GitHub repository},
  year={{n.d.}},
  url={https://github.com/OpenEarthLab/FengWu},
  note={Accessed 2026-09-26}
}

@article{lavers2022era5precipitation,
  title={An evaluation of {ERA5} precipitation for climate monitoring},
  author={Lavers, David A. and Simmons, Adrian and Vamborg, Freja and Rodwell, Mark J.},
  journal={Quarterly Journal of the Royal Meteorological Society},
  volume={148},
  number={748},
  pages={3152--3165},
  year={2022},
  doi={10.1002/qj.4351}
}

@misc{ecmwf2024era5uncertainty,
  author={{ECMWF}},
  title={{ERA5}: Uncertainty Estimation},
  howpublished={Copernicus Knowledge Base},
  year={2024},
  url={https://confluence.ecmwf.int/spaces/CKB/pages/111158117/ERA5+uncertainty+estimation},
  note={Updated 2024-09-17; accessed 2026-11-13}
}
